\documentclass{article}

\usepackage{PRIMEarxiv}

\usepackage[utf8]{inputenc} 
\usepackage[T1]{fontenc}    
\usepackage{hyperref}       
\usepackage{url}            
\usepackage{booktabs}       
\usepackage{amsfonts}       
\usepackage{amsmath,amssymb} 
\usepackage{threeparttable}
\usepackage{nicefrac}       
\usepackage{microtype}      
\usepackage{lipsum}
\usepackage{fancyhdr}       
\usepackage{graphicx}       

\usepackage[round]{natbib} 
\title{A photochemical model of Triton's atmosphere with an uncertainty propagation study
\thanks{Supplementary material related to this article is available at: \url{https://doi.org/10.13140/RG.2.2.12820.99203}}}

\author{
  \large{B. Benne$^{1,*}$, M. Dobrijevic$^1$, T. Cavalié$^{1,2}$, J-C. Loison$^3$, K. M. Hickson$^3$} \\
  ~ \\
  \scriptsize{$^1$ Laboratoire d’Astrophysique de Bordeaux, Univ. Bordeaux, CNRS, B18N, Allée Geoffroy Saint-Hilaire, 33615 Pessac, France} \\
  \scriptsize{$^2$ LESIA, Observatoire de Paris, Université PSL, CNRS, Sorbonne Université, Univ. Paris Diderot, Sorbonne Paris Cité, 5 place Jules Janssen, 92195 Meudon, France} \\
  \scriptsize{$^3$ Institut des Sciences Moléculaires, CNRS, Univ. Bordeaux, 351 Cours de la Libération, 33400 Talence, France} \\
  \scriptsize{$^*$E-mail adress: benjamin.benne@u-bordeaux.fr} \\
}

\begin{document}
\maketitle

\begin{abstract}
   Triton is the largest satellite of Neptune and probably a Kuiper Belt Object that was captured by the planet. It has a tenuous nitrogen atmosphere similar to the one of Pluto and may be an ocean world. The Neptunian system has only been visited by Voyager 2 in 1989. Over the last few years, the demand for a new mission to the Ice Giants and their systems has increased so that a theoretical basis to prepare for such a mission is important. 

    We aim to develop a photochemical model of Triton’s atmosphere with an up-to-date chemical scheme, as previous photochemical models date back to the post-flyby years. This is done to better understand the mechanisms governing Triton’s atmospheric chemistry and highlight the critical parameters having a significant impact on the atmospheric composition. We also study model uncertainties to find what chemical studies are necessary to improve the modeling of Triton’s atmosphere.

    We adapted a model of Titan’s atmosphere from \citet{dobrijevic_1d-coupled_2016} to Triton’s conditions. We first used Titan’s chemical scheme before updating it to better model Triton’s atmospheric conditions. Once the nominal results were obtained, we studied model uncertainties with a Monte-Carlo procedure, considering the reaction rates as random variables. Then, we performed global sensitivity analyzes to identify the reactions responsible for model uncertainties.

    With the nominal results, we determined the composition of Triton’s atmosphere and studied the production and loss processes for the main atmospheric species. We highlighted key chemical reactions that are the most important for the overall chemistry. We also identified some key parameters having a significant impact on the results. Uncertainties are large for most of the main atmospheric species as the atmospheric temperature is very low. We identified key uncertainty reactions that have the largest impact on the results uncertainties. These reactions must be studied in priority in order to improve the significance of our results by decreasing these uncertainties. 
\end{abstract}


\section{Introduction}
\label{Intro}

Triton is the largest satellite of Neptune. Its orbit is inclined and retrograde, suggesting that it is a former Kuiper Belt Object (KBO) that was captured by Neptune \citep{mckinnon_origin_1995,agnor_neptunes_2006}. This belief is reinforced by the similarities observed with Pluto. 
Triton was visited by Voyager 2 in August 1989, the only mission so far to have studied the Neptunian system. The flyby allowed to observe and characterize the surface ices, composed of N$_2$, CO$_2$, H$_2$O, CH$_4$ and CO \citep{brown_surface_1995,yelle_lower_1995}, and to discover the presence of plumes of organic material and hazes \citep{herbert_ch4_1991,yelle_energy_1991,krasnopolsky_properties_1992,yelle_lower_1995}. 
A study of the atmosphere was performed by occultations and the measurement of its airglow \citep{broadfoot_ultraviolet_1989}. It showed that the atmosphere is mainly composed of N$_2$ and traces of CH$_4$ were found near the surface. The presence of atomic nitrogen and atomic hydrogen was also deduced from these measurements. 
The surface pressure and temperature were determined using the radio data of Voyager \citep{tyler_voyager_1989} and were found to be 16$\pm$3 $\mu$bar and 38\,K respectively. 
It is considered that the atmosphere is formed by sublimation of the surface ices and that it is at vapor pressure equilibrium with those ices \citep{yelle_lower_1995}. CO was not detected during this mission but was observed from Earth \citep{lellouch_detection_2010}. 
A dense ionosphere was also detected with a peak concentration of electrons of about 10$^4$ cm$^{-3}$ \citep{tyler_voyager_1989} and the thermospheric temperature was measured as 95$\pm$5\,K \citep{broadfoot_ultraviolet_1989}. 
A review of the knowledge acquired about Triton during the mission can be found in \citet{cruikshank_neptune_1995}. 

The chemistry in the lower atmosphere is mainly triggered by the photolysis of CH$_4$ by Lyman-$\alpha$ photons coming from solar irradiation and from the interplanetary medium \citep{strobel_photochemistry_1990,herbert_ch4_1991,krasnopolsky_properties_1992,krasnopolsky_temperature_1993,krasnopolsky_photochemistry_1995,strobel_tritons_1995,strobel_comparative_2017}, while at higher altitudes it is governed by the photolysis of N$_2$ by solar EUV radiation ($\lambda<100$ nm) and by its interaction with energetic electrons from Neptune's magnetosphere \citep{strobel_magnetospheric_1990,krasnopolsky_temperature_1993,krasnopolsky_photochemistry_1995,strobel_tritons_1995,strobel_comparative_2017}. 

Apart from this, very little is known about Triton, as no mission has been sent to the Neptunian system since Voyager 2. This is why the demand for a new mission to the Ice Giants is currently growing in the community. 
Also, Triton is thought to be an ocean world (such as Titan, Enceladus, Europa and Ganymede) meaning that it may have a liquid ocean under its icy surface, heated by obliquity tides \citep{rymer_neptune_2021,fletcher_ice_2020}. It makes it a high priority target to study the possibility of developing life in the outer worlds of the Solar System \citep{committee_on_the_planetary_science_and_astrobiology_decadal_survey_origins_2022}. Hence, a mission to the Neptunian system would allow studies across a very large spectrum of disciplines. In order to prepare such a mission, it is important to develop photochemical models of Triton's atmosphere, as this will give a theoretical basis to develop the instruments and anticipate future in-situ measurements.
\newline

Due to the scarcity of data available after the Voyager flyby, few articles about the photochemistry of Triton's atmosphere have been published: \citet{majeed_ionosphere_1990,strobel_photochemistry_1990,lyons_solar_1992,krasnopolsky_temperature_1993,krasnopolsky_photochemistry_1995,strobel_tritons_1995}. 
Significant improvements in the modeling of the photochemistry of Titan's atmosphere have been made thanks to the Cassini-Huygens data, in particular in the determination of the chemical scheme. A lot of models of this atmosphere were developed and refined, and are now quite robust \citep[e.g.][]{dobrijevic_1d-coupled_2016,loison_photochemical_2019,nunez-reyes_low_2019,nunez-reyes_rate_2019,hickson_kinetic_2020,vuitton_simulating_2019}. They can be used as a starting point for the development of a new photochemical model of Triton's atmosphere since it is composed of N$_2$ and CH$_4$, which happen to be also the main constituents of Titan's atmosphere. Recent 1D-photochemical models use thousands of chemical reactions and consider hundreds of species including neutral and ionic compounds. Using a more complete chemical scheme could change the vision and the understanding of the chemical mechanisms governing Triton's atmosphere. 
It is also important to take into account ground-based observations such as those of \citet{lellouch_detection_2010} which measured the abundance of CO.
\newline

As the temperature of Triton's atmosphere is particularly low (<100\,K at all altitudes), we expect to have large uncertainties with regard to the chemistry. Indeed, reaction rates are mostly measured or calculated at room temperature. Hence, their values may be wrong in Triton's conditions, even if these rates are given with an uncertainty factor which accounts for errors within the experiments or the computations. This problem was presented in \citet{hebrard_how_2009}. To see the impact of these uncertainties on our results, we used a Monte-Carlo procedure over all reaction rates, as done in \citet{dobrijevic_effect_1998,dobrijevic_effect_2003,hebrard_photochemical_2007,dobrijevic_key_2010} and following papers. Along with this study, we also performed global sensitivity analyzes to determine which reactions had the strongest impact on chemical uncertainties, which we call key uncertainty reactions. The determination of these reactions indicate which reactions need to be measured in priority by new chemical studies. 
\newline

The aim of the present work was to develop a new photochemical model of Triton's atmosphere and determine the key uncertainty reactions that must be studied in priority in order to reduce the uncertainties on model results. 
Our atmospheric model is presented in Sect. \ref{atm_model}, 
our photochemical model in Sect. \ref{photohem_model}, our updated chemical scheme in Sect. \ref{update_chem_schm}, our results for the nominal chemistry with this updated scheme in Sect. \ref{results}, our study of chemical uncertainties and the determination of key uncertainty reactions in Sect. \ref{section_chem_uncert} and our conclusions in Sect. \ref{conclu}.

\section{Atmospheric model}
\label{atm_model}

In this section, we present all the basic inputs of our model. These inputs are the temperature, pressure and density profiles, the altitude grid, the boundary conditions, the diffusion coefficients and the atmospheric escape rates. All these inputs are independent from the chemical scheme and from the photochemical parameters. 

\subsection{Atmospheric profiles and altitude grid}

With the measurement of the surface temperature, the thermospheric temperature $T_{th}$ was inferred by using the number density of N$_2$ and assuming hydrostatic equilibrium. It gave $T_{th} = 95\pm$ 5\,K \citep{broadfoot_ultraviolet_1989}, but the complete temperature profile could not be determined and was the subject of several subsequent studies \citep[see][]{yelle_energy_1991,stevens_thermal_1992,krasnopolsky_temperature_1993,strobel_comparative_2017}. 

Due to the presence of plumes (that were observed up to 8\,km above the surface) and clouds, it is thought that the temperature gradient near the surface is negative, indicating the presence of a troposphere \citep{yelle_energy_1991,yelle_lower_1995}. Energy is brought to the atmosphere by solar Extreme Ultraviolet (EUV) photons and by precipitating electrons from Neptune's magnetosphere \citep{strobel_magnetospheric_1990,yelle_energy_1991,stevens_thermal_1992,strobel_tritons_1995,krasnopolsky_photochemistry_1995,strobel_comparative_2017}. Energy is then transferred through the atmosphere by conduction \citep{yelle_energy_1991,yelle_lower_1995,strobel_tritons_1995}. 
Magnetospheric electrons (ME) have not always been taken into account, some models considering the Sun and the interplanetary radiation flux as the only energy sources, as in \citet{lyons_solar_1992}.
However, \citet{strobel_magnetospheric_1990,stevens_thermal_1992,krasnopolsky_photochemistry_1995,strobel_comparative_2017} showed that they are necessary to explain the thermospheric temperature measured by Voyager. 
Another critical parameter is the abundance of CO because of its cooling capabilities \citep{stevens_thermal_1992,krasnopolsky_temperature_1993,strobel_comparative_2017}. As its abundance was not measured by Voyager, it was adjusted to fit the measured tangential N$_2$ column densities \citep{stevens_thermal_1992}. \citet{krasnopolsky_temperature_1993} tried different values of the initial abundance of CO but were unable to constrain its value from thermal balance calculations. This abundance was measured as (2-18)$\times 10^{-4}$ by high-resolution spectroscopic observations in the 2.32-2.37 $\mu$m range, using the CRyogenic high-resolution InfraRed Echelle Spectrograph (CRIRES) at the Very Large Telescope (VLT) \citep{lellouch_detection_2010} and is consistent with the upper limit inferred by Voyager data (i.e. <1\% \citealt{broadfoot_ultraviolet_1989}).

In a more recent paper, taking advantage of the similarities between Pluto and Triton, \citet{strobel_comparative_2017} adapted the thermal model of Pluto from \citet{zhu_density_2014} to Triton. The main differences between the two atmospheres are the mole fraction of CH$_4$ (higher on Pluto) and the energy supply from magnetospheric electrons from Neptune's magnetosphere. They used the abundance of CO determined in \citet{lellouch_detection_2010} and studied three different models to see the impact of magnetospheric electrons on the thermal profile: two models without magnetospheric electrons and with different CH$_4$ abundances and a third with magnetospheric electrons. Their conclusion was that magnetospheric heating is necessary to retrieve N$_2$ tangential column number densities comparable to the measurements from Voyager 2. 
\newline

In our model, we used data from their Triton-3 model, which considers  precipitations of magnetospheric electrons.
Thus, we used the associated temperature and pressure profiles. It has to be noted that this thermal profile does not consider a troposphere as the temperature gradient is always positive, in contrast to the work of \citet{krasnopolsky_temperature_1993}, \citet{krasnopolsky_photochemistry_1995}, \citet{yelle_energy_1991} and \citet{yelle_lower_1995}.

The maximum altitude for this model is 1026\,km. The temperature varies between 37.8 and 90.3\,K from the surface to the upper end of the atmosphere, the pressure between 16 and 2.8$\times 10^{-8}$ $\mu$bar and the number density is computed following the ideal gas law. 

\begin{figure}[!h]
   \resizebox{\hsize}{!}
            {\includegraphics{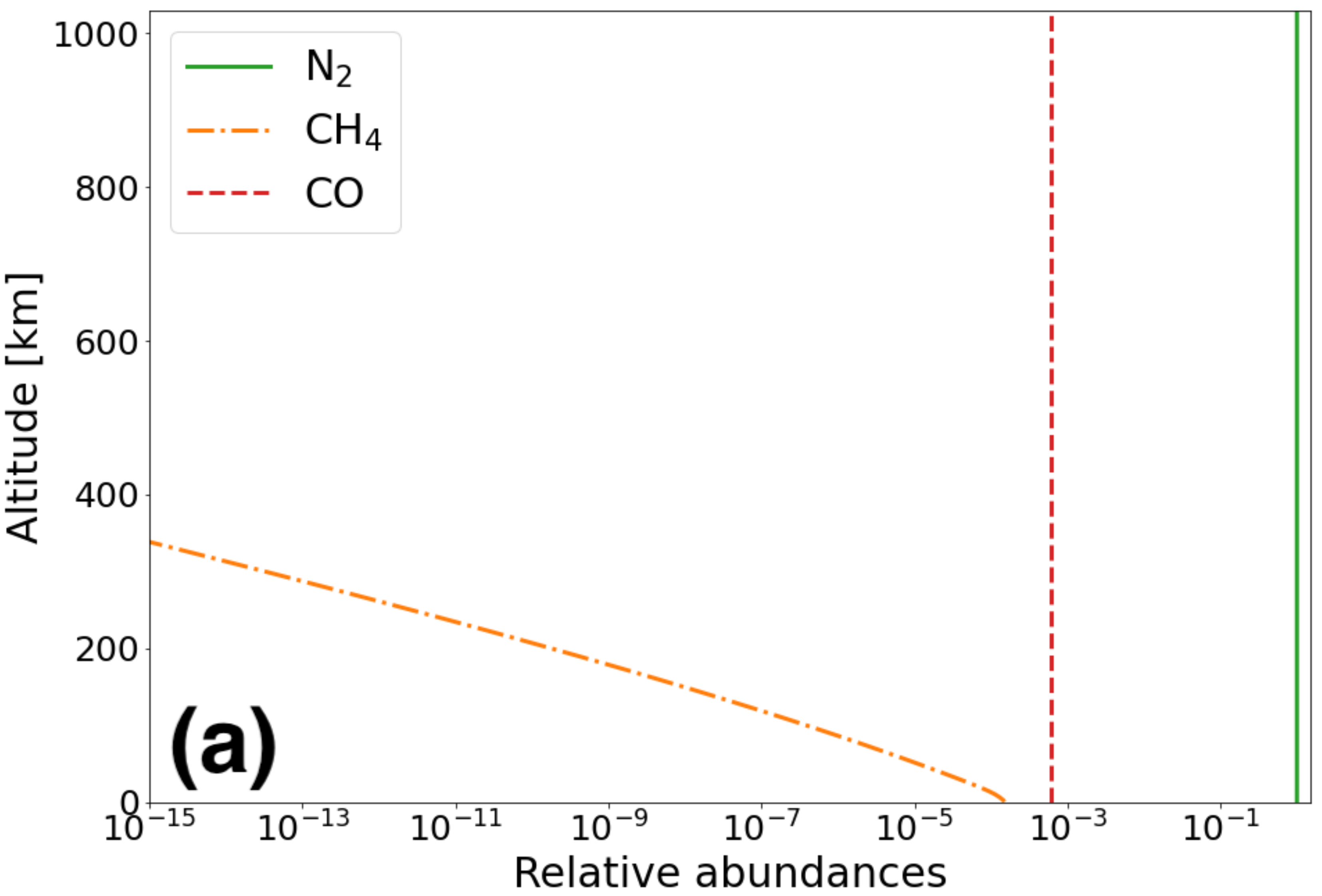}
            \includegraphics{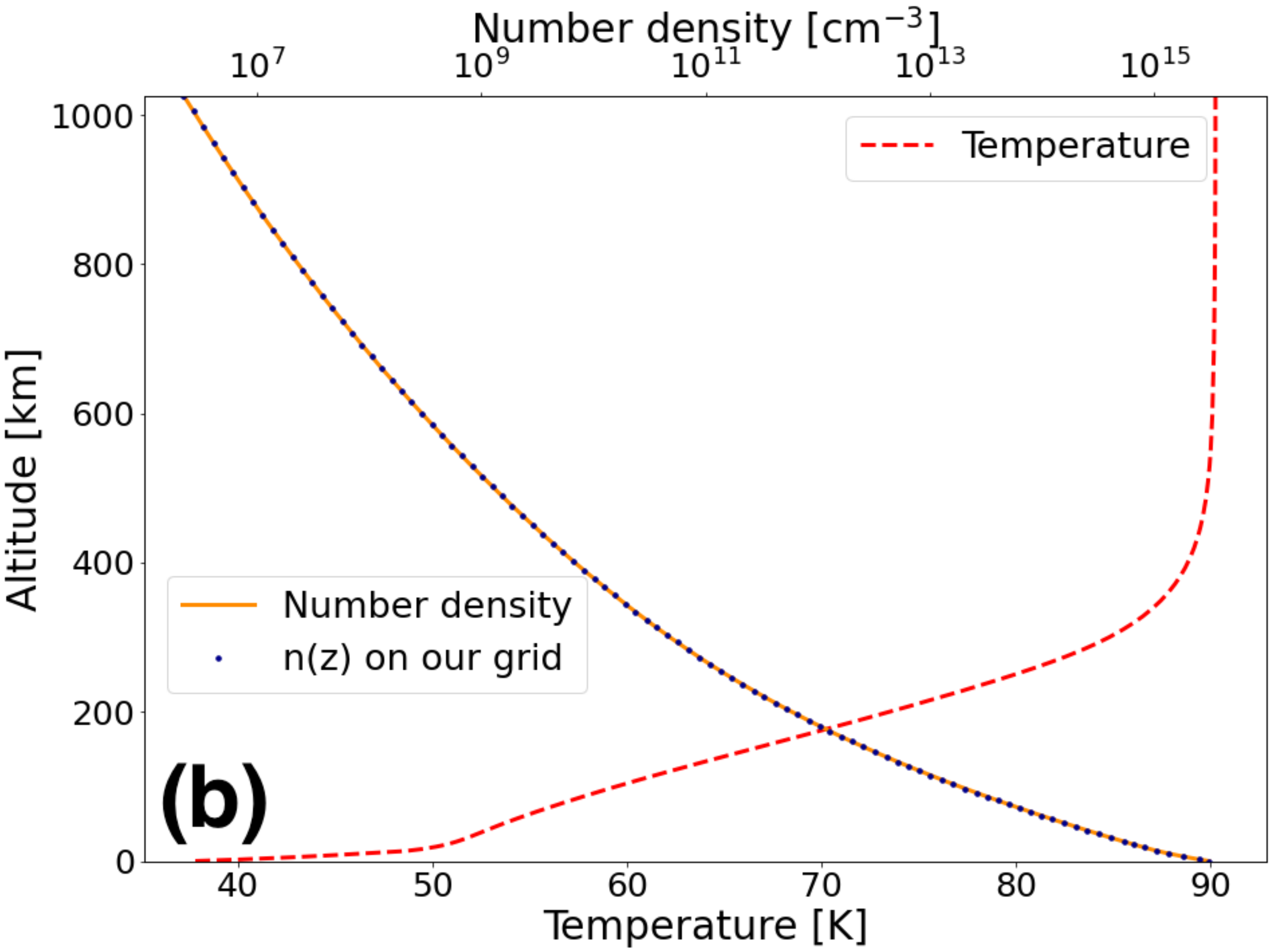}
            }
      \caption{\textbf{(a): }Initial abundance profiles of N$_2$ (green solid line), CH$_4$ (orange dash-dotted line) and CO (red dashed line). The CH$_4$ abundance decreases exponentially with a scale height of 9\,km. \textbf{(b): }Temperature (red dashed line) and number density (orange solid line) profiles from the Triton-3 case of \citet{strobel_comparative_2017} that are used as inputs in our model. The altitudinal grid is composed of 96 levels and the total number density is plotted with blue dots.
              }
         \label{Profil_Tn_n(levels)_init_profiles}
   \end{figure}

We sampled the altitude grid with $H$/5 steps, where $H(z) = \frac{k_{\text{B}}T(z)}{\Bar{m}(z)g(z)}$ is the scale height of the atmosphere at altitude $z$ ($k_{\text{B}}$ is the Boltzmann constant, $T$, $\Bar{m}$ and $g$ are respectively the temperature, mean mass and gravitational acceleration at altitude $z$). Using these criteria, our altitude levels are spaced by 2\,km near the surface and by 21\,km near the top of the atmosphere, giving a 96 level grid.
Temperature and number density profiles are shown in Fig. \ref{Profil_Tn_n(levels)_init_profiles}, along with the altitude levels.

The electronic temperature profile is a parameter required to compute the reaction rates of dissociative recombination reactions. As it has never been measured, we considered that the electronic temperature is equal to the neutral temperature at all altitudes.

\subsection{Initial and boundary conditions}

The initial abundance of a given species corresponds to the value taken at the beginning of the program, before the photochemical calculations. To be consistent with the use of the thermal, pressure and number density profiles of \citet{strobel_comparative_2017}, we also used 
their initial abundance of CO: $y_0$(CO)=$6.0\times 10^{-4}$, which corresponds to the measurement made by \citet{lellouch_detection_2010}. The initial abundance of CO is constant throughout the atmosphere. The initial mole fraction of CH$_4$ was taken equal to the $\frac{P_v(\text{CH$_4$})}{P}$ ratio at the surface, $P_v$ being the vapor pressure and $P$ the total pressure. With the formula of \citet{fray_sublimation_2009}, this corresponds to $y_0$(CH$_4$)=0.89$\times$10$^{-4}$. Thus, the initial abundance is 40\% lower than the value of \citet{strobel_comparative_2017}, which is $y_0$(CH$_4$)=$1.5\times10^{-4}$. 

We chose to take an exponentially decreasing profile for CH$_4$ as the initial condition, with a scale height of 9\,km corresponding to Voyager's observations \citep{strobel_photochemistry_1990}.
Then, at all altitudes, we simply filled the rest of the atmosphere with N$_2$ by taking $y_{\text{N$_2$}}(z)=1-$[$y_{\text{CH$_4$}}(z)$+$y_{\text{CO}}(z)$]. The initial profiles of these 3 compounds are plotted in Fig. \ref{Profil_Tn_n(levels)_init_profiles}.


\subsection{Eddy diffusion coefficient}

\begin{figure}[!h]
   \centering
   \includegraphics[width=10cm]{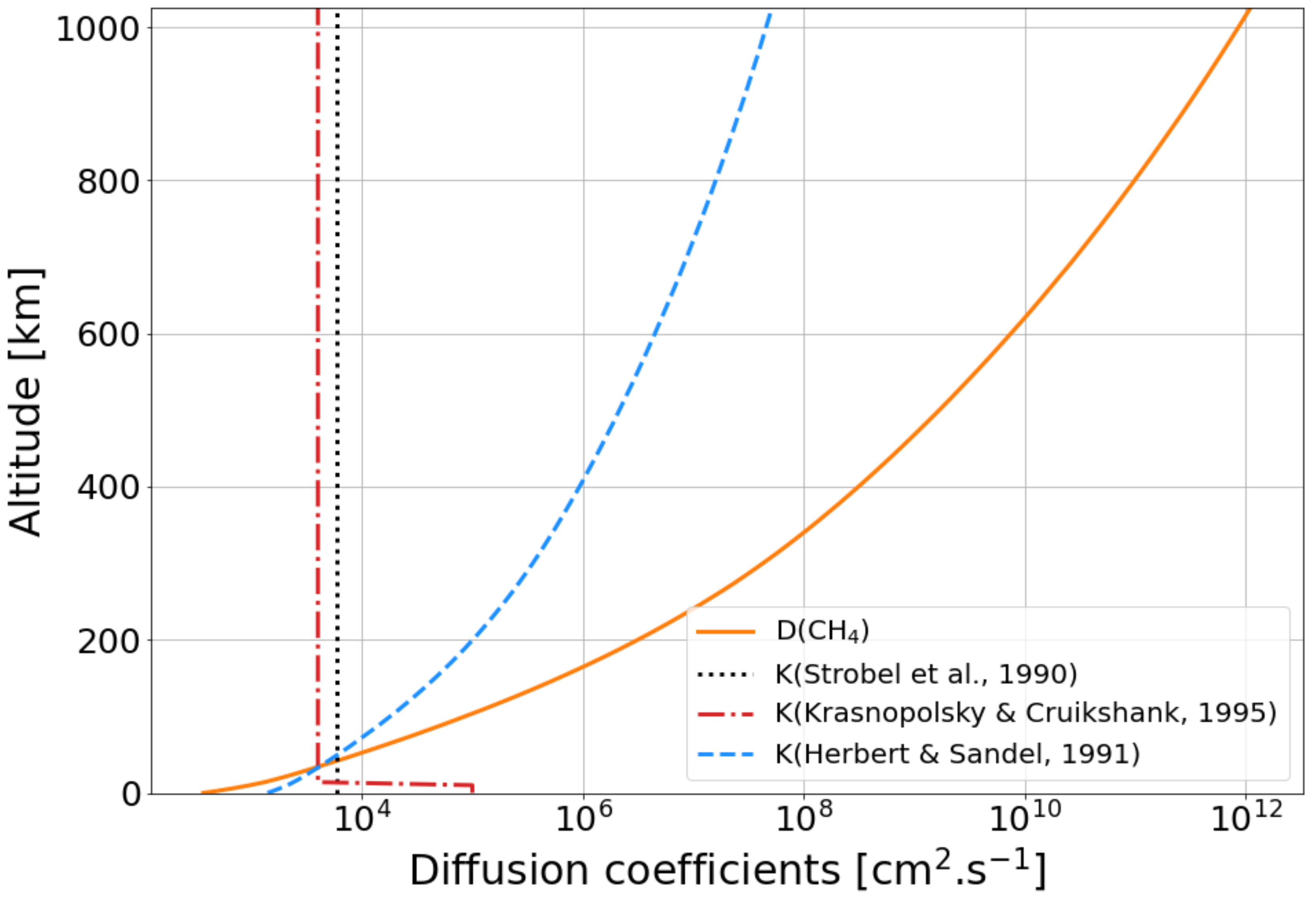}
      \caption{Comparison between the eddy diffusion coefficients $K_{zz}$ of \citet{strobel_photochemistry_1990} (black dotted line), \citet{herbert_ch4_1991} (blue dashed line) and \citet{krasnopolsky_photochemistry_1995} (red dash-dotted line). For the \citet{strobel_photochemistry_1990} profile, we plotted $K_{zz} = 6.0\times 10^3$ cm$^2$.s$^{-1}$ and for the \citet{herbert_ch4_1991} profile $K_{zz} = 1.4\times10^3\left ( \frac{\left [ \text{N}_2 \right]_0}{\left [ \text{N}_2 \right](z)} \right )^{0.5}$ cm$^2$.s$^{-1}$, the averages of the values for the summer and winter hemispheres. The homopause of CH$_4$ is located at the altitude where the molecular diffusion coefficient $D_{\text{CH$_4$}}$ of this species (orange solid line) is equal to $K_{zz}$.
              }
         \label{Profils_Kzz}
   \end{figure}

The eddy diffusion coefficient $K_{zz}$ is a critical parameter of the model. 
In previous articles about the photochemistry of Triton's atmosphere, this coefficient was adapted to match the CH$_4$ profile measured by Voyager 2 \citep{strobel_photochemistry_1990,herbert_ch4_1991,krasnopolsky_photochemistry_1995}. All the profiles inferred in these studies are different, as shown in Table \ref{Compar_Kzz} and Fig. \ref{Profils_Kzz}.

\begin{table}[!h]    
\begin{center}
   \begin{tabular}{c c c}     
\hline                    
Study  & $K_{zz}$($z$) {[}cm$^2$.s$^{-1}${]} & Homopause {[}km{]} \\ \hline
\citet{strobel_photochemistry_1990} & {[}4-8{]}$\times$10$^3$ & {[}35-47{]}\\
\citet{herbert_ch4_1991} & {[}1.2-1.6{]}$\times$10$^3\left ( \frac{\left [ \text{N}_2 \right]_0}{\left [ \text{N}_2 \right](z)} \right )^{0.5}$ & 35 \\
\citet{krasnopolsky_photochemistry_1995} & 10$^5$ in the troposphere, 4.10$^3$ above  & 35 \\
\hline                  
\end{tabular}
\end{center}
\caption{Comparison of the different $K_{zz}$($z$) used in previous Triton  photochemical models since the flyby of Voyager 2 in 1989.} 
\label{Compar_Kzz}  
\end{table}

We tested the different profiles from Table \ref{Compar_Kzz} and obtained the best agreement with observations using the profile from \citet{herbert_ch4_1991} and thus kept it for the rest of our work.

\subsection{Molecular diffusion coefficient}

The other main type of diffusion in planetary atmospheres is molecular diffusion. It is used to describe the diffusion of minor species in one or more major species. This type of diffusion occurs when the number density of the minor species deviates from its distribution at hydrostatic equilibrium. A coefficient is linked to this type of diffusion and we computed it with formulas \eqref{coeff_diff_mol} and \eqref{pré_coeff_diff_mol} taken from \citet{Poling}, as we considered molecular diffusion in the two main species of Triton's atmosphere. In this case, one of the terms of \eqref{coeff_diff_mol} always depends on N$_2$ as it is the most abundant species throughout the atmosphere but the second main species varies with altitude. The latter is noted $j_2$ in \eqref{coeff_diff_mol} describing the molecular diffusion coefficient of species $i$ at altitude $z$: 

\begin{equation}
      D_i(z) = \frac{1}{\frac{y_{\text{N$_2$}}(z)}{D_{i,\text{N$_2$}}(z)}+\frac{y_{j_2}(z)}{D_{i,j_2}(z)}}
\label{coeff_diff_mol}
\end{equation}
where $D_{i,j}$ is the diffusion coefficient of the minor species $i$ in a single major species $j$ whose relative abundance is $y_j$. $D_{i,j}$ is computed with: 

\begin{equation}
      D_{i,j}(z)=\frac{0.00143\times T(z)^{\frac{1}{3}}}{P(z)M_{i,j}^{\frac{1}{2}}\left [ (\Sigma_v)_i^{\frac{1}{3}}+(\Sigma_v)^{\frac{1}{3}}_j \right ]^2}
\label{pré_coeff_diff_mol}
\end{equation}
where $T$ is the temperature, $P$ the pressure, $M_{i,j} = \mu_{i,j}\times 2$ with $\mu_{i,j}$ the reduced mass of species $i$ and $j$ and $\Sigma_v$ is the diffusion volume. 

\subsection{Atmospheric escape}

Atmospheric escape of neutral and ionized species is considered in many papers about Triton, such as \citet{summers_tritons_1991} or \citet{krasnopolsky_photochemistry_1995}. It is thought that this mass load in Neptune's magnetosphere could affect Neptune's auroras \citep{broadfoot_ultraviolet_1989}. 

In our model, we simply considered Jeans thermal escape from the top of the atmosphere, which is at 1026\,km. The flux is computed for each neutral species $i$ following Eq. \eqref{echappement}:
\begin{equation}
      J_i = n_i\times v_{\text{lim$_i$}} =  n_i\times \sqrt{\frac{k_{\text{B}}\times T(z_{\text{max}})}{2\pi m_i}}\times \exp{\left( -q \right)}\times (1 + q)
\label{echappement}
\end{equation}
with $v_{\text{lim}_i}$ the escape velocity of species $i$ at the top of the atmosphere, corresponding to the altitude $z_{\text{max}}$, $n_i$ its number density, $m_i$ its mass and $k_{\text{B}}$ the Boltzmann constant. $T(z_{\text{max}})$ is the neutral temperature at this altitude level and $q$ is computed as follows, with $R_T$ the radius of Triton:
\begin{equation}
      q = \frac{(R_T+z_{\text{max}})\times m_i\times g(z_{\text{max}})}{k_{\text{B}}\times T(z_{\text{max}})}
\label{q_echap}
\end{equation}
Unlike \citet{krasnopolsky_photochemistry_1995}, we did not consider ion escape and did not scale our electronic profile on this escape above 600\,km.

\section{Photochemical model}
\label{photohem_model}

\subsection{Baseline chemical scheme}
\label{base_chem_model}

As our model is one-dimensional, we could use a complex chemical scheme without suffering excessively long computation times. 
Capitalizing on the similarities between the major constituents of Triton's and Titan's atmospheres,
we used the chemical scheme of Titan's atmosphere presented in \citet{dobrijevic_1d-coupled_2016} as a basis for our work. This chemical scheme was updated in \citet{loison_photochemical_2019,nunez-reyes_low_2019,nunez-reyes_rate_2019,hickson_kinetic_2020}. The number of reactions and atmospheric species used in this scheme is presented in Table \ref{compar_chemical_Schemes}. 

Although the initial composition of the atmospheres of Titan and Triton are quite similar, differences have to be noted as they could have an impact on the results. Firstly, CH$_4$ is only a trace species on Triton, whereas its abundance on Titan reaches 20\% at the top of the atmosphere. Thus, some reactions could be less important on Triton due to the absence of methane in the upper atmosphere. Conversely, some reactions that do not have a great impact on Titan could be crucial on Triton. Secondly, the temperature and pressure are much lower on Triton, and this directly impacts the reaction rates and the condensation of some species, such as hydrocarbons. 
Thus, this initial scheme was modified after the first results were obtained, following the methodology presented in Sect. \ref{update_chem_schm}.

\subsection{Generalities about calculations}

Our photochemical model solves the continuity Eq. \eqref{continuity_eq} for all the considered species $i$ at all the altitude levels. At altitude $z$, it gives:
\begin{equation}
    \frac{\partial n_i(z)}{\partial t} = -\text{div}~\overrightarrow{\Phi_i}(z) + P_i(z) - n_i(z)L_i(z)
\label{continuity_eq}
\end{equation}
where $n_i$ is the number density of the species $i$, $P_i$ is the chemical production term in cm$^{-3}$.s$^{-1}$ and $L_i$ the chemical loss term in s$^{-1}$. $\overrightarrow{\Phi_i}$ is the particle vertical flux computed by:

\begin{equation}
    \begin{split}
        \Phi_i(z) = -D_i(z)n_i(z)\left[ \frac{1}{y_i(z)}\frac{\partial y_i(z)}{\partial z}+\frac{1}{H_i(z)}-\frac{1}{H(z)} \right ] - K_{zz}(z)n_i(z)\left [ \frac{1}{y_i(z)}\frac{\partial y_i(z)}{\partial z} \right ]
    \end{split}
\label{particle_flux}
\end{equation}
where $D_i$ is the molecular diffusion coefficient, $K_{zz}$ the eddy diffusion coefficient, $y_i$ the abundance, $H_i$ the scale height of species $i$ and $H$ the atmospheric scale height. Here, thermal diffusion is neglected.
\newline

Equation \eqref{continuity_eq} is integrated over time until steady state is reached, that is when $\frac{\Delta n_i(z)}{\Delta t}$ is below a given threshold. This ratio is computed at the end of each time step.
In our case, the threshold was fixed at 10\%, because such a variation was small in comparison to model result uncertainties caused by chemical uncertainties. 

To compute the abundance profiles of all the considered atmospheric species, we operated in steps.
At the start, we used an atmosphere only composed of N$_2$, CH$_4$ and CO and computed the chemical rates and the actinic flux. Then, we calculated chemical and transport terms to integrate the continuity equation and determine the abundance profiles of all the species. When convergence was reached, we computed again the chemical rates and the actinic flux with the newly obtained abundance profiles and ran the integration again. This pattern was repeated until the difference between the results of two successive steps was weak compared to the model uncertainties. 
In our case, three iterations were needed to reach steady state. 

\noindent In the following, we describe each parameter that is useful to compute all the terms of Eqs. \eqref{continuity_eq} and \eqref{particle_flux}. 

\subsection{Energy sources}

\subsubsection{Solar flux}
\label{solar_flux_subsec}

Triton is 30 AU away from the Sun. Consequently, the solar flux it receives is 900 times lower than on Earth and so approximately 10 times lower than on Titan. Despite this, the ionosphere of Triton is denser than the one of Titan. 

We used different sources of data for the solar flux, allowing us to consider different solar activity cases. For the low activity case, we used a high resolution composite spectrum built with data from \citet{curdt_sumer_2001,curdt_sumer_2004} that has a resolution of 0.004 nm between 67 and 160 nm and from \citet{thuillier_solar_2004}. This spectrum is the same as the one used in \citet{dobrijevic_1d-coupled_2016}.

For medium and high solar activity, we used low resolution spectra with a resolution of 1 nm. 
As the flyby of Triton in 1989 occurred at a maximum solar activity, we used the corresponding low resolution file between 1 and 730 nm in our calculations. 

\subsubsection{Magnetospheric electrons}

As the solar flux seemed too weak to explain the presence of a dense ionosphere, a second source of energy was hypothesized \citep{majeed_ionosphere_1990}. 
The most obvious candidate was the precipitation of energetic electrons from Neptune's magnetosphere, as energetic electrons were observed with the Low-Energy Charged Particles (LECP) instrument aboard Voyager 2 \citep{krimigis_hot_1989}. 

These measurements were used in \citet{strobel_magnetospheric_1990} to calculate the production rates of N$_2^+$ and N$^+$ in Triton's atmosphere. They show that without electron precipitation, the predicted electronic peak derived using only the solar flux does not correspond to the one observed by Voyager, as it is weaker and at a higher altitude. Adding magnetospheric electrons, they find a more reliable profile but at an altitude lower than expected. Thus, their ionization profile has to be moved up by two scale heights in order to find an electronic peak that fits with the observations \citep{summers_tritons_1991}. Finally, as the incident electronic flux used for the calculations was measured when Triton was close to Neptune's magnetic equator, the ionization profile has to be adapted to represent mean orbital conditions as done in \citet{strobel_magnetospheric_1990,stevens_thermal_1992,krasnopolsky_photochemistry_1995,strobel_tritons_1995}. 

In our model, we need an electron production rate to compute the reaction rates of the ionizations and dissociations of N$_2$ by magnetospheric electrons. Thus, we used the ionization profile of \citet{strobel_magnetospheric_1990}, moved it up by two scale heights and divided it by 6 in order to represent mean conditions, as done in \citet{krasnopolsky_photochemistry_1995}. 

The reaction rates $k$(ME,N$_2$) for the interaction between magnetospheric electrons and N$_2$ are then computed with Eq. \eqref{krate_ME}.

\begin{equation}
    k(\text{ME},\text{N}_2) = \frac{prod(z)\times br}{n_{\text{N}_2}(z)}
\label{krate_ME}
\end{equation}
where $prod(z)$ is the production rate of electrons at altitude $z$, $br$ is the branching ratio of the reaction and $n_{\text{N}_2}(z)$ is the number density of N$_2$ at the altitude $z$. 

We considered three different reactions between magnetospheric electrons and N$_2$ for which branching ratios are respectively 0.8, 0.2 and 0.6 \citep{fox_electron_1988}: 
\begin{align*}
    \text{N}_2 + \text{ME} &\longrightarrow \text{N}_2^+ + e^-\\
    \text{N}_2 + \text{ME} &\longrightarrow \text{N}_2^+ + \text{N(}^2\text{D)} + e^-\\
    \text{N}_2 + \text{ME} &\longrightarrow \text{N(}^4\text{S)} + \text{N(}^2\text{D)}
\end{align*}

\subsubsection{Interplanetary flux}

We also took into account the interplanetary radiation flux. As stated in \citet{strobel_photochemistry_1990}, at Triton's distance from the Sun, this radiation is not negligible with a flux $F$ at Lyman-$\alpha$ (121.6 nm) of 340\,R \citep{broadfoot_ultraviolet_1989} (1\,R = $\frac{10^{6}}{4\pi}$ photons.cm$^{-2}$.s$^{-1}$.sr$^{-1}$), equivalent to a 8.5$\times$10$^7$ photons.cm$^{-2}$.s$^{-1}$ flux, to be compared to a 3.1$\times$10$^8$ photons.cm$^{-2}$.s$^{-1}$ solar flux at Lyman-$\alpha$ with a maximum solar activity. In addition, we also considered two additional interplanetary radiation fluxes: one at Lyman-$\beta$ (102.5 nm) with a ratio $\frac{F(Ly-\alpha)}{F(Ly-\beta)}=360$ and one at the Helium line (58.4 nm) with a ratio $\frac{F(Ly-\alpha)}{F(Helium)}=170$, as done in \citet{dobrijevic_etude_1996}.

\subsection{Condensation}

As the temperature is very low in the lower atmosphere of Triton, condensation occurs for several species. It is consistent with observations of hazes in the lowest 30\,km by Voyager 2. This haze is thought to be composed of hydrocarbons that are the products of CH$_4$ photolysis \citep{strobel_photochemistry_1990,herbert_ch4_1991,krasnopolsky_properties_1992}. 

In our model, we used a simplified consideration of the condensation, by fixing the abundance of the condensing species at $y_i(z) = \frac{P_v(z)}{P(z)}-1.0\times10^{-10}$ if the abundance of the considered species exceeds the $\frac{P_v(z)}{P(z)}$ ratio.

Our formulas to compute the vapor pressure $P_v$ of the different species come from different sources, the main ones being \citet{lara_vertical_1996}, \citet{fray_sublimation_2009}, the NIST database and \citet{haynes_crc_2012}.

It has to be noted that as the temperature near the surface of Triton is very low, small differences in the vapor pressure formulas in the temperature range where they are commonly measured could lead to significant differences in the final abundance profiles, as these abundances are restricted by the $\frac{P_v(z)}{P(z)}$ ratio.

\section{Update of the chemical scheme}
\label{update_chem_schm}

\begin{figure}[!h]
   \centering
   \includegraphics[width=10cm]{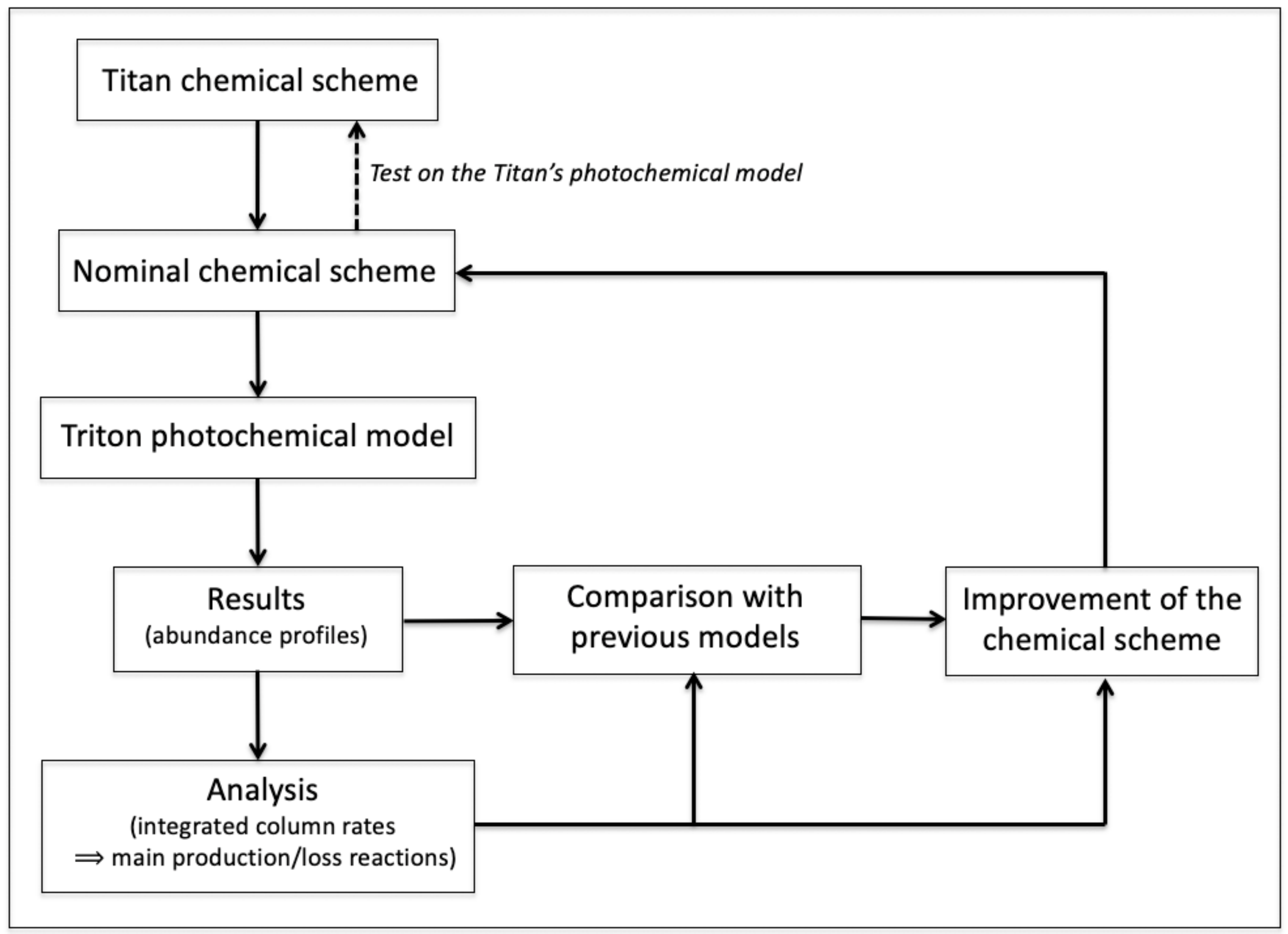}
      \caption{Methodology used to adapt Titan's chemical scheme to Triton.
              }
         \label{modif_chem_scheme}
   \end{figure}

As said in \ref{base_chem_model}, we expected some chemical differences to emerge between the Titan and Triton models, forcing us to modify the initial chemical scheme.
To do this, we first performed a run under the conditions of Triton's atmosphere. A very important difference between the atmospheres of Triton and Titan arises from the absence of CH$_4$ in the higher atmosphere of Triton, which impacts the overall chemistry. To complete our chemical network, we focused on the species that became much more abundant in the atmosphere of Triton. This is the case in particular for some atomic species such as N($^4$S), N($^2$D), C and C$^+$ as already noted before by \citet{krasnopolsky_photochemistry_1995}. The low abundance of CH$_4$ in the upper atmosphere of Triton induces low abundances of hydrocarbons and hydrocarbon radicals (in particular CH$_3$). As a result, association reactions with N$_2$ become much more important, such as the C + N$_2$ $\longrightarrow$ CNN reaction which is the main loss reaction for atomic carbon in the new model. It is thus also necessary to include these new species, such as CNN, and to introduce their chemical network. The high abundance of atomic carbon associated with its low ionization energy makes charge transfer reactions efficient. This leads to a high abundance of ionized atomic carbon which becomes the main ion above 175\,km; an important difference to the atmosphere of Titan. Once the new reactions to be included in the network were identified, the rate constants and branching ratios were chosen mainly from literature searches (e.g. \citet{husain_kirsch_1971} for the new critical reaction C + N$_2$ or \citet{anicich_index_2003} for the C$^+$ reactions). When no study existed we followed the same methodology as in our previous studies on Titan’s chemistry \citep{hebrard_neutral_2012,loison_neutral_2015}. As some reactions require the introduction of new species, some iterations were necessary to converge on a nominal chemical scheme. We also took care to compare our final network with that of \citet{krasnopolsky_photochemistry_1995}, allowing us to identify several important differences (see later). We also compared our results with data derived from the Voyager 2 observations presented in \citet{broadfoot_ultraviolet_1989}, \citet{tyler_voyager_1989}, \citet{herbert_ch4_1991} and with results from previous photochemical models such as \citet{krasnopolsky_photochemistry_1995} and \citet{strobel_tritons_1995}.
This methodology is presented in Fig. \ref{modif_chem_scheme}.

After having modified the chemical scheme, we ended up with a nominal chemical scheme consisting of 220 atmospheric species and 1764 reactions, as described in Table \ref{compar_chemical_Schemes}. 
A file containing a list of all the atmospheric species and their masses is available as supplementary material. 

\begin{table}[!h]
   \begin{center}
      \begin{tabular}{c c c}
        \hline
   & \begin{tabular}[c]{@{}c@{}}Initial chemical\\ scheme\end{tabular} & \begin{tabular}[c]{@{}c@{}}Revised chemical\\ scheme\end{tabular} \\ \hline
Neutral species  & 99   & 131   \\
Ionic species  & 83  & 89  \\
\begin{tabular}[c]{@{}c@{}}Neutral-neutral\\ reactions\end{tabular}   & 419  & 710   \\
\begin{tabular}[c]{@{}c@{}}Ion-neutral\\ reactions\end{tabular}       & 468  & 582 \\
Photodissociations  & 124  & 170  \\
Photoionizations  & 25 & 32 \\
\begin{tabular}[c]{@{}c@{}}Dissociative\\ recombinations\end{tabular} & 236 & 264 \\
\begin{tabular}[c]{@{}c@{}}ME/GCR\\ reactions\end{tabular}   & 6 & 6 \\
\begin{tabular}[c]{@{}c@{}}Total number\\ of reactions\end{tabular}   & 1278 & 1764  \\ \hline
\end{tabular}
   \end{center}
\caption{Comparison between the initial and updated chemical schemes. The initial chemical scheme comes from the model of Titan's atmosphere from \citet{dobrijevic_1d-coupled_2016} that we used as a basis for our work. The revised chemical scheme is used in our actual model of Triton's atmosphere. "ME" is used for Magnetospheric Electrons and "GCR" for Gamma Cosmic Rays. "Neutral-neutral reactions" regroups two-body, three-body, bimolecular and termolecular reactions.}
\label{compar_chemical_Schemes}
\end{table}

\section{Nominal results with the updated chemical scheme}
\label{results}

For the nominal model, we used the nominal values of the chemical reaction rates, meaning that we did not consider any uncertainty in their computation. By doing this, we only had to run the program once. As described in Sect. \ref{photohem_model}, we did three steps before reaching steady abundance profiles. In the following, we present the abundance profiles of the main neutral species and of the main ions. 
We detail the main production and loss processes for each important species in order to better understand the main mechanisms of Triton's atmospheric chemistry and why they are different to those found for Titan. Tables containing all these reactions and plots with their reaction rates depending on altitude are given in the supplementary material associated with this paper.
We also aim to identify the parameters that have the greatest impact on the final abundance profiles and compare our results with observations at the end of this section. 

\subsection{Neutral atmosphere}

\subsubsection{Main species}

\begin{figure}[!h]
   \resizebox{\hsize}{!}
            {\includegraphics{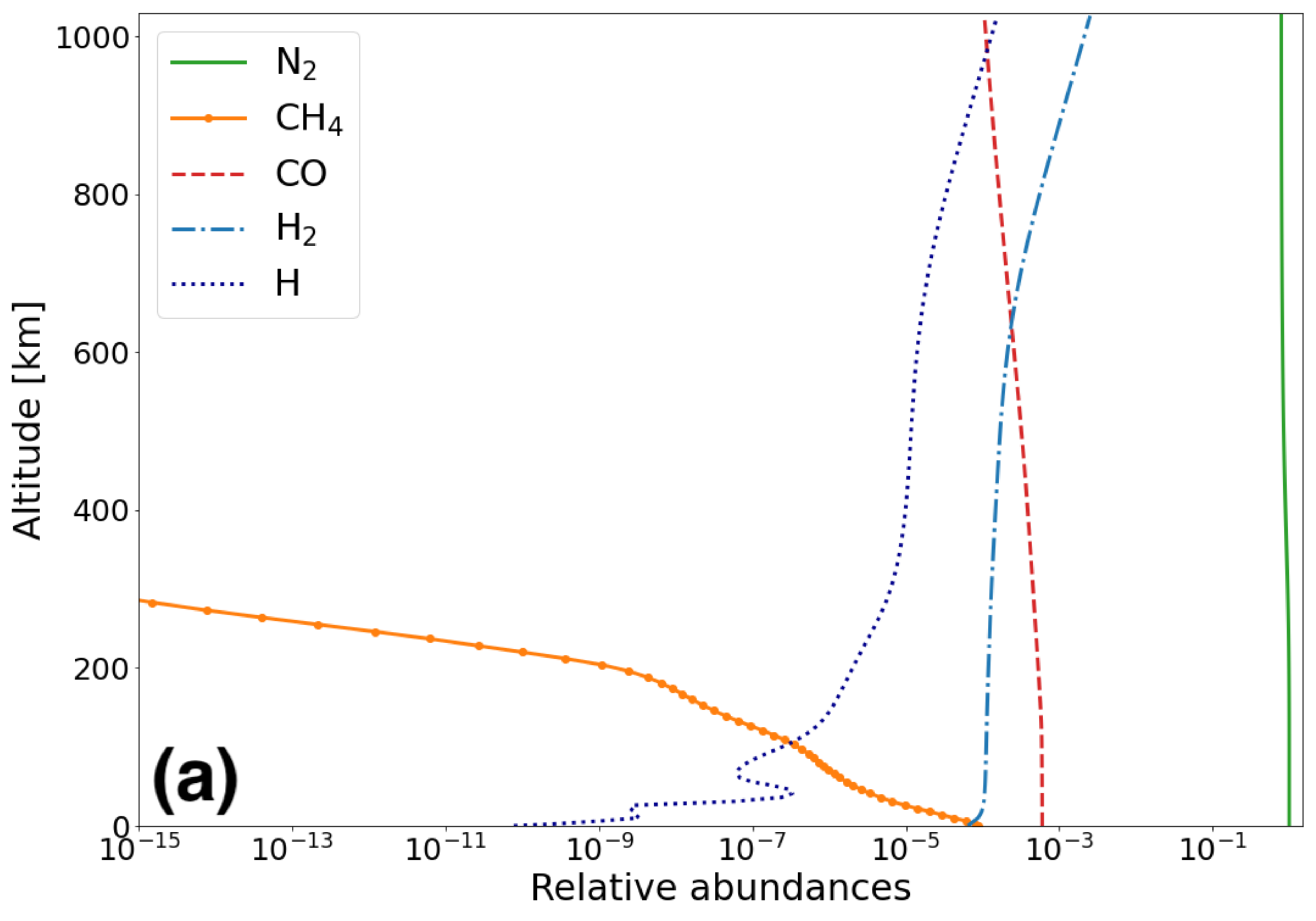}
            \includegraphics{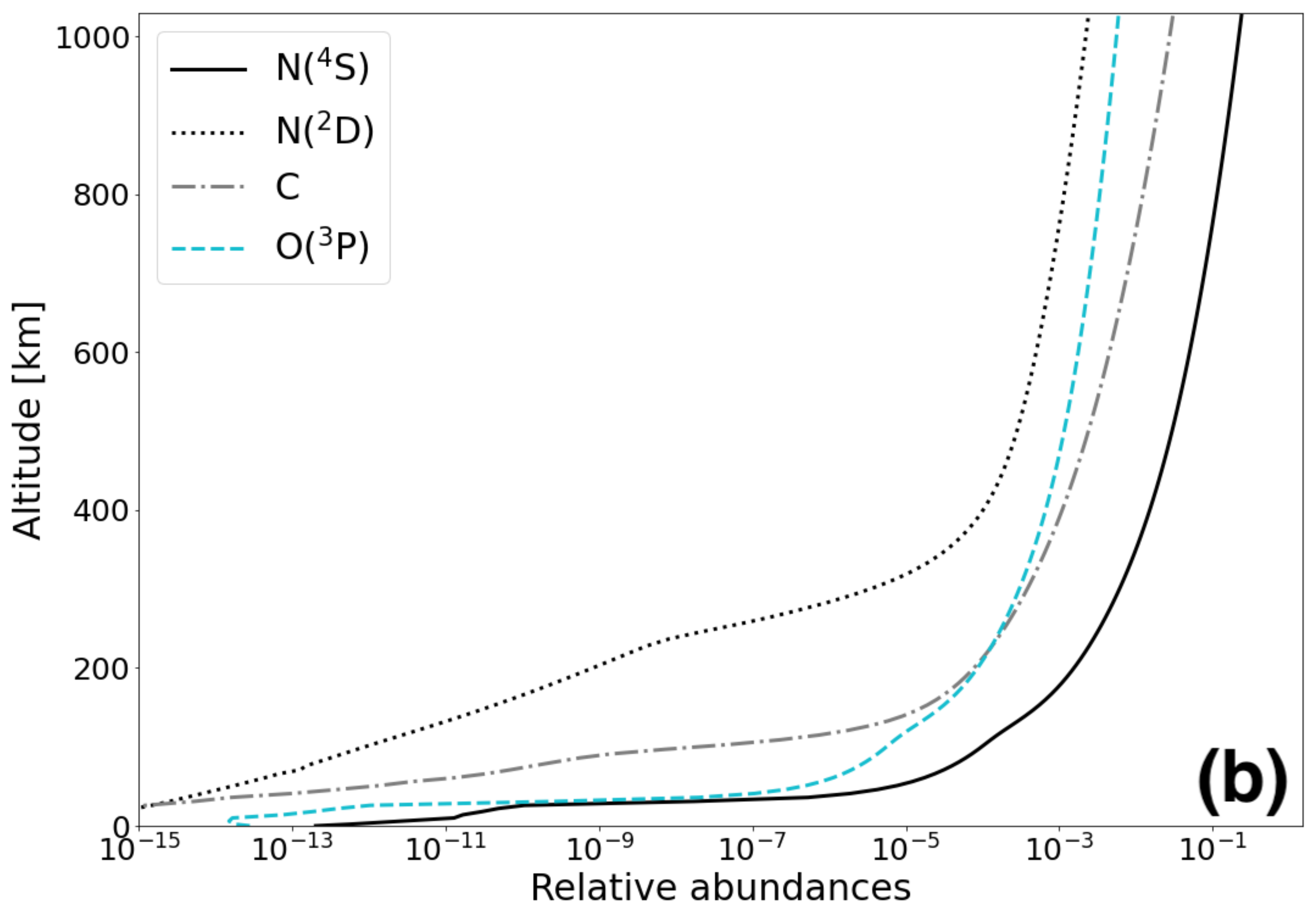}
            }
      \caption{Relative abundances of the main neutral species in the atmosphere of Triton. \textbf{(a):} Relative abundance of N$_2$ (green solid line), CH$_4$ (orange solid line with circles), CO (red dashed line), H$_2$ (blue dash-dotted line) and H (dark blue dotted line). \textbf{(b):} Relative abundances of N($^4$S) (black solid line), N($^2$D) (black dotted line), C (gray dash-dotted line) and O($^3$P) (cyan dashed line). We only give the O($^3$P) profile as the abundance of O($^1$D) is negligible.
              }
         \label{fm_main_neutrals}
   \end{figure}

The most abundant neutral species are N$_2$, N (N($^4$S) and N($^2$D)), C, CO, H, H$_2$ and O($^3$P). Their abundance profiles are given in Fig. \ref{fm_main_neutrals}. 
N($^4$S) corresponds to the ground state of atomic nitrogen and N($^2$D) is its first excited state. We only consider O($^3$P) here as the abundance of O($^1$D) remains negligible.

We can observe that N$_2$ remains the main atmospheric  throughout the atmosphere. Near the surface, CO, CH$_4$ and H$_2$ are the most abundant trace species. The abundance of CH$_4$ decreases quickly at higher altitude due to its photolysis by Lyman-$\alpha$ radiation. Above 50\,km, we see a large increase in the abundances of atomic species N, C, O and H. N becomes the second most abundant species and C the third. 

In the following paragraphs, we detail the main production and loss processes for each of these species in order to understand these evolutions (we precise that the third body of all the three-body reactions is N$_2$, thus, it is not precised in the rest of the paper). All the reactions used in this model and their integrated column rates are given as supplementary material. 

\paragraph{N$_2$}
~
\newline

N$_2$ being the main species of Triton's atmosphere, it is used as a reservoir in our model. Therefore, its abundance is not renormalized at each time step. It is destroyed by photodissociation, photoionization and interaction with magnetospheric electrons. These reactions produce atomic nitrogen N($^4$S) and N($^2$D) and N$_2^+$ and N$^+$ ions. The loss rate by photointeraction is of the order of one third of that by electrointeraction. This is consistent with the input energy flux, the energy carried by magnetospheric electrons being higher than the one from the solar flux. More details are given about this in Sect. \ref{photoionisation_and_ME}. 
The interaction with magnetospheric electrons is the second most important loss process for N$_2$, the first one being the three-body reaction with C giving CNN. N$_2$ also reacts with CH, which is a product of methane photolysis, to produce HCNN, with a loss rate half that of the previous reaction. Photoionization and reactions with magnetospheric electrons reach their maximum rate in the ionosphere, at 390 and 345\,km respectively, while other loss reactions mainly occur below 200\,km.
\newline

N$_2$ is mostly produced through the CNN cycle:

\begin{table}[!h]
\begin{center}
\begin{tabular}{l l l}
N$_2$ + C               & $\longrightarrow$ & CNN        \\
CNN + N($^4$S)          & $\longrightarrow$ & N$_2$ + CN \\
CN + N($^4$S)           & $\longrightarrow$ & N$_2$ + C  \\ \hline
net N($^4$S) + N($^4$S) & $\longrightarrow$ & N$_2$     
\end{tabular}    
\end{center}
\end{table}

The peak rate of these reactions occurs at 121\,km.
The reaction between H and HCNN giving $^1$CH$_2$ + N$_2$ has an integrated production rate four times lower than the ones of the CNN cycle but is the main production process below 50\,km, which is logical as the reaction producing HCNN reaches its maximum rate at 10\,km. 
At altitudes higher than 250\,km, dissociative recombination of N$_2$H$^+$ is the main source of N$_2$. 

\paragraph{CH$_4$}
~
\newline

CH$_4$ is very important for Triton's atmospheric chemistry as its photolysis is a source of the CH, $^3$CH$_2$, $^1$CH$_2$ and CH$_3$ radicals, of H and H$_2$. It also leads to the production of more complex hydrocarbons. 
Its chemistry is triggered by photodissociation and photoionization by Lyman-$\alpha$ radiation from the Sun and from the interstellar medium (ISM). Photodissociation reactions account for 71\% of the total loss of CH$_4$.

CH$_4$ also reacts with CH to produce C$_2$H$_4$. This reaction contributes for 29\% of the total loss of CH$_4$, explaining why C$_2$H$_4$ is the most abundant hydrocarbon in the lower atmosphere (cf Sect. \ref{results_C2Hx}). These results are consistent with the description of \citet{strobel_photochemistry_1990}. 

\begin{figure}[!h]
   \centering
   \includegraphics[width=10cm]{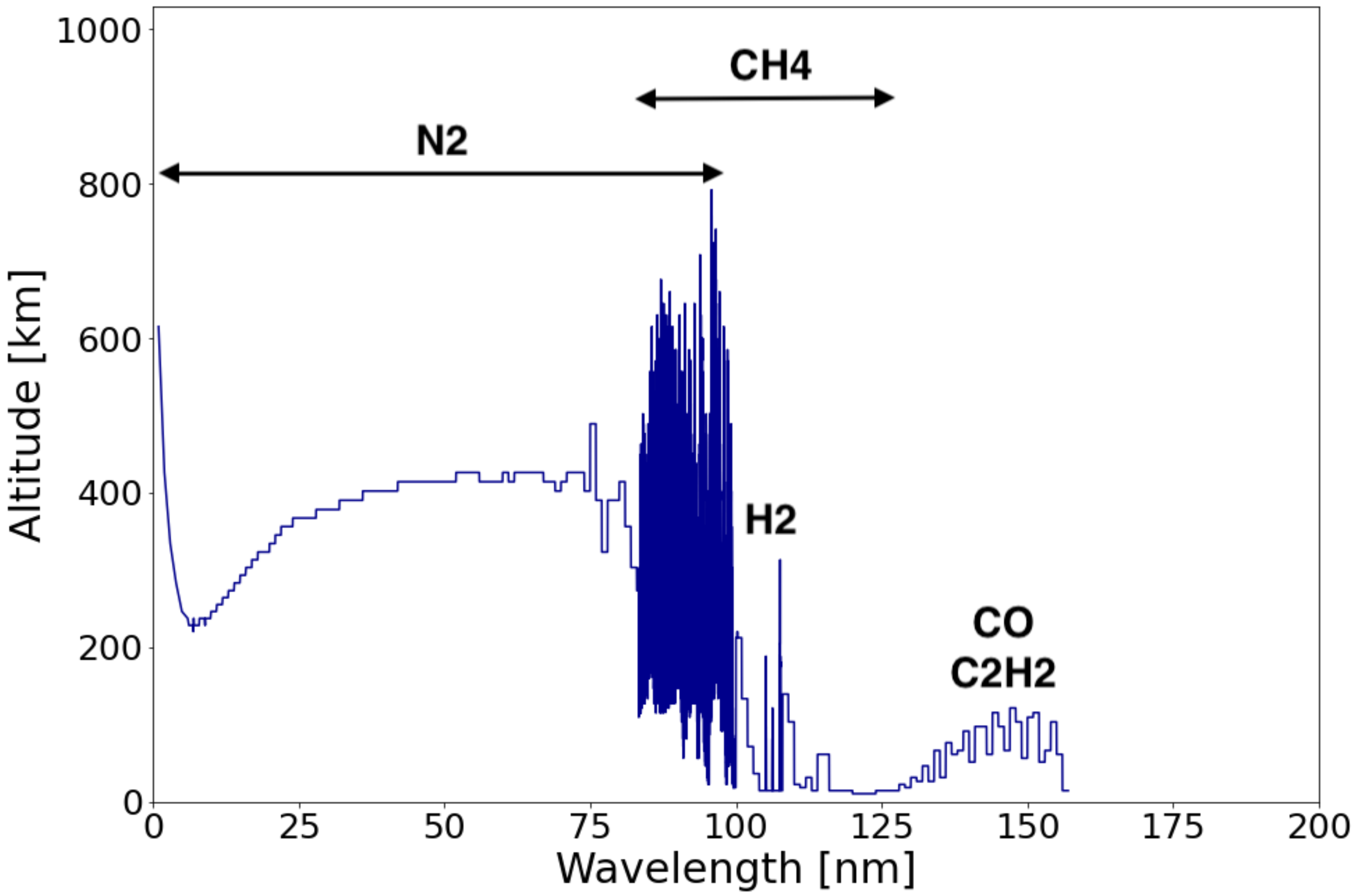}
      \caption{Altitude at which the optical thickness of the atmosphere is 1 depending on the wavelength of the incident radiation. The plot is cut at 200 nm as the atmosphere is not optically thick for all wavelengths above 160 nm, but the initial solar flux goes from 0 to 730 nm.
              }
         \label{penetration_0_200nm}
   \end{figure}

All these reactions reach their maximum rate at 10\,km, which corresponds to the altitude where the atmosphere becomes optically thick at the Lyman-$\alpha$ wavelength, as shown in Fig. \ref{penetration_0_200nm}.

Almost all CH$_4$ production comes from the three-body reaction between CH$_3$ and H, in agreement with \citet{strobel_photochemistry_1990}. This reaction accounts for 94.5\% of the integrated production and peaks at 10\,km, again at the photolysis maximum. 
At altitudes higher than 75\,km, CH$_4$ is produced by diverse ion-neutral reactions, the main one being CH$_5^+$ + CO $\longrightarrow$ CH$_4$ + HCO$^+$.

\paragraph{N($^4$S) and N($^2$D)}
~
\newline

Atomic nitrogen is the second most abundant species between 155\,km and the top of the atmosphere. In our chemical scheme, we consider two distinct states of atomic nitrogen: the ground state N($^4$S) and the first excited state N($^2$D). 

N($^2$D) is produced through reactions between N$_2$ and magnetospheric electrons (32\%), but also by photodissociation (12.5\%) and photoionization (1\%) of this species.
However, the main production process is the dissociative recombination of N$_2^+$ (54.5\%). The production peaks of all these reactions are located around 350\,km, except for the photodissociation of N$_2$ giving N($^4$S) + N($^2$D) whose peak is at 71\,km (the other channel producing 2N($^2$D) peaks at 378\,km). 

Then, N($^2$D) is quenched to ground state N($^4$S) through collisions with CO (75.5\%), C (15\%) and O($^3$P) (9.5\%), whose loss rates are maximum at 334\,km for the former reaction and 356\,km for the others. These reactions account for 47.5, 9.5 and 6\% of the integrated production of N($^4$S) respectively. This species is also produced by dissociative recombination of N$_2^+$ (11\%) and dissociation of N$_2$ by magnetospheric electrons (13\%). Photodissociation of N$_2$ accounts for 4.5\%. The CN + O($^3$P) $\longrightarrow$ N($^4$S) + CO reaction is the main production process of N($^4$S) around 120\,km where the production rate of this reaction is maximum. It contributes for 2.3\% of the integrated production of N($^4$S). 
\newline

As said in \citet{krasnopolsky_photochemistry_1995}, the CNN cycle is an important loss process for atomic nitrogen. In our case, it contributes for 81.5\% of the integrated loss of N($^4$S). Around 35\,km, N($^4$S) also reacts with species from methane photolysis such as H, CH$_3$ and $^3$CH$_2$, producing NH, H$_2$CN + H and HCN + H respectively. The rates of these reactions are at least one order of magnitude lower than the ones of the CNN cycle.  

\paragraph{H$_2$ and H}
~
\newline

The direct photolysis of CH$_4$ only accounts for 38.5\% of the integrated production of H$_2$ (considering the reactions giving directly H$_2$ from CH$_4$). H$_2$ is also produced through other reactions involving products of CH$_4$ photolysis such as H, $^3$CH$_2$ or CH$_3$. The main one is H + $^3$CH$_2$ which gives H$_2$ + CH (51\% of the integrated production). Consequently, H$_2$ is mainly produced around 10\,km, the altitude where the photolysis loss rate of CH$_4$ is maximum. 
\newline

Losses of H$_2$ mainly occur at higher altitude, through ion-neutral reactions with N$_2^+$ giving N$_2$H$^+$ + H (50\% of the integrated loss, maximum loss rate at 303\,km), with N$^+$ producing NH$^+$ + H (13\%, maximum at 356\,km), with CO$^+$ which gives (HCO$^+$, HOC$^+$) + H (14\%, maximum at 127\,km) and with CH$^+$ and CH$_2^+$ giving CH$_2^+$ + H and CH$_3^+$ + H respectively (12\% and 7\%, maxima at 127\,km). The reaction with N$^+$ is the main loss process above 550\,km.
\newline

H is also produced directly by CH$_4$ photolysis (43.5\%) and through the reaction CH + CH$_4$ $\longrightarrow$ C$_2$H$_4$ + H (28\%) near the methane photolysis peak. 
In the ionosphere, it is mainly produced through the N$_2^+$ + H$_2$ $\longrightarrow$ N$_2$H$^+$ + H reaction and by the dissociative recombination of the latter ion (each reaction contributes for 3.5\% of the integrated production of H). 
\newline 

H is mainly lost through reactions with $^3$CH$_2$ (56\% of the integrated loss) and HCNN (28.5\%) to produce CH + H$_2$ and $^1$CH$_2$ + N$_2$, respectively. These reactions are important in the lower atmosphere as they involve products of methane photolysis (HCNN is produced by N$_2$ + CH $\longrightarrow$ HCNN whose production rate peaks at 10\,km). 
H is also converted to H$_2$ through the following cycle:
\begin{align*}
    \text{H + C$_3$ }&\longrightarrow \text{ (c-C$_3$H,l-C$_3$H)  (1.30\%, 1.30\%)} \\
    \text{(c-C$_3$H,l-C$_3$H) + H }&\longrightarrow \text{ C$_3$ + H$_2$  (1.63\%, 1.02\%)}
\end{align*}

The three-body reaction H + H $\longrightarrow$ H$_2$ is an important loss process for H in \citet{krasnopolsky_photochemistry_1995} but this reaction is much less noticeable in our case, as it contributes  for only 0.025\% on the integrated loss. 

\paragraph{CO}
~
\newline

Losses of CO mostly occur in the ionosphere where it reacts with N$^+$, which explains the decrease of its relative abundance observed in Fig. \ref{fm_main_neutrals}. It also photodissociates and photoionizes in the lower atmosphere, with maximum rates reached at 181 and 127\,km respectively. It has to be noted that in our model CO absorbs solar radiation up to 163\,nm, but is only photoionized by radiation with $\lambda<89$\,nm and photodissociated by radiation with $\lambda \in [89,108[$\,nm. Thus, CO absorbs radiation between 108 and 163\,nm without being dissociated. This leads to an attenuation of the solar flux at these wavelengths, thus impacting the photolysis of hydrocarbons such as C$_2$H$_2$ (see Fig. \ref{penetration_0_200nm}). We should consider that CO molecules re-emit absorbed photons at these wavelengths in every direction, thus contributing to the photolysis of other species but this is not done in our model.  
\newline

CO is mainly produced through reactions of O($^3$P) with CNN and CN, which produce respectively N$_2$ + CO and N($^4$S) + CO, the latter having a slightly higher production rate. 

\paragraph{C}
~
\newline

We find a higher relative abundance of C than \citet{krasnopolsky_photochemistry_1995} throughout the atmosphere. The peak concentration is 5.2$\times$10$^7$ cm$^{3}$ at 167\,km compared with $\sim$1.5$\times$10$^7$ cm$^{3}$ at $\sim$130\,km in the cited article.
This species is mainly produced through the reaction N($^4$S) + CN $\longrightarrow$ N$_2$ + C (68.5\% of the integrated production of C), which is part of the CNN cycle converting atomic nitrogen into N$_2$. But this production is counterbalanced by the N$_2$ + C $\longrightarrow$ CNN reaction whose integrated rate is 9\% higher and contributes for 76\% of the integrated loss of C, being therefore the main loss process. The rates of these reactions are maximum at 121\,km. Various ionic reactions also produce C, such as the dissociative recombination of CO$^+$ (8\% of the integrated production), the radiative recombination of C$^+$ (1.5\%) or the ion-neutral reaction CO + N$^+$ $\longrightarrow$ C + NO$^+$ (1\%). 

Apart from the reaction with N$_2$, C is lost through various reactions with ions in the ionosphere, but these reactions are at least two orders of magnitude less significant than the former reaction.

\paragraph{O($^3$P)}
~
\newline

O($^3$P) is also an abundant species in our model. As in \citet{krasnopolsky_photochemistry_1995}, the peak concentration of O($^3$P) is about 10$^8$ cm$^{-3}$. We recall that we do not consider O($^1$D) here because its abundance is negligible in comparison to O($^3$P). 
This species is mainly produced by dissociative recombination of the CO$^+$ and NO$^+$ ions (the latter contributing approximately 8 times less than the former). These reactions are important in the ionosphere as their peak rate is reached at 345\,km. At lower altitudes, O($^3$P) is produced through the reaction N($^4$S) + NO $\longrightarrow$ O($^3$P) + N$_2$ with an integrated rate one order of magnitude lower than for the dissociative recombinations. 
\newline

O($^3$P) is mainly lost through reactions with CNN and CN as discussed above and whose maximum rates are reached at 121\,km. 
Below this altitude, O($^3$P) reacts with various species such as CH$_3$, NH or H$_2$CN, giving respectively (H$_2$CO + H, CO + H$_2$ + H), NO + H and (OH + HCN, HCNO + H), but the integrated rates of these reactions are one order of magnitude lower than those of the reactions with CN and CNN.

\subsubsection{Hydrocarbons and HCN}
\label{results_C2Hx}

\begin{figure}[!h]
   \resizebox{\hsize}{!}
            {\includegraphics{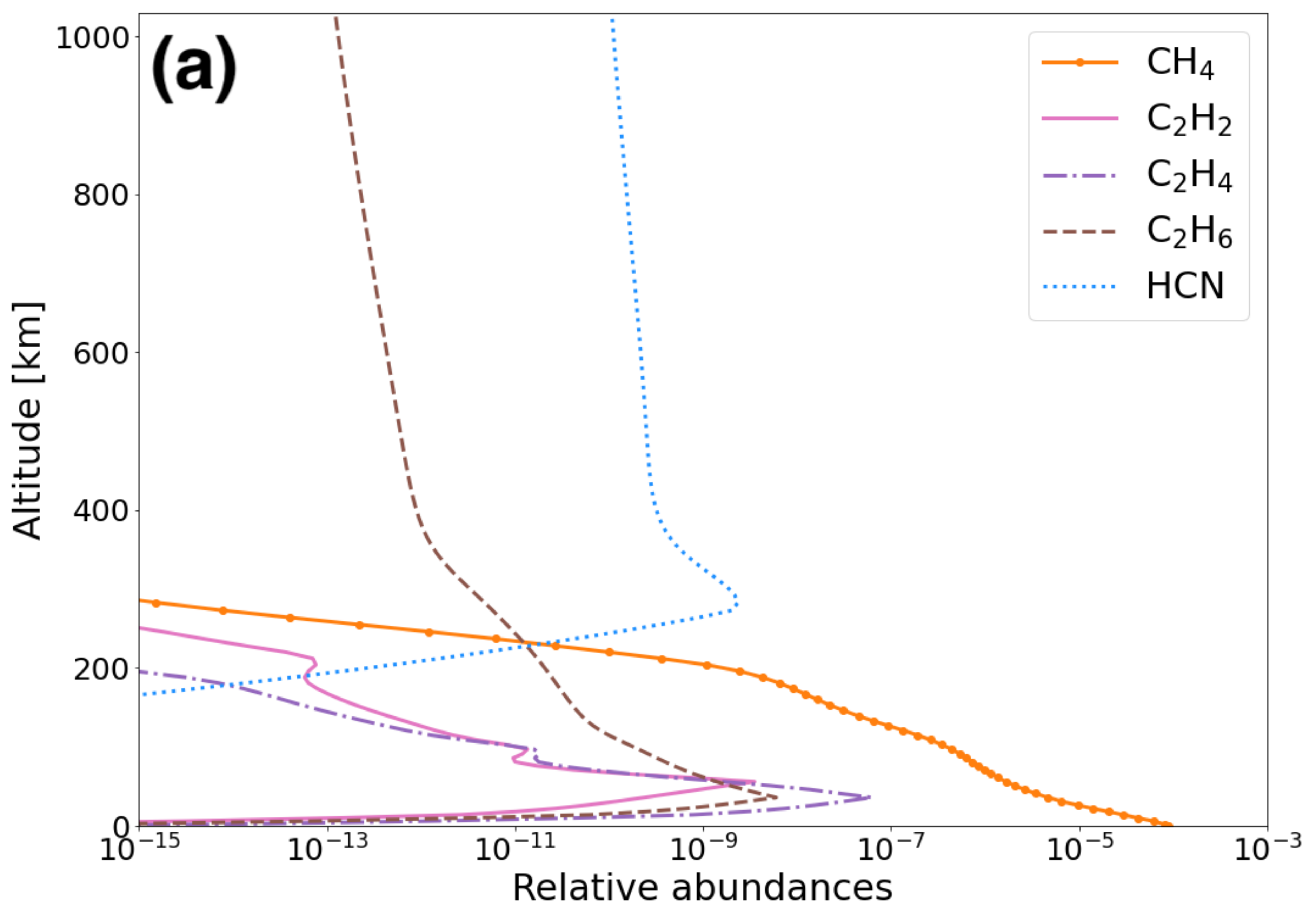}
            \includegraphics{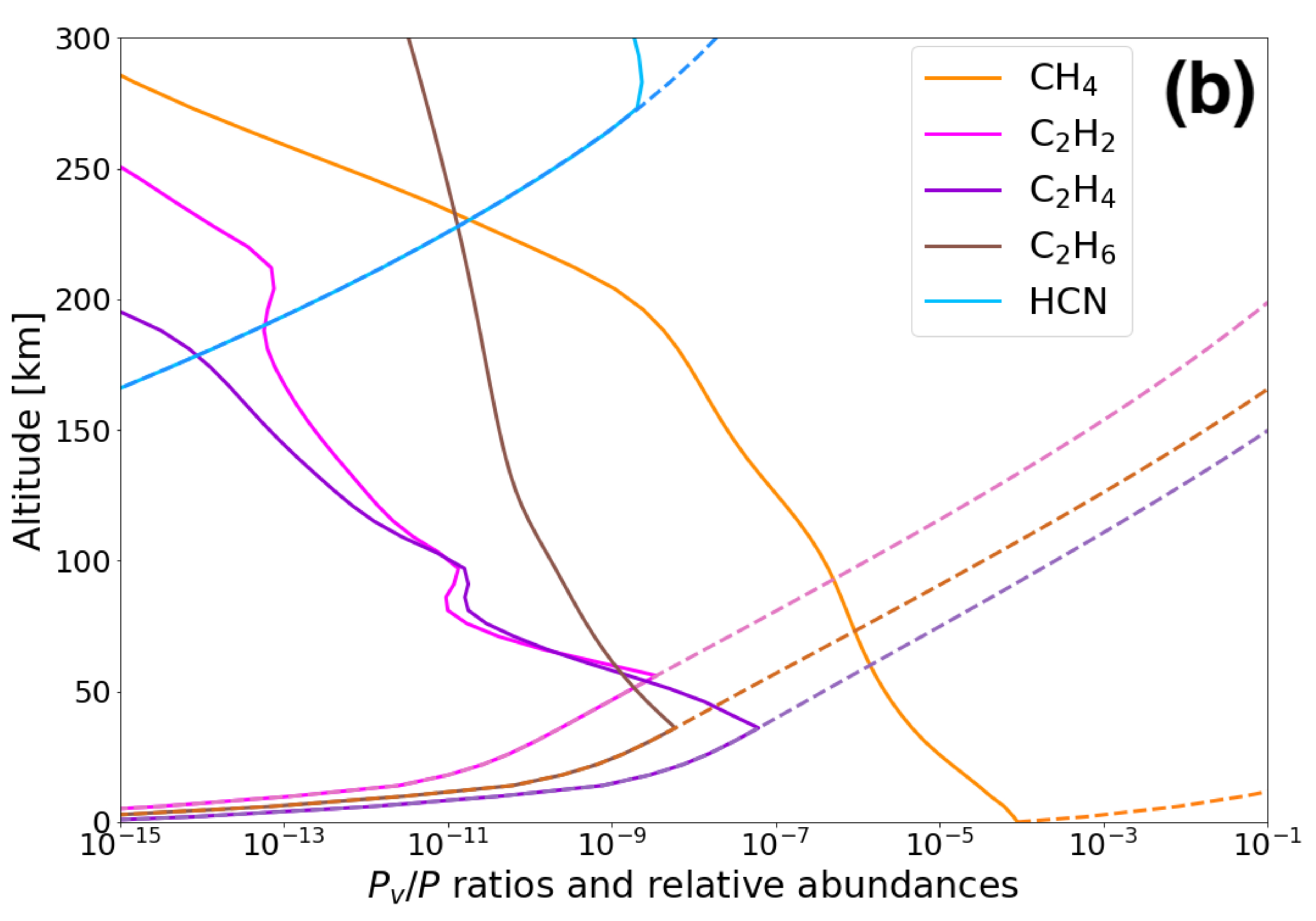}
            }
      \caption{\textbf{(a):} Relative abundances of the main hydrocarbons (orange solid line with circles for CH$_4$, solid pink line for C$_2$H$_2$, purple dash-dotted line for C$_2$H$_4$ and brown dashed line for C$_2$H$_6$) and HCN (blue dotted line). \textbf{(b):} Relative abundances (solid lines) compared to the ratio of the vapor pressure $P_v$ and the pressure $P$ (dashed lines) depending on altitude for CH$_4$, the main C$_2$H$_x$ and HCN. When the curves of the $P_v/P$ ratio and the abundance merge, this means that the species is condensing, as its abundance cannot be higher than this ratio.
              }
         \label{fm_main_c2hx+vap}
\end{figure}

The abundances of the main hydrocarbons and of HCN are presented in Fig. \ref{fm_main_c2hx+vap}. 

In agreement with \citet{krasnopolsky_photochemistry_1995}, the most abundant hydrocarbon in our model is C$_2$H$_4$. However, it is half as abundant in their model as in ours, with a peak concentration of 5.1$\times$10$^6$ cm$^{-3}$ at 47\,km where we have 2.0$\times$10$^7$ cm$^{-3}$ at 36\,km. 
We also find that C$_2$H$_6$ is more abundant than C$_2$H$_2$ as the peak concentration of the former is 1.9$\times$10$^6$ cm$^{-3}$ at 36\,km compared with 4.4$\times$10$^5$ cm$^{-3}$ at 56\,km for the latter. In \citet{krasnopolsky_photochemistry_1995}, these two species have approximately the same peak number density (1.3$\times$10$^6$ cm$^{-3}$ and 1.4$\times$10$^6$ cm$^{-3}$ respectively). These differences come from the different vapor pressure formula used in our model. In comparison, for the summer and winter hemispheres respectively, \citet{strobel_tritons_1995} find (2.6-1.6)$\times$10$^7$ cm$^{-3}$ at (17-30)\,km for C$_2$H$_4$ and (3.3$\times$10$^5$-1.3$\times$10$^6$) cm$^{-3}$ at (100-104)\,km for C$_2$H$_2$. 
In addition, our HCN abundance is much lower, its peak number density being 1.8$\times$10$^2$ cm$^{-3}$ against 3$\times$10$^6$ cm$^{-3}$ for \citet{krasnopolsky_photochemistry_1995} (while the peak concentration of HCN is nearly 10$^7$ cm$^{-3}$ in \citealt{strobel_tritons_1995}). This difference comes from the vapor pressure of HCN that is much lower than the ones of the studied hydrocarbons, as shown in the right-hand panel of Fig. \ref{fm_main_c2hx+vap}, and which forces the number density of this species to drop below 275\,km.
As for neutral species, we discuss the main production and loss processes for these compounds in the following paragraphs. 

\paragraph{C$_2$H$_2$}
~
\newline

C$_2$H$_2$ is the least abundant of the three studied C$_2$H$_x$ in the lower atmosphere, as its vapor pressure is lower than those of C$_2$H$_4$ and C$_2$H$_6$.  
Its production relies almost exclusively on methane photolysis as it is produced at 53.5\% through $^3$CH$_2$ + $^3$CH$_2$ $\longrightarrow$ C$_2$H$_2$ + H$_2$ and 39\% through CH$_3$ + HCNN $\longrightarrow$ C$_2$H$_2$ + H$_2$ + N$_2$. The remaining production comes from C$_2$H$_4$ photolysis (4.5\%). The two former reactions reach their peak production rate at 10\,km whereas C$_2$H$_4$ photolysis peaks at 36\,km. 

C$_2$H$_2$ is mainly lost around 56\,km where it reacts with C to form C$_3$ + H$_2$ and c-C$_3$H + H (37 and 59.5\%) or is photodissociated, which gives C$_2$H + H (2\%). The C + C$_2$H$_2$ $\longrightarrow$ l-C$_3$H + H channel also exists but its integrated rate is two orders of magnitude lower than the one of the c-C$_3$H channel. 
C$_2$H$_2$ condenses below 60\,km, as the temperature falls when approaching the surface, as shown in Fig. \ref{fm_main_c2hx+vap}. 
The integrated condensation rate of C$_2$H$_2$ is 1.1$\times$10$^7$ cm$^{-2}$.s$^{-1}$, which corresponds to a mass condensation rate of 4.6$\times$10$^{-16}$ g.cm$^{-2}$.s$^{-1}$. This value is one order of magnitude higher than the one of \citet{krasnopolsky_photochemistry_1995}, which is 4$\times$10$^{-17}$ g.cm$^{-2}$.s$^{-1}$. This difference comes from the use of a lower vapor pressure profile and a different chemical scheme, leading to a higher integrated production rate for this compound compared to \citet{krasnopolsky_photochemistry_1995}. 
\newpage

\paragraph{C$_2$H$_4$}
~
\newline

C$_2$H$_4$ is the most abundant C$_2$H$_x$ species. 
\citet{strobel_photochemistry_1990} stated that C$_2$H$_4$ is produced through the reaction $^3$CH$_2$ + $^3$CH$_2$ $\longrightarrow$ C$_2$H$_4$ and also through CH + CH$_4$ $\longrightarrow$ C$_2$H$_4$ + H. 
In our case, we effectively find that the latter is responsible for 84\% of the integrated production but the former is not taken into account, as we consider $^3$CH$_2$ + $^3$CH$_2$ $\longrightarrow$ C$_2$H$_2$ + H$_2$. Instead, 15.5\% of the production is due to the reaction $^3$CH$_2$ + CH$_3$ $\longrightarrow$ C$_2$H$_4$ + H. The production rates of these reactions are maximum at 10\,km, which is consistent with the fact that C$_2$H$_4$ derives from methane photolysis, which is maximum at this altitude. 

C$_2$H$_4$ is lost by photodissociation (22.5\%), yielding C$_2$H$_2$, C$_2$H$_3$, H$_2$ and H. It also reacts with C (9.5\%) to produce C$_3$H$_3$ and H, but the main loss process is the three-body reaction with H giving C$_2$H$_5$ (60.5\%). The photodissociation peak is located at 36\,km, as the maximum rate of the three-body reaction, whereas the reaction with C peaks at 46\,km. 
C$_2$H$_4$ condenses below 40\,km (Fig. \ref{fm_main_c2hx+vap}). 
The integrated condensation rate of C$_2$H$_4$ is 9.0$\times$10$^7$ cm$^{-2}$.s$^{-1}$, which corresponds to a mass condensation rate of 4.2$\times$10$^{-15}$ g.cm$^{-2}$.s$^{-1}$. \citet{krasnopolsky_photochemistry_1995} find 4.3$\times$10$^{-15}$ g.cm$^{-2}$.s$^{-1}$, which is very close to our value. 

\paragraph{C$_2$H$_6$}
~
\newline

This species is entirely produced by a three-body reaction between two CH$_3$ (99.98\%), with a maximum production rate at 10\,km, again due to methane photolysis. 

It is destroyed by photolysis (97.5\%) and reaction with CN (2\%) and condenses below 40\,km (Fig. \ref{fm_main_c2hx+vap}). All these reactions reach their maximum loss rate at 36\,km.
The integrated condensation rate of C$_2$H$_6$ is 2.0$\times$10$^7$ cm$^{-2}$.s$^{-1}$, which corresponds to a mass condensation rate of 1.0$\times$10$^{-15}$ g.cm$^{-2}$.s$^{-1}$. The value given in \citet{krasnopolsky_photochemistry_1995} is 8.9$\times$10$^{-16}$ g.cm$^{-2}$.s$^{-1}$, which is slightly lower than ours. The ratio between these two rates is nearly the same as the ratio of our integrated production rates, thus the difference comes from the chemical scheme. 

\paragraph{HCN}
~
\newline

HCN is mostly produced through reactions involving H$_2$CN (produced through the reaction N($^4$S) + CH$_3$ $\longrightarrow$ H$_2$CN + H), which reacts with H and O($^3$P) to give respectively HCN + H$_2$ (47\% of the integrated production) and HCN + OH (6.5\%). But it is also produced through the following reactions involving N($^4$S):
\begin{align*}
    \text{N($^4$S) + $^3$CH$_2$ } &\longrightarrow \text{ HCN + H   (30\%)}\\
    \text{N($^4$S) + HCNN } &\longrightarrow \text{ HCN + N$_2$   (13.5\%)}
\end{align*}
The maximum rates of these reactions are reached between 31 and 51\,km.

HCN is lost in the ionosphere through reactions with N($^2$D) giving CH and N$_2$ (19\%) and with H$^+$ giving HNC$^+$ and H (80.5\%). These reaction rates peak at 303\,km and 293\,km respectively.
This species condenses below 280\,km (Fig. \ref{fm_main_c2hx+vap}), so at much higher altitude than the C$_2$H$_x$ species. 

\subsubsection{Radicals}

\begin{figure}[!h]
   \centering
   \includegraphics[width=10cm]{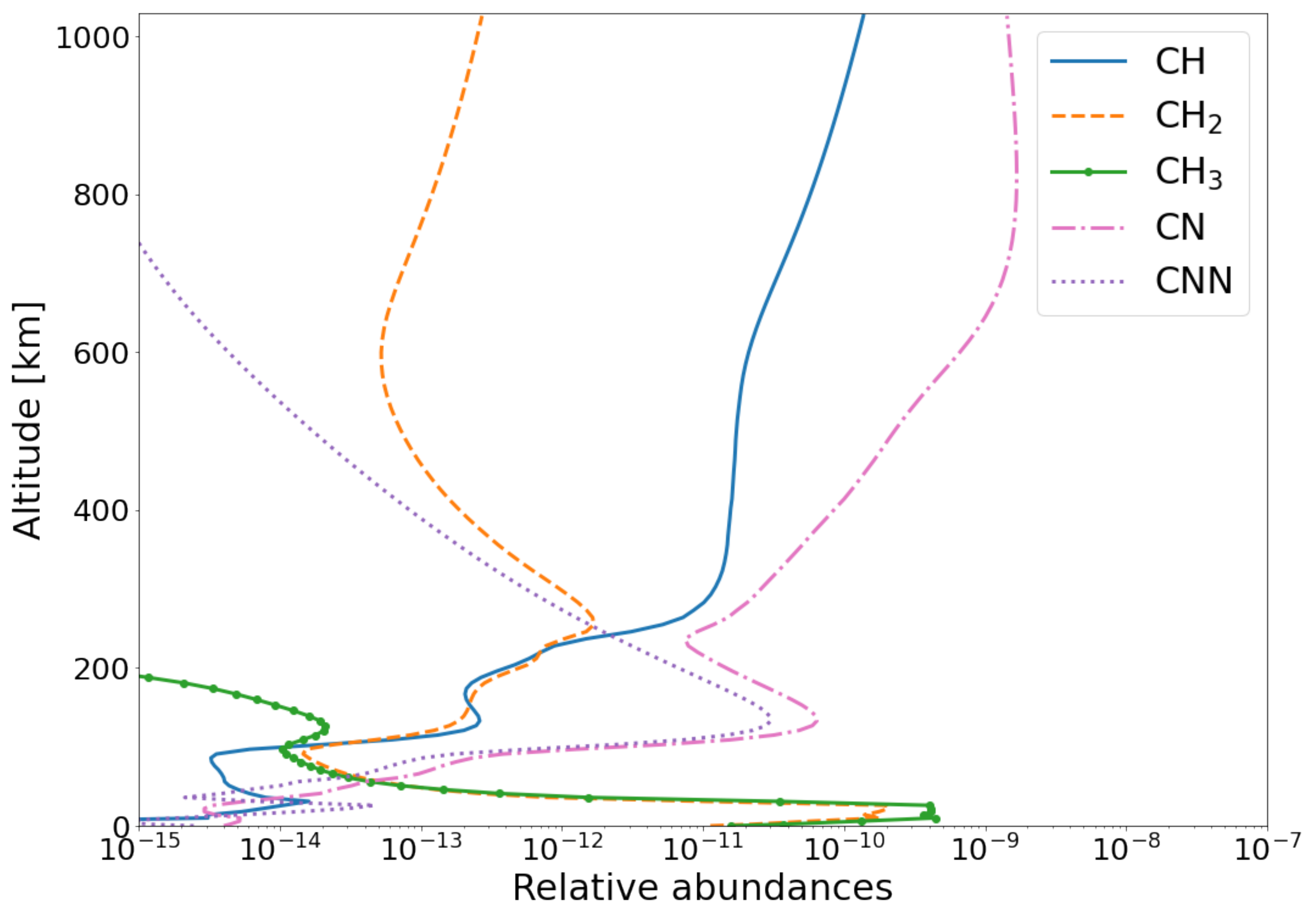}
      \caption{Relative abundances of the radicals that appeared in the key chemical reactions of the main neutral species (solid blue line for CH, orange dashed line for CH$_2$, green solid line with circles for CH$_3$, dash-dotted pink line for CN and purple dotted line for CNN). The CH$_2$ profile corresponds to $^3$CH$_2$, as the relative abundance of $^1$CH$_2$ is negligible.
              }
         \label{fm_main_radicals}
   \end{figure}

As radicals were evoked during our study of key chemical reactions for the main neutral species, we give their relative abundances in Fig. \ref{fm_main_radicals}. $^1$CH$_2$ is completely converted to $^3$CH$_2$ through collisions with N$_2$. Therefore, its abundance is negligible and we focus on $^3$CH$_2$ in the following. 

In agreement with what we discussed above, the production of CH$_3$ and $^3$CH$_2$ is maximum at 10\,km as it depends on methane direct photolysis, this process contributing for respectively 97\% and 8.5\% of the integrated production of these species. The remaining production of $^3$CH$_2$ comes from collisions between $^1$CH$_2$ and N$_2$ (91\%). $^1$CH$_2$ is also produced by direct methane photolysis (54.5\%) and through H + HCNN $\longrightarrow$ $^1$CH$_2$ + N$_2$ (45\%), thus depending on H which is also a photolysis product.

CH$_3$ mainly reacts with other CH$_3$ radicals to produce C$_2$H$_6$ (55.5\% of the integrated loss), but also with $^3$CH$_2$ (19\%) and N($^4$S) (13.5\%), which yields C$_2$H$_4$ + H and H$_2$CN + H respectively.
The reaction between CH$_3$ and $^3$CH$_2$ accounts for 8.5\% of the integrated loss of $^3$CH$_2$. The latter compound reacts mainly with H to produce CH + H$_2$ (80\% of the integrated loss). It also reacts with other $^3$CH$_2$ radicals to form C$_2$H$_2$ + H$_2$ (7\%). All these reactions reach their maximum rate at 10\,km apart from the N($^4$S) + CH$_3$ reaction whose maximum is at 31\,km. 
\newline

CH is mainly produced through the $^3$CH$_2$ + H $\longrightarrow$ CH + H$_2$ reaction (87\%) and direct methane photolysis (12\%). It is mainly lost through CH + CH$_4$ $\longrightarrow$ C$_2$H$_4$ + H (49.5\%) and CH + N$_2$ $\longrightarrow$ HCNN (49\%). All these reactions also reach their maximum rate at 10\,km.
\newline

CN and CNN are mostly produced and lost through the CNN cycle that converts atomic nitrogen to N$_2$, as seen above. Thus, CN is almost exclusively produced through the reaction N($^4$S) + CNN $\longrightarrow$ CN + N$_2$ (96\%) and CNN through C + N$_2$ $\longrightarrow$ CNN (100\%). CN is then mostly lost through N($^4$S) + CN $\longrightarrow$ C + N$_2$ (94\%) and CNN through N($^4$S) + CNN $\longrightarrow$ CN + N$_2$ (92.5\%). These reactions all reach their maximum rate at 121\,km.

In addition, CN and CNN react with O($^3$P) to yield CO +  N($^4$S) and CO + N$_2$ respectively, these reactions accounting for 6 and 5\% of the integrated loss. 

\subsubsection{Heavier C$_x$-compounds}

The most abundant C$_3$H$_x$ species is C$_3$ with a peak relative abundance of 1.4$\times$10$^{-7}$ at 103\,km. It intervenes in a cycle converting atomic hydrogen into molecular hydrogen, which we have discussed above. Reactions of this cycle account for 98.25\% of the integrated production of C$_3$ and 96.28\% of its integrated loss. The integrated rates of the production reactions are slightly higher than the ones of the loss reactions. Apart from this cycle, C$_3$ is produced through the reaction C + C$_2$H$_2$ $\longrightarrow$ C$_3$ + H$_2$ and lost by photodissociation producing $^3$C$_2$ + C. 
\newline

Aside this species, the other neutral C$_3$-compounds are much less abundant. For example, the second most abundant is C$_3$H$_3$CN and the third is c-C$_3$H$_2$ with respective peak relative abundances of 1.7$\times$10$^{-10}$ at 46\,km and 2.5$\times$10$^{-11}$ at 181\,km. 
C$_3$H$_3$CN is mostly produced through the reaction CN + CH$_3$CCH $\longrightarrow$ C$_3$H$_3$CN + H (11\% of the integrated production) and H + CH$_2$C$_3$N $\longrightarrow$ C$_3$H$_3$CN (79\%). It is lost at 77.2\% by photodissociation, producing CH$_2$C$_3$N + H and C$_3$H$_3$ + CN. It also reacts with C$_2$H$_5^+$ to produce CH$_3$C$_3$NH$^+$ + C$_2$H$_4$ (15\% of the integrated loss). 
c-C$_3$H$_2$ is mainly produced through H + (c-C$_3$H + l-C$_3$H) $\longrightarrow$ c-C$_3$H$_2$ (72\%) and dissociative recombinations of c-C$_3$H$_3^+$ (5.9\%), l-C$_3$H$_3^+$ (3.7\%) and C$_3$H$_5^+$ (9.2\%).  c-C$_3$H$_2$ is almost exclusively lost through a three-body reaction with H producing C$_3$H$_3$ (97.7\% of the integrated loss).
\newline

l-C$_3$H$_3^+$ is the most abundant C$_3$H$_x$ ion with a peak relative abundance of 2.0$\times$10$^{-12}$ at 153\,km. It is mainly produced through C$_3$H$_4^+$ + H $\longrightarrow$ l-C$_3$H$_3^+$ + H$_2$ (12.7\%).
It is lost by dissociative recombination (91.5\%) and reactions with C$_2$H$_4$ producing heavier ions C$_5$H$_5^+$ and C$_2$H$_7^+$ (3.5\% of the integrated loss for each channel). 

The second most abundant C$_3$H$_x$ ion is C$_3$H$_5^+$ with a peak relative abundance of 7.9$\times$10$^{-13}$ at 109\,km. It is produced through CH$_4$ + (C$_2$H$_2^+$, C$_2$H$_3^+$) $\longrightarrow$ C$_3$H$_5^+$ + (H, H$_2$) (10\%, 83\%) and C$_2$H$_4$ + (C$_2$H$_4^+$, C$_2$H$_5^+$) $\longrightarrow$ C$_3$H$_5^+$ + (CH$_3$, CH$_4$) (5\%, 1.5\%). It is mainly lost by reacting with H (19\%), which gives C$_2$H$_3^+$ + CH$_3$ or by dissociative recombination (68\%). 
It also reacts with C$_2$H$_4$ to produce C$_5$H$_7^+$ + H$_2$ (10\%). 
\newline

In total, eight neutral C$_3$-compounds and eight C$_3$H$_x^+$ ions have a relative abundance higher than 10$^{-15}$. In the same interval, we find seven neutral C$_4$-compounds, the most abundant being nC$_4$H$_8$ with a peak abundance of 6.5$\times$10$^{-13}$.  
We also identify six heavier ions C$_8$H$_{11}^+$, C$_5$H$_5^+$, C$_7$H$_7^+$, C$_5$H$_7^+$, C$_5$H$_9^+$ and C$_6$H$_7^+$ with peak abundances ranging from 1.6$\times$10$^{-14}$ to 2.2$\times$10$^{-15}$ (species are given in order of decreasing abundance).

\subsubsection{Main ions}
\label{subsec_ionsmaj}

\begin{figure}[!h]
   \resizebox{\hsize}{!}
            {\includegraphics{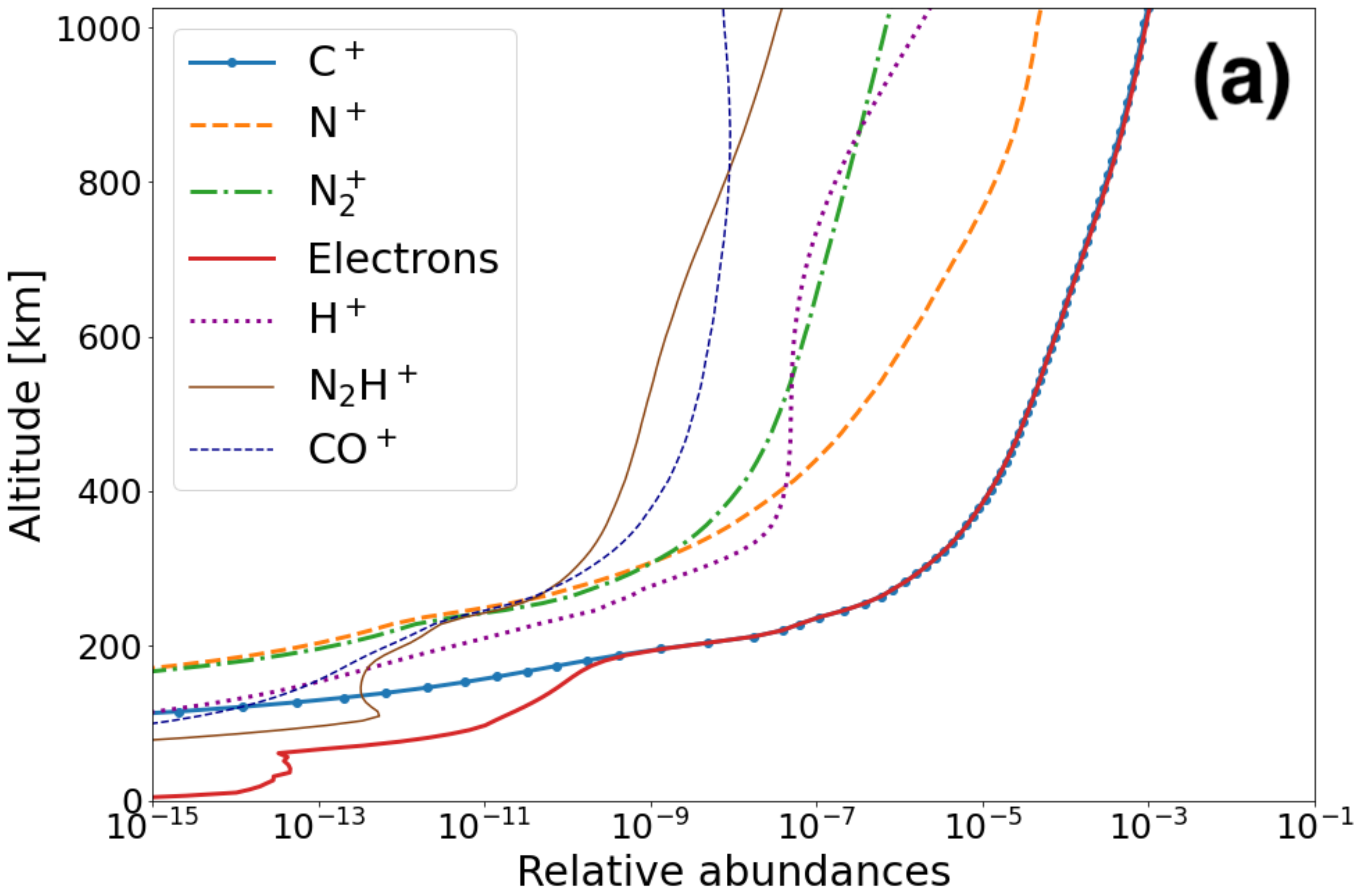}
            \includegraphics{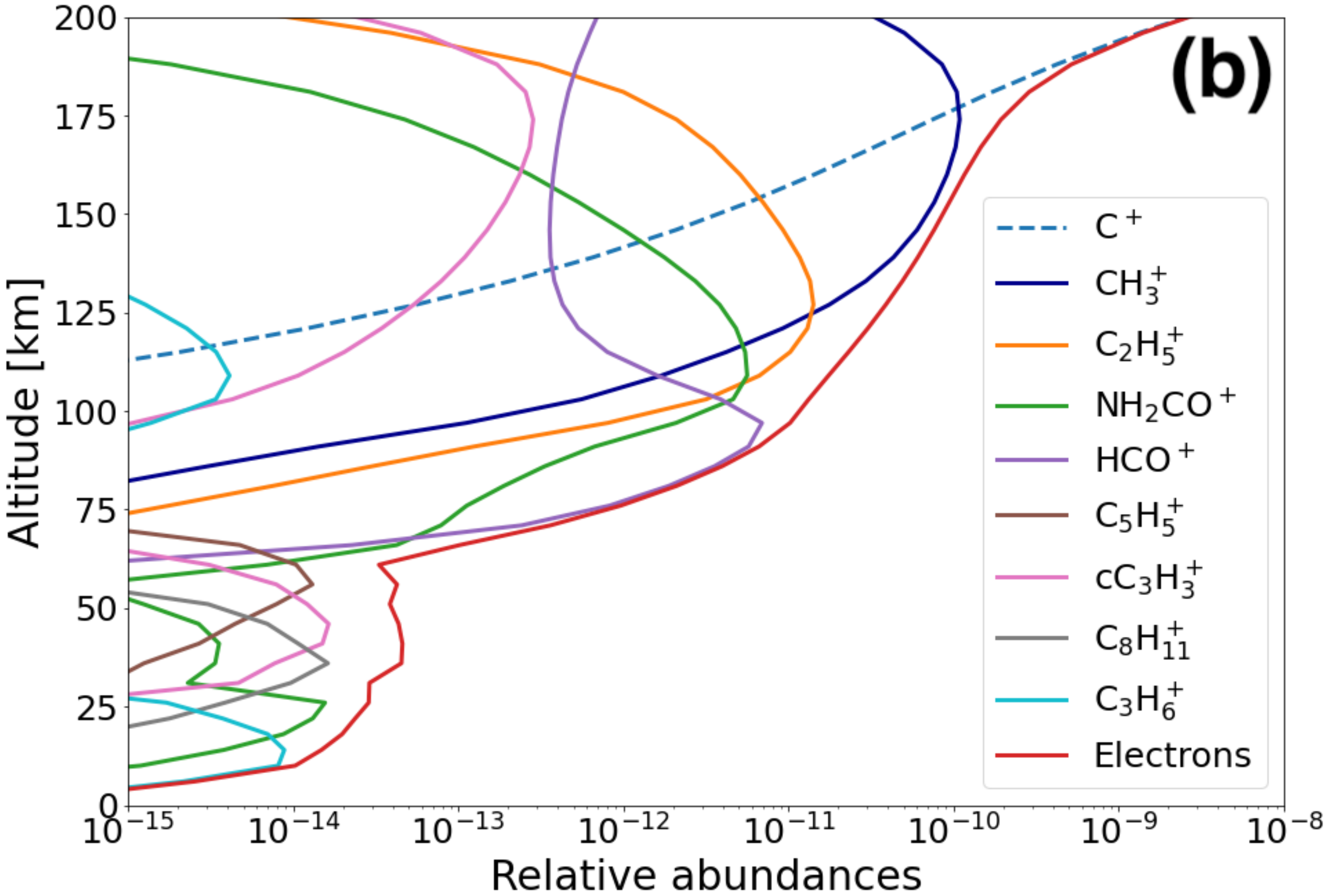}
            }
      \caption{Relative abundances of the main ions of the atmosphere of Triton. \textbf{(a):} Most abundant ions, which we focus on (blue solid line with circles for C$^+$, orange dashed line for N$^+$, green dash-dotted line for N$_2^+$, purple dotted line for H$^+$, thin brown solid line for CO$^+$, thin dark blue dashed line for N$_2$H$^+$ and red solid line for electrons). \textbf{(b):}  The most abundant ions below 175\,km.
              }
         \label{fm_main_ions}
   \end{figure}

Using nominal chemical reaction rates, we find that the main ions of Triton's ionosphere are C$^+$, N$^+$, H$^+$ and N$_2^+$, as shown in Fig. \ref{fm_main_ions}. 

The electronic number density is maximum at 334\,km, which is close to the interval (340-350)\,km given in \citet{tyler_voyager_1989}. 
We can see that the electronic number density quickly increases above 175\,km, where the concentration of C$^+$ varies strongly. 
It nearly corresponds to the sharp ionospheric cutoff around 200\,km observed in Voyager's data and shown in \citet{tyler_voyager_1989}. 
In \citet{krasnopolsky_photochemistry_1995}, the main ions were C$^+$ and N$^+$, H$^+$ being only the sixth most abundant ion. Another difference is that in their model C$^+$ and N$^+$ abundances tend to converge after the electronic peak, but this behavior in less pronounced in our results.

In panel (b) of Fig. \ref{fm_main_ions}, we present the most abundant ions below 175\,km. We can observe that the higher mass ions reach their peak relative abundance at lower altitude than the lower mass ones, with CH$_3^+$ being the main ion between 175 and 125\,km. Then C$_2$H$_5^+$, NH$_2$CO$^+$ and HCO$^+$ are most abundant between 125 and 60\,km and finally C$_5$H$_5^+$, c-C$_3$H$_3^+$, C$_8$H$_{11}^+$, C$_3$H$_6^+$ and NH$_2$CO$^+$ dominate below 60\,km. 
This is consistent with the fact that heavier species are abundant in the lower atmosphere only (e.g. hydrocarbons) whereas lighter species (e.g. atomic species as C, O and N) are dominant at higher altitudes. However, the relative abundances of these heavy ions remain low in comparison to the lighter ions in the upper atmosphere. Therefore, we do not focus on the lower atmosphere ions in the rest of our study.   

\subsubsection{Photoionization and interaction with magnetospheric electrons}
\label{photoionisation_and_ME}

The photoionization reactions with the highest integrated column rates are listed in Table \ref{photoion_rates}. These reactions contribute for 99.98\% of the total photoionization integrated column rate. 
For the interaction with magnetospheric electrons, the main reactions and their integrated column rates are given in Table \ref{ME_rates}.
Unsurprisingly, ionization of N$_2$ dominates as it is the main atmospheric species. These reactions are sources of ions N$_2^+$, N$^+$ but also of atomic nitrogen. Dissociation of N$_2$ by magnetospheric electrons is also a source of the latter species. 
We note that the peak of these reactions is located slightly above the electronic peak, which is located at 334\,km. 

\begin{table}[!h]
   \begin{center}
      \begin{tabular}{c c c}
        \hline
Photoionization reaction & \begin{tabular}[c]{@{}c@{}}Integrated \\ colum rate \\ {[}cm$^{-2}$.s$^{-1}${]}\end{tabular} & \begin{tabular}[c]{@{}c@{}}Maximum rate\\  altitude {[}km{]}\end{tabular} \\ \hline
N$_2$ + $h\nu \longrightarrow$ N$_2^+$ + $e^-$  & 2.7$\times$10$^7$  & 390 \\ 
N$_2$ + $h\nu \longrightarrow$ N$^+$ + N($^2$D) + $e^-$ & 2.2$\times$10$^6$ & 345 \\
CO + $h\nu \longrightarrow$ CO$^+$ + $e^-$ & 1.3$\times$10$^6$ & 127 \\ \hline
\end{tabular}
   \end{center}
\caption[]{Photoionization reactions with the highest integrated column rates.}
\label{photoion_rates}
\end{table}

\begin{table}[!h]
   \begin{center}
      \begin{tabular}{c c c}
        \hline
Reaction with ME  & \begin{tabular}[c]{@{}c@{}}Integrated\\  colum rate\\  {[}cm$^{-2}$.s$^{-1}${]}\end{tabular} & \begin{tabular}[c]{@{}c@{}}Maximum \\ rate altitude \\ {[}km{]}\end{tabular} \\ \hline
N$_2$ + ME $\longrightarrow$ N$_2^+$ + $e^-$  & 6.6$\times$10$^7$  & 345  \\ 
N$_2$ + ME $\longrightarrow$ N$^+$ + N($^2$D) + $e^-$  & 1.6$\times$10$^7$  & 345 \\
N$_2$ + ME $\longrightarrow$ N($^4$S) + N($^2$D) + $e^-$ & 4.9$\times$10$^7$ & 345 \\ \hline
\end{tabular}
   \end{center}
\caption[]{Ionization and dissociation reactions by magnetospheric electrons (ME) and their integrated column rates.}
\label{ME_rates}
\end{table}

We can see that ionization by magnetospheric electrons is more important than photoionization. The ratio between the rates for photoionization and the rates for magnetospheric interaction is 3/8, which is comparable to the ratio given in \citet{krasnopolsky_photochemistry_1995} of 0.5. 

\subsubsection{Production and loss processes}

We detail here the main production and loss processes for the main ions of the ionosphere of Triton. 

\paragraph{C$^+$}
~
\newline

C$^+$ is the most abundant ionospheric ion in \citet{strobel_tritons_1995} and \citet{krasnopolsky_photochemistry_1995}. \citet{lyons_solar_1992} were the first to consider C$^+$ as an abundant ion after using a charge exchange reaction between N$_2^+$ and C. 
In our model, this reaction is the main source of C$^+$, accounting for 74.5\% of the integrated production. This ion is also produced by two other charge exchange reactions between C and N$^+$ or CO$^+$, with respective contributions of 11 and 14.5\%. The maximum production rate of the reaction between C and N$_2^+$ is located at 334\,km, which corresponds to the electronic concentration peak. The production peak for charge exchange with N$^+$ is located at 414\,km and the one for charge exchange with CO$^+$ at 313\,km. 
\newline

C$^+$ is almost exclusively lost by radiative recombination (98\%) whose rate is maximum at 334\,km also.
In \citet{krasnopolsky_photochemistry_1995}, the main chemical process for loss of C$^+$ is by reacting with CH$_4$, but in our case, this reaction has an integrated loss rate 10$^4$ times lower than the radiative recombination reaction mentioned before. This is due to the very low number density of CH$_4$ at ionospheric altitudes. 
Moreover, we do not consider atmospheric escape for ions, which is the main loss process in \citet{krasnopolsky_photochemistry_1995}. This may explain why we have a higher number density of C$^+$. 

\paragraph{N$^+$}
~
\newline

N$^+$ is the second main ion of the ionosphere, as in \citet{krasnopolsky_photochemistry_1995}. On the other hand, in \citet{strobel_tritons_1995}, N$^+$ is the second main ion between 250 and 550\,km and then becomes the most abundant ion. In our case, N$^+$ was the main ion with the initial chemical scheme but this changed with the updated one where the N$_2^+$ + C $\longrightarrow$ N$_2$ + C$^+$ reaction was added, making C$^+$ the main ion. 
In our updated chemical scheme, we also added reactions between N$^+$ and CO, based on \citet{anicich_index_2003}. These reactions became important for N$^+$, as they account for 87.5\% of its integrated loss. Their loss rate is maximum at 345\,km, where the ionization reactions of N$_2$ giving N$^+$ are maximum.  
N$^+$ also reacts with H$_2$ to produce NH$^+$ + H. The loss rate of this reaction is maximum at 356\,km and accounts for 11\% of the integrated loss of N$^+$. 
\newline

The ionization of N$_2$ by magnetospheric electrons contributes for 75\% of the integrated production of N$^+$. Photoionization contributes for 10\% and charge exchange between N$_2^+$ and atomic nitrogen for 15\%. The first two reactions reach the production peak at 345\,km whereas the latter reaches it at 334\,km.

\paragraph{N$_2^+$}
~
\newline

Even though N$_2$ is the main atmospheric species and the ionization reactions giving N$_2^+$ have a higher branching ratio compared to the ones giving N$^+$, N$_2^+$ is only the third or fourth most abundant ion of Triton's ionosphere depending on altitude. It is produced by photoionization and electron impact ionization of N$_2$ (respectively 29 and 71\% of the integrated production) but it quickly recombines with electrons to produce atomic nitrogen (82\% of the integrated loss). It also reacts with H$_2$ to produce N$_2$H$^+$ (10.5\%) and with C, N($^4$S) and CO through charge exchange reactions (7\% in total). The loss rate for dissociative recombination is maximum at 356\,km, whereas the one for charge exchange with C and N($^4$S) is at 334\,km (293\,km for charge exchange with CO). For the reaction with H$_2$, the maximum rate is reached at 303\,km.  

\paragraph{H$^+$}
~
\newline

H$^+$ is mostly produced through a charge exchange reaction between CO$^+$ and H whose maximum rate is reached at 293\,km and accounts for 96\% of the integrated production. Photoionization of H contributes for 3\% and is maximum at 146\,km. 

H$^+$ is lost by radiative recombination (29\% of the integrated loss) around 345\,km, but mostly by reacting with HCN and HNC to produce HNC$^+$ + H (45.88\%), whose rates are maximum at 293 and 264\,km respectively. It also appears in charge exchange reactions with C and C$_3$ (4.5 and 17\%) and reacts with CH$_4$ to produce CH$_3^+$ + H$_2$ (2.5\%).

\subsection{Key chemical reactions for the main species}

We studied in the previous sections the main production and loss processes for the main species of Triton's atmosphere. The reactions associated to these processes thus contribute significantly to the production or loss of these species. We call these reactions key chemical reactions. All these reactions are given as supplementary material. Table \ref{key_react_chimie_final_table} displays the reactions that contribute significantly to the production or loss of several of the main atmospheric species. 

\begin{table}[!h]     
   \begin{center}
      \begin{tabular}{l l l}     
   \hline                    
   Reaction & Species (production) & Species (loss) \\ \hline
   CH$_4$ + $h\nu$ $\longrightarrow$ CH$_3$  + H  & H (26\%) & CH$_4$ (27\%)  \\
   CH$_4$ + $h\nu$ $\longrightarrow$ $^1$CH$_2$ + H$_2$ & H$_2$ (32\%)                      & CH$_4$ (31\%)               \\
   H + $^3$CH$_2$ $\longrightarrow$ CH + H$_2$          & H$_2$ (51\%)                      & H (56\%)                    \\
   CH$_4$ + CH $\longrightarrow$ C$_2$H$_4$ + H         & H (28\%) ; C$_2$H$_4$ (84\%)      & CH$_4$ (29\%)               \\
   H + HCNN $\longrightarrow$ $^1$CH$_2$ + N$_2$        & N$_2$ (12\%)                      & H (29\%)                    \\
   N$_2$ + $h\nu$ $\longrightarrow$ N$_2^+$ + $e^-$         & N$_2^+$ (29\%) ; $e^-$ (24\%)         &                             \\
   N$_2$ + ME $\longrightarrow$ N$_2^+$ + $e^-$             & N$_2^+$ (71\%) ; $e^-$ (59\%)         & N$_2$ (12\%)                \\
   N$_2$ + ME $\longrightarrow$ N$^+$ + N($^2$D) + $e^-$    & N$^+$ (75\%) ; $e^-$ (15\%)           &                             \\
   N$_2$ + ME $\longrightarrow$ N($^4$S) + N($^2$D)     & N($^4$S) (13\%) ; N($^2$D) (24\%) &                             \\
   N$^+$ + H$_2$ $\longrightarrow$ H + NH$^+$           &                                   & H$_2$ (13\%) ; N$^+$ (11\%) \\
   N$_2^+$ + H$_2$ $\longrightarrow$ N$_2$H$^+$ + H     &                                   & N$_2$ (10\%) ; H$_2$ (50\%) \\
   N$_2^+$ + $e^-$ $\longrightarrow$ N($^4$S) + N($^2$D)    & N($^4$S) (11\%); N($^2$D) (20\%)  & N$_2^+$ (44\%) ; $e^-$ (37\%)   \\
   N$_2^+$ + $e^-$ $\longrightarrow$ N($^2$D) + N($^2$D)    & N($^2$D) (34\%)                   & N$_2^+$ (38\%) ; $e^-$ (31\%)   \\
   N($^4$S) + CN $\longrightarrow$ N$_2$ + C            & N$_2$ (25\%) ; C (69\%)           & N($^4$S) (40\%)             \\
   N($^4$S) + CNN $\longrightarrow$ N$_2$ + CN          & N$_2$ (25\%)                      & N($^4$S) (41\%)             \\
   N$_2$+ C $\longrightarrow$ CNN                       &                                   & N$_2$ (27\%) ; C (76\%)     \\
   N($^2$D) + CO $\longrightarrow$ N($^4$S) + CO        & N($^4$S) (47\%)                   & N($^2$D) (75\%)             \\
   CO$^+$ + $e^-$ $\longrightarrow$ C + O($^3$P)            & O($^3$P) (37\%)                   & $e^-$ (14\%)                    \\
   H$^+$ + HCN $\longrightarrow$ HNC$^+$ + H            &                                   & H$^+$ (20\%) ; HCN (81\%)   \\ \hline
   \end{tabular}
   \end{center}     
   \caption{Key chemical reactions for the main atmospheric species. These reactions contribute for at least 10\% of the integrated production or loss of at least two of the main species. The contribution of the reaction to the total integrated loss or production of a species is noted next to the species name.} 
   \label{key_react_chimie_final_table} 
   \end{table}

We can identify several groups of reactions in this table: 
\begin{itemize}
    \item CH$_4$ photolysis: unsurprisingly, we find that photolysis of CH$_4$ is important, as stated in \citet{strobel_photochemistry_1990,krasnopolsky_photochemistry_1995,strobel_tritons_1995}. Its products also appear in other key chemical reactions. It triggers the chemistry in the lower atmosphere, the photolysis peak being located at 10\,km. It is a source of H, H$_2$ and radicals. 
    \item Photoionization of N$_2$ and reactions with magnetospheric electrons: these reactions impact logically the number density of N$_2$, N$_2^+$, N$^+$, electrons and atomic nitrogen. These products appear in numerous reactions of the table (as presented in the following points), thus the former reactions are important for the atmospheric chemistry in general. 
    \item N$_2^+$ dissociative recombinations: these reactions have a significant impact on the electronic and N$_2^+$ loss as these species recombine together quickly. This gives atomic nitrogen in the ground or first excited state. 
    \item Atomic nitrogen: several reactions involve atomic nitrogen. We can in particular identify a cycle involving CNN that regenerates N$_2$ from N($^4$S). N($^2$D) is quenched to ground state N($^4$S) through collisions with CO, O($^3$P) and C (the main channel being the one with CO). 
\end{itemize}

\subsection{Discussion}

As the data available about Triton mostly come from Voyager 2, we have very few data points to validate our results. As given in \citet{krasnopolsky_photochemistry_1995}, we know that the number density of N$_2$ at 575\,km is (4$\pm$0.4)$\times$10$^8$ cm$^{-3}$, that the concentration of atomic nitrogen at 400\,km is (1$\pm$0.25)$\times$10$^8$ cm$^{-3}$ and (5$\pm$2.5)$\times$10$^8$ cm$^{-3}$ at 200\,km. 
Our N$_2$ profile is in agreement with the data as we find 3.7$\times$10$^8$ cm$^{-3}$ at 571\,km. Likewise for the atomic nitrogen data at 400\,km as we find 1.0$\times$10$^8$ cm$^{-3}$ at 402\,km. But our value at 200\,km is slightly above the corresponding interval, as we have 9.6 and 8.8$\times$10$^8$ cm$^{-3}$ at 196 and 204\,km respectively. 
At this level, we expect chemical uncertainties to be quite significant, possibly explaining the departure of our nominal value from the observed range.
The peak concentration of atomic nitrogen is 2.0$\times$10$^9$ cm$^{-3}$ at 80\,km, which is close to values from \citet{krasnopolsky_photochemistry_1995} and \citet{strobel_tritons_1995} which are about $\sim$1$\times$10$^9$ cm$^{-3}$ at $\sim$115\,km.
\newline

\begin{figure}[!h]
   \resizebox{\hsize}{!}
            {\includegraphics{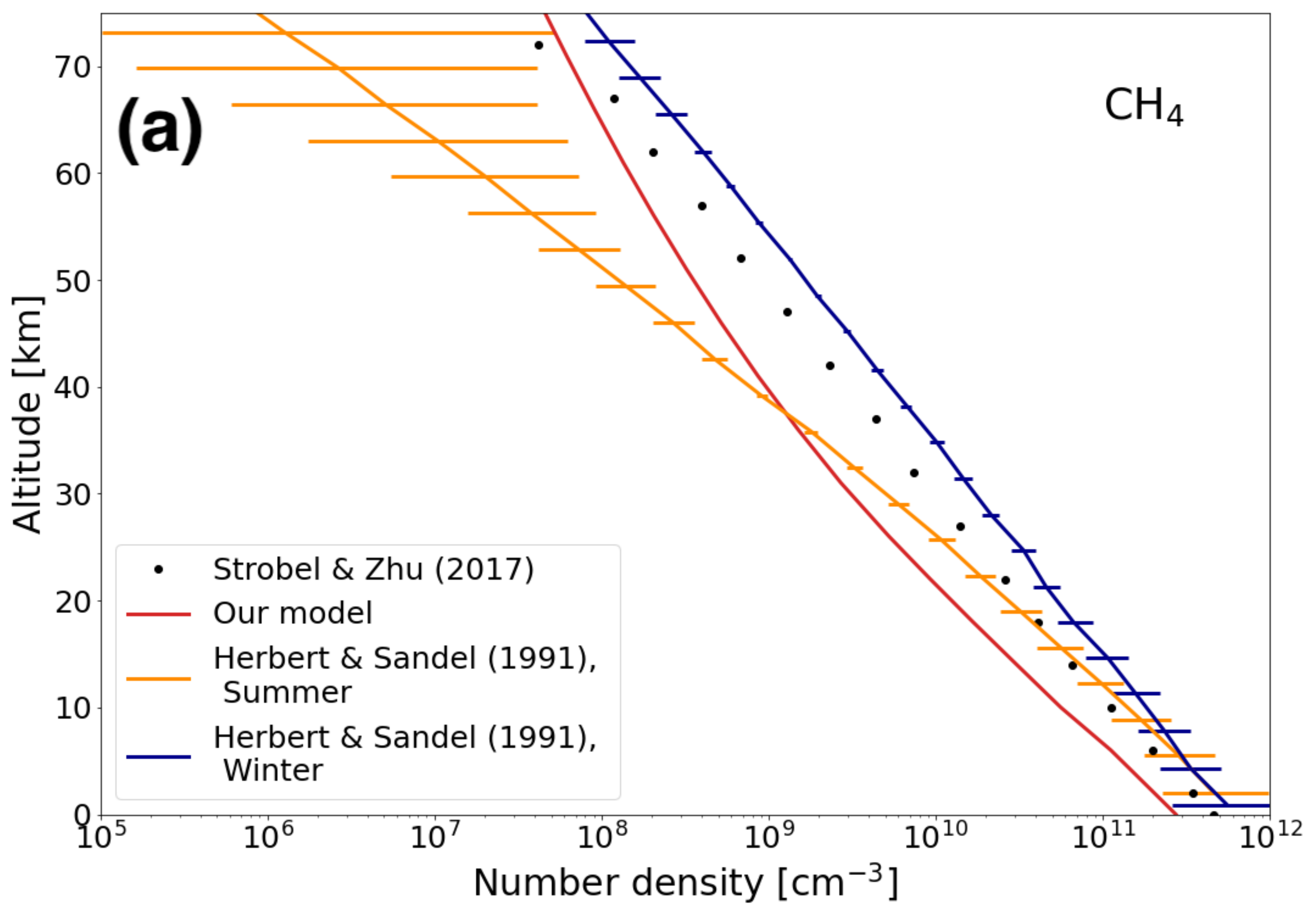}
            \includegraphics{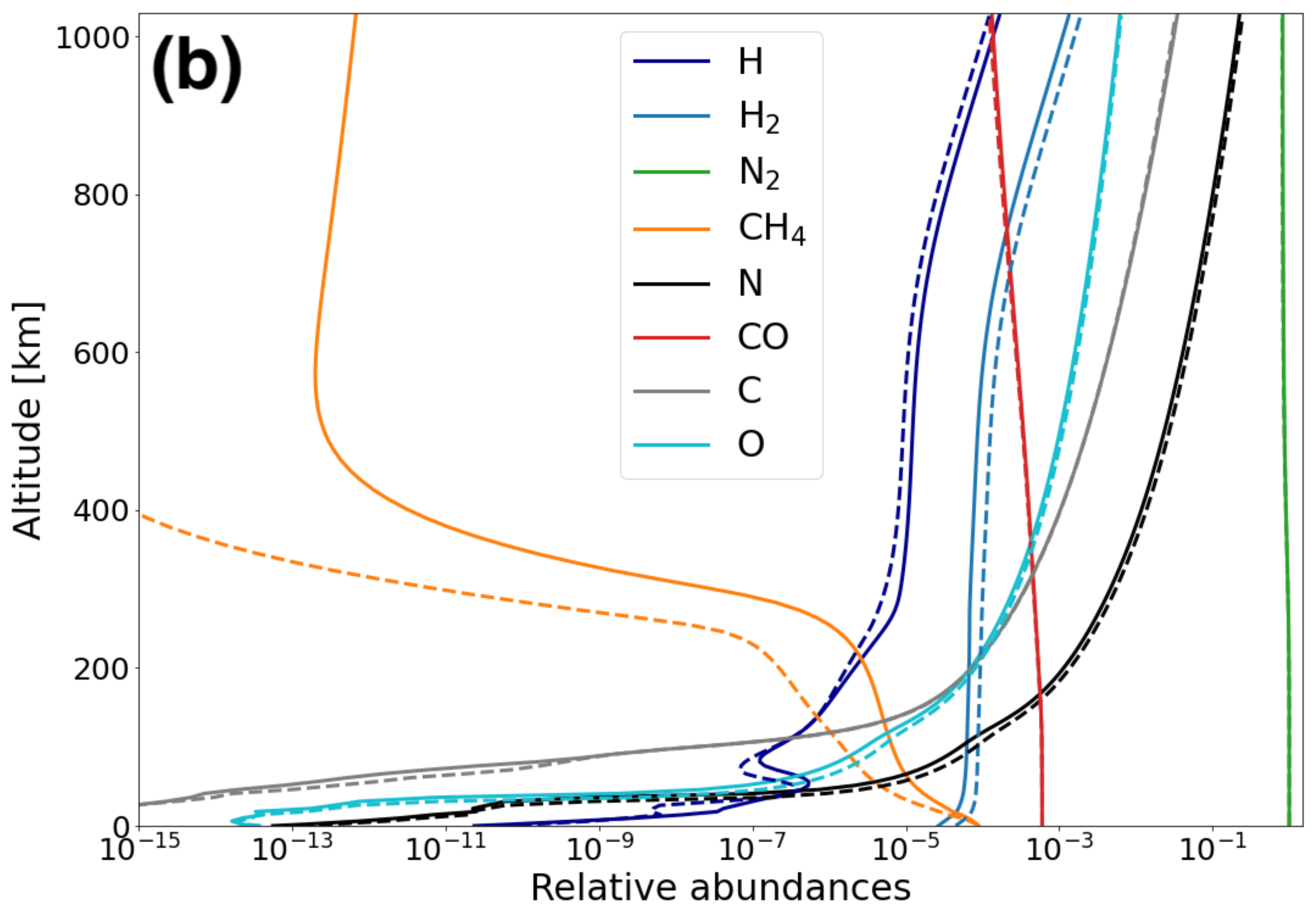}
            }
      \caption{\textbf{(a):} CH$_4$ number density profiles near the surface derived from Voyager 2 data and presented in \citet{herbert_ch4_1991} (dark blue and orange), data points from \citet{strobel_comparative_2017} (black dots) and result from our model using nominal reaction rates (red), i.e. without considering any uncertainty. \textbf{(b):} Comparison of the relative abundances of the main neutral species when using a high resolution solar spectrum (solid lines) versus a low resolution spectrum (dashed lines). These results are obtained with solar fluxes corresponding to low solar activity.
              }
         \label{ch4_near_surf+solar_resolution}
   \end{figure}

We also have the CH$_4$ number density profiles near the surface for the two solar occultation points from \citet{herbert_ch4_1991}. With the actual model and nominal reaction rates, we are nearly in agreement with these profiles, as shown in panel (a) of Fig. \ref{ch4_near_surf+solar_resolution}. The differences at low altitude are only due to our lower CH$_4$ abundance at the surface
coming from the use of the vapor pressure formula of \citet{fray_sublimation_2009}. 
We also note that we were unable to match the data if we used a different $K_{zz}$ profile or if we used a solar flux not corresponding to a maximum solar activity. Therefore, these two parameters seem to be critical for the modeling of Triton's atmosphere. As $K_{zz}$ impacts strongly our results, it could be important to better determine its profile as the one we use was inferred by using the CH$_4$ number density near the surface only \citep{herbert_ch4_1991}. 

The spectral resolution of the solar flux was found to have a non negligible impact on the abundance profiles of CH$_4$ as shown in the right-hand panel of Fig. \ref{ch4_near_surf+solar_resolution}. Following these observations, we may need a high resolution spectrum for high solar activity in order to obtain more representative results.

If we sum the mass condensation rates for the three C$_2$H$_x$ studied in this section, we find a total mass condensation rate of 5.7$\times$10$^{-15}$ g.cm$^{-2}$.s$^{-1}$, which fits the aerosol production rate interval of [4-8]$\times$10$^{-15}$ g.cm$^{-2}$.s$^{-1}$ given in \citet{strobel_tritons_1995} for the winter and summer hemispheres respectively.
\newline

\begin{figure}[!h]
   \centering
   \includegraphics[width=10cm]{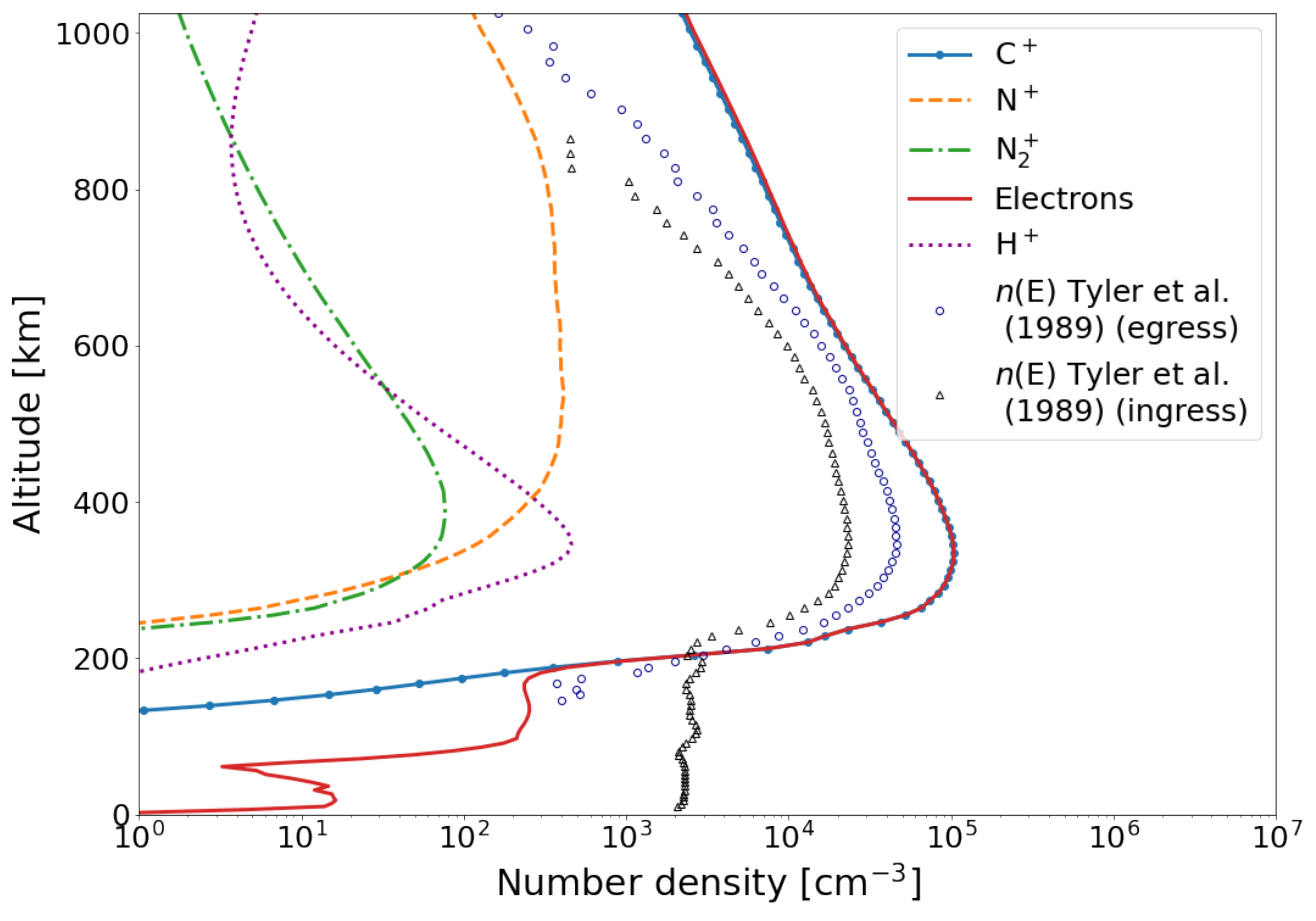}
      \caption{Number densities of the main ions in Triton's atmosphere. The black triangles and dark blue circles represent the electronic profiles measured by Voyager 2 at the two occultations points \citep{tyler_voyager_1989}.
               }
         \label{conc_ionsmaj}
\end{figure}

For ions, we first examine if our electronic profile corresponds to the profiles presented in \citet{tyler_voyager_1989} and derived from Voyager data. These profiles are shown in Fig. \ref{conc_ionsmaj}. We can observe that our electronic peak is located at 334\,km, which is slightly lower than the altitudinal range (340-350)\,km determined from Voyager measurements and given in \citet{tyler_voyager_1989}. Also, our electronic peak concentration is 1.0$\times$10$^5$ cm$^{-3}$, which is higher than the interval of (3.5$\pm$1)$\times$10$^4$ cm$^{-3}$ given in \citet{krasnopolsky_photochemistry_1995}. We show in Sect. \ref{section_chem_uncert} the impact of chemical uncertainties on the electronic profile. But as we found that reactions with magnetospheric electrons had a large impact on the atmospheric chemistry, these results could change significantly if we take another electronic production profile. Also, we modified the ionization profile from \citet{strobel_magnetospheric_1990} in a rather arbitrary way, following the manipulations made in \citet{summers_tritons_1991} and \citet{krasnopolsky_photochemistry_1995}. Thus, changing these arbitrary values could also impact our results in a significant way. In order to model the interaction between magnetospheric electrons and Triton's atmosphere better, we recommend using an electron transport code in further studies. 
\newline

We also recall that we considered the electronic temperature to be equal to the neutral temperature in all the atmosphere because this parameter was not measured by Voyager. But we observed that dissociative recombination reactions of N$_2^+$ were among the most important key chemical reactions for the main species. As the rates of these reactions are computed using the electronic temperature, having a good estimation or measurements of this parameter seems mandatory in order to improve the confidence in our model. 

In the same way, recent occultation measurements presented in \citet{oliveira_constraints_2022} indicate that the thermal profile in the lower atmosphere could be quite different from the profile that we use here, with a strong positive gradient near the surface and the potential presence of a mesosphere. If correct, this could greatly impact the profiles of condensable species such as methane and hydrocarbons.

\section{Chemical uncertainties}
\label{section_chem_uncert}

As the temperature is very low on Triton, we expect to have large uncertainties on our abundance profiles. 
Indeed, chemical reaction rates are determined experimentally or theoretically but always with an uncertainty. It is expressed with two different factors: the temperature dependent uncertainty factor $F_i(T)$ and $g_i$, a coefficient that is used to extrapolate $F_i$ depending on temperature. The uncertainty factor is computed from Eq. \eqref{calc_Fi(T)} \citep[see for instance][]{sander_notitle_2006,hebrard_photochemical_2006,hebrard_photochemical_2007,hebrard_how_2009}:

\begin{equation}
    F_i(T) = F_i(300\text{K})\times \exp{\left \{ g_i \left [ \frac{1}{T}-\frac{1}{300} \right ] \right \}}
\label{calc_Fi(T)}
\end{equation}
where $F_i(300\text{K})$ is the uncertainty factor at 300\,K, which is commonly given with the reaction rates, as they are mainly measured around room temperature. This is why we expect uncertainties to be large: the temperature on Triton being below 100\,K, we use rate formulas that are in general not known in these conditions and we extrapolate the associated uncertainty factors, about which we have a very limited knowledge. This generally leads to a greater uncertainty for most of the reaction rates that subsequently propagates into the model. Thus, it is necessary to examine how these uncertainties propagate during the calculations and their impact on the results and thus on the number density profiles of the different species. 
\newline

To study the propagation of chemical uncertainties in our model, we use a Monte-Carlo simulation. After the model was run with nominal reaction rates, that is without considering any uncertainty (as done in the previous section), we compute again all the reaction rates using the uncertainty factors $F_i$ and $g_i$, considering each rate as a random variable $k_i$ with a log-normal distribution centered on the nominal rate $k_{0_i}$ and with a standard deviation $\log{F_i}$ \citep{hebrard_photochemical_2007,dobrijevic_experimental_2008}. For two-body reactions, $k_i$ is obtained from Eq. \eqref{calc_ki}:

\begin{equation}
    \log(k_i) = \log(k_{0_i}) + \varepsilon_i \log\left [F_i(T)\right ]
\label{calc_ki}
\end{equation}
$\varepsilon_i$ is a random number with a normal distribution centered on zero and with a standard deviation of one. 
With this, we have a 68.3\% probability to find $k_i$ in the interval $\left [ \frac{k_{0_i}}{F_i}, k_{0_i}\times F_i \right ]$. 
To avoid considering extreme values of $k_i$, we only use values of $k_i$ computed with $\lvert \varepsilon \rvert$ <2.

For three-body reactions the reaction rate is given by:

\begin{equation}
    k_i(z) = \frac{k_0\times [M]\times \chi + k_r}{1 + \frac{k_0[M]}{k_{\infty}}}
\label{three_body_ki}
\end{equation}
$[M]$ being the number density of the third body, $k_0$ the reaction rate for low pressure conditions, $k_{\infty}$ the reaction rate for high pressure conditions and $k_r$ the rate for recombination. $\chi$ is the uncertainty factor of Troe which is computed with (for all three-body reactions except H + C$_2$H$_2$ and H + C$_4$H$_2$ that have their own formulas):

\begin{equation}
    \chi = \frac{\log(0.64)}{1 + \left [ \log \left ( \frac{k_0[M]}{k_{\infty}} \right ) \right ]^2}
\label{Troe}
\end{equation}
Thus, we have to compute an uncertainty factor for $k_0$, $k_{\infty}$ and $k_r$ by using formula \eqref{calc_Fi(T)}, with a $F(300\text{K})$ and a $g$ for each. Then, each $k_i$ is recalculated using equation \eqref{calc_ki}. 
\newline

For photodissociation, photoionization and electron-impact dissociation/ionization, the corresponding reaction rates do not depend on temperature. In this case, we assume a constant uncertainty factor $F_i = 1.2$ (which may be underestimated) for all these reactions. Reaction rates are then computed directly with Eq. \eqref{calc_ki}. 

For ion-neutral and dissociative recombination reactions, branching ratios are applied on reaction rates to express the probability that the reaction gives a specific set of products. These branching ratios are also measured or computed theoretically and thus have an associated uncertainty. To account for it, we also generate for each run of the program a new branching ratio $br_i$, randomly generated between $\left [ \frac{br_{0_i}}{F_{br_i}}, br_{0_i}\times F_{br_i} \right ]$ using a Dirichlet uniform distribution (cf \citealt{carrasco_influence_2007}), $F_{br_i}$ being the uncertainty factor for the considered branching ratio and $br_{0_i}$ the nominal branching ratio. The chemical reaction rate of each branch is then multiplied by $\frac{br_i}{br_{0_i}}$.

\subsection{Results}

\begin{figure}[!h]
            {\includegraphics[width=0.5\hsize]{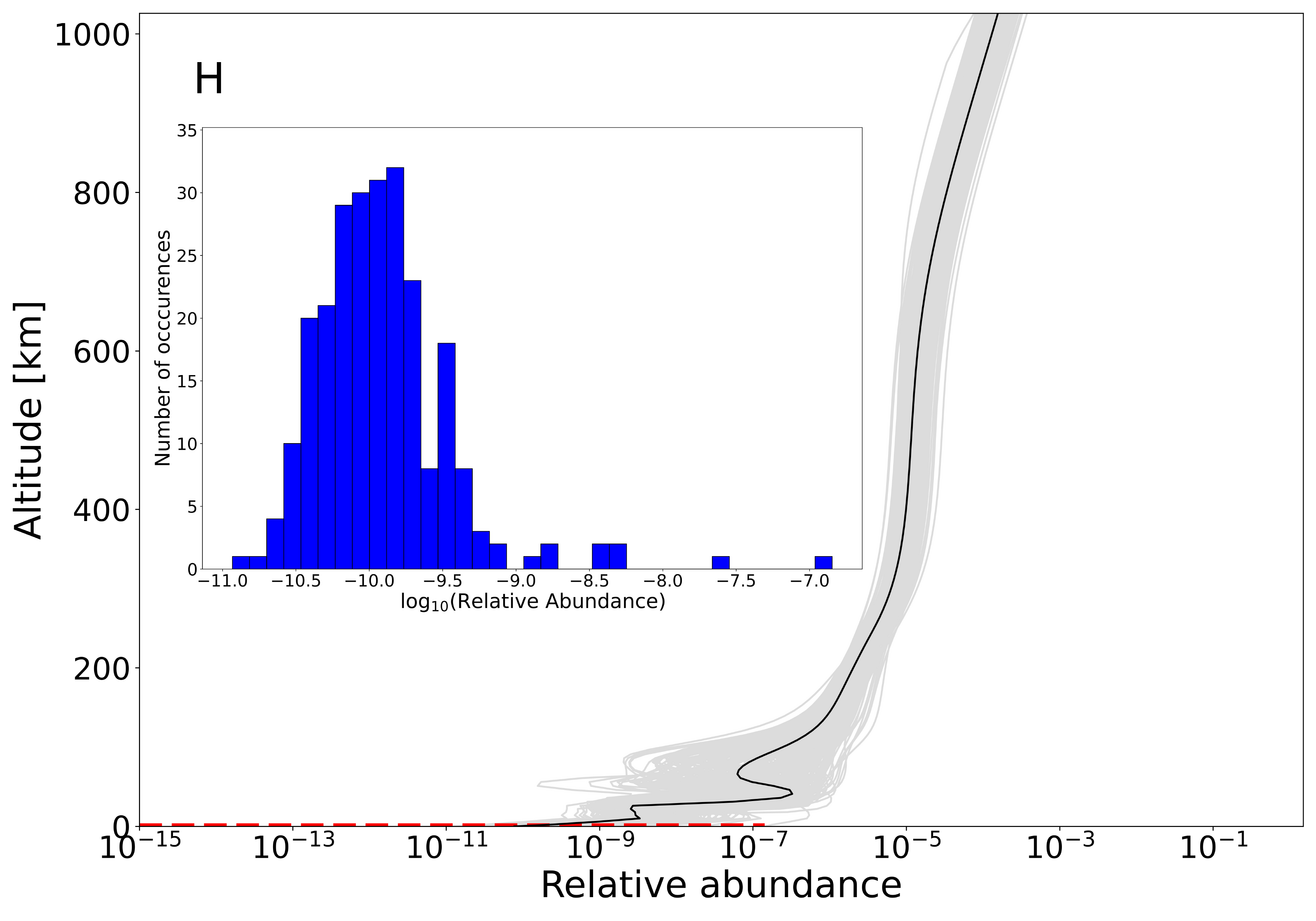}
            \includegraphics[width=0.5\hsize]{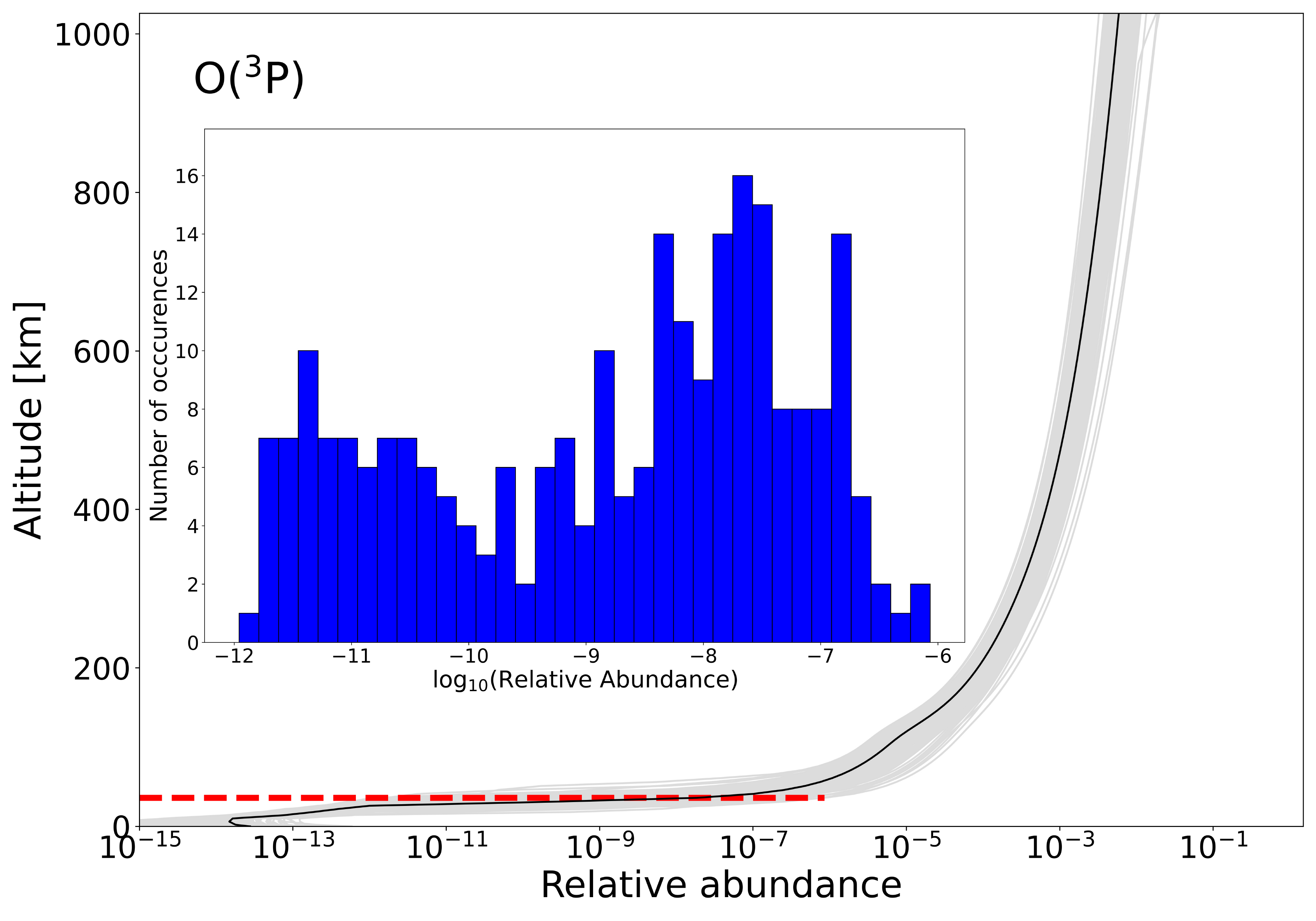}
            \includegraphics[width=0.5\hsize]{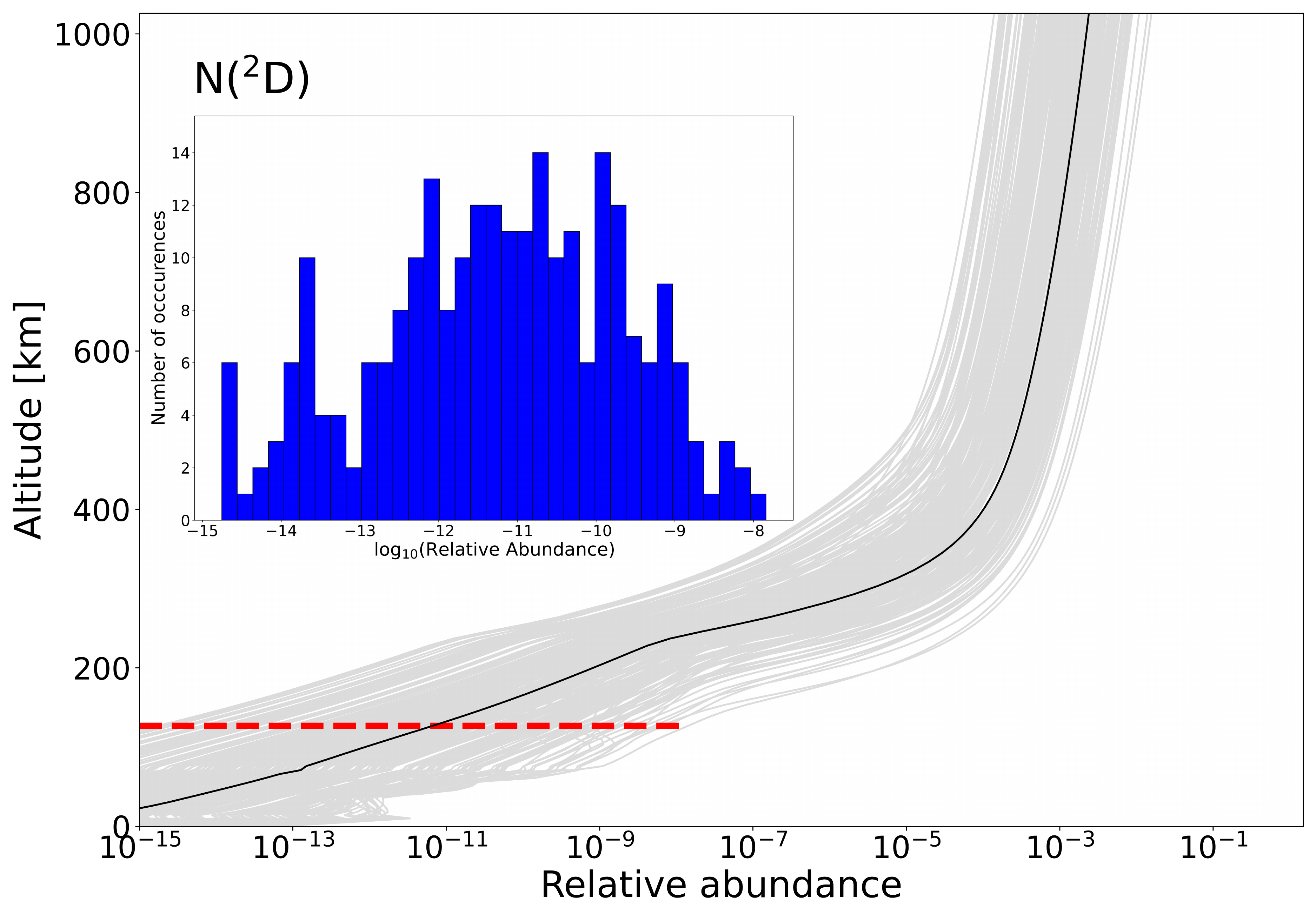}
            \includegraphics[width=0.5\hsize]{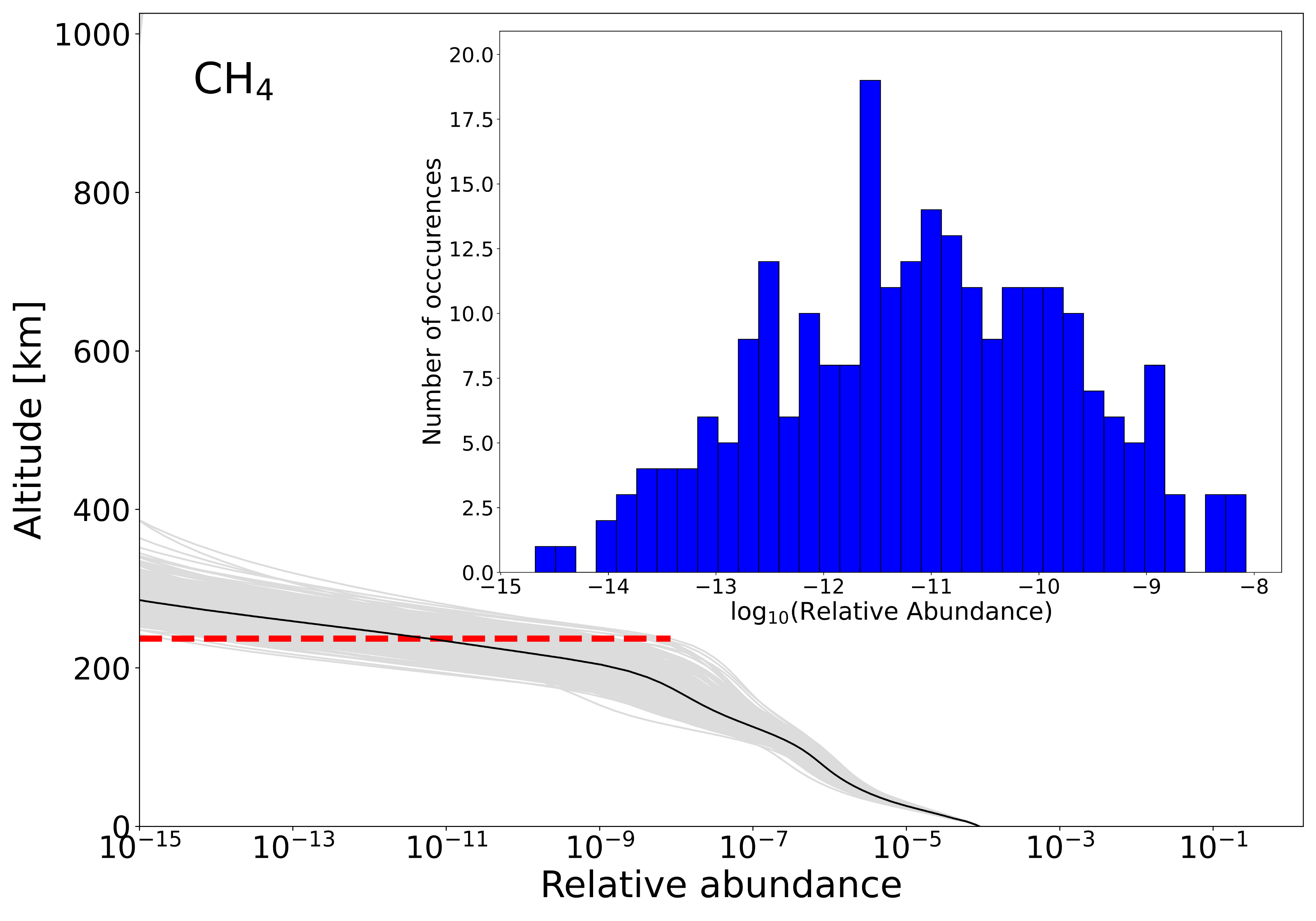}
            \includegraphics[width=0.5\hsize]{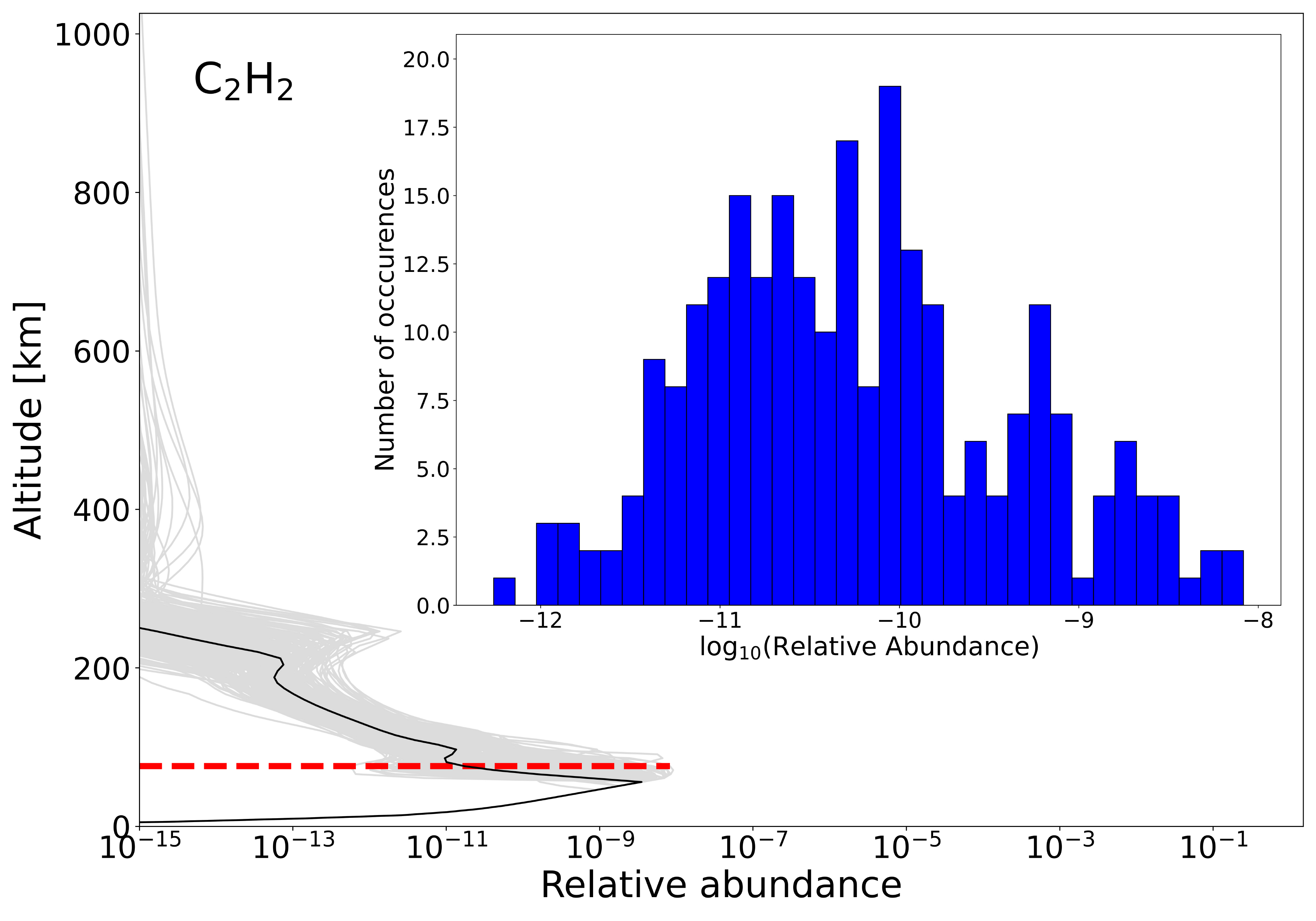}
            \includegraphics[width=0.5\hsize]{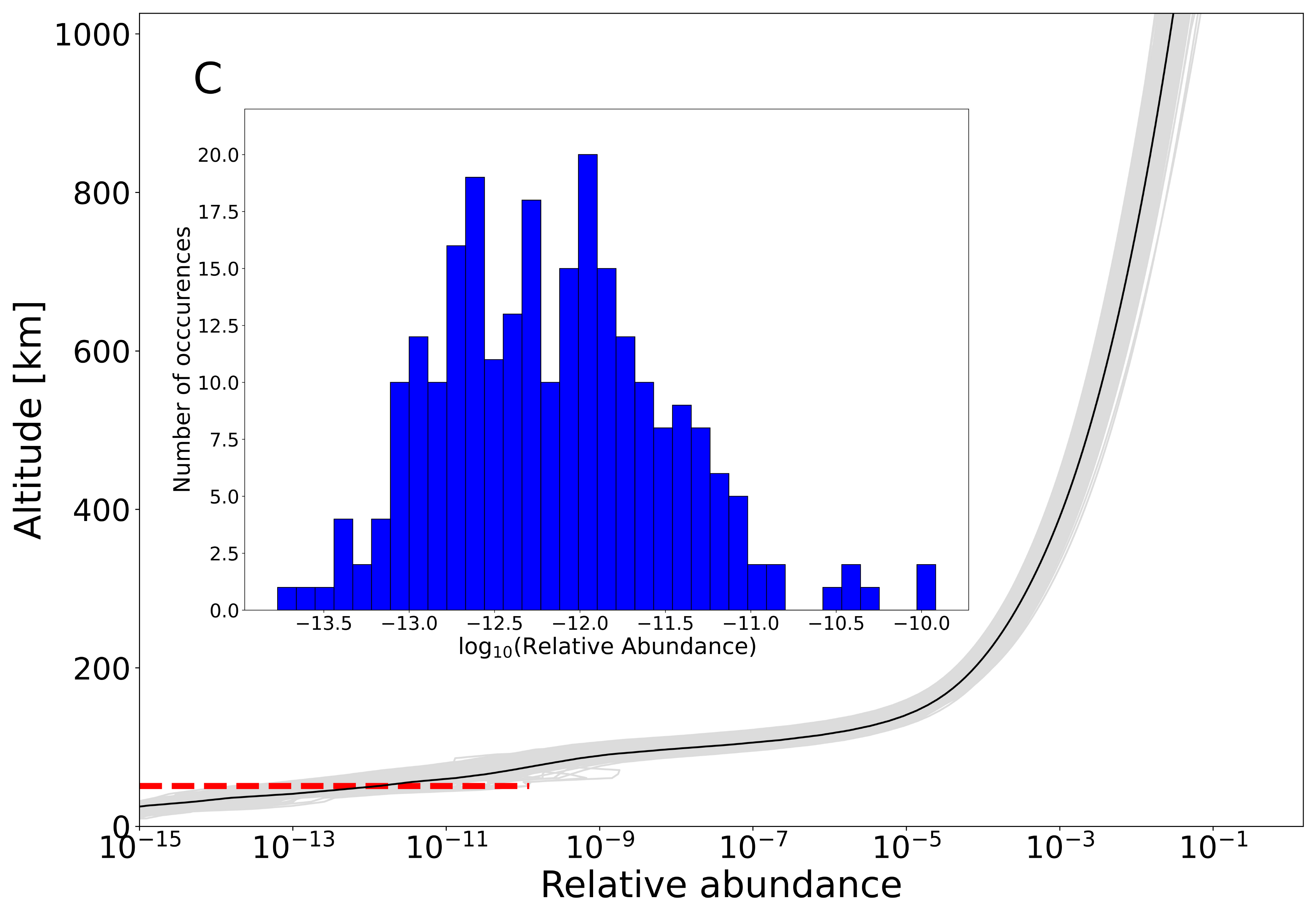}
            }
      \caption{Abundance profiles and histograms of the logarithm of the abundance of H, O($^3$P), N($^2$D), CH$_4$, C$_2$H$_2$ and C for 250 runs of the Monte-Carlo procedure. Histograms are plotted at the altitude where the uncertainty is maximum for each species (red dashed line).
              }
         \label{histos_fm_250tirages}
   \end{figure}

We performed 250 iterations of the Monte-Carlo procedure. 
In Fig. \ref{histos_fm_250tirages}, we present the nominal mole fraction profiles of six species alongside the 250 profiles generated by the procedure. Histograms of mole fractions at the altitude where the associated uncertainty is maximum are also plotted. As the reaction rates have a log-normal distribution, we would expect to find normal distributions with reasonable standard deviations, as shown for H in Fig. \ref{histos_fm_250tirages}. But in Triton's low temperature atmosphere, we find large uncertainties for the majority of the studied species, meaning a high standard deviation values.
In Table \ref{table_means&F_250tirages} we give the mean abundances and the standard deviation of these abundance distributions, expressed by an uncertainty factor $F(\Bar{y_i})$ at the altitude where the uncertainty is maximum.

\begin{table}[!h]
   \begin{center}
      \begin{tabular}{l c l l}
         \hline
\multicolumn{1}{c}{Species} & \multicolumn{1}{c}{\begin{tabular}[c]{@{}c@{}}Altitude \\ {[}km{]}\end{tabular}} & \multicolumn{1}{c}{$\Bar{y_i}$} & \multicolumn{1}{c}{$F(\Bar{y_i})$} \\ \hline
N$_2$      & 1026 & 7.8$\times$10$^{-1}$  & 1.02  \\
CH$_4$     & 237  & 7.6$\times$10$^{-12}$ & 23.59 \\
N($^4$S)   & 31   & 3.0$\times$10$^{-9}$  & 33.07 \\
N($^2$D)   & 127  & 5.5$\times$10$^{-12}$ & 36.35 \\
H$_2$      & 903  & 1.1$\times$10$^{-3}$  & 1.17  \\
H          & 0    & 1.2$\times$10$^{-10}$ & 2.97  \\
CO         & 1026 & 1.0$\times$10$^{-4}$  & 1.15  \\
C          & 51   & 6.9$\times$10$^{-13}$ & 4.69  \\
O($^3$P)   & 36   & 1.4$\times$10$^{-9}$  & 37.85 \\
C$_2$H$_2$ & 76   & 5.6$\times$10$^{-11}$ & 7.19  \\
C$_2$H$_4$ & 66   & 4.0$\times$10$^{-10}$ & 6.02  \\
C$_2$H$_6$ & 174  & 2.0$\times$10$^{-11}$ & 3.71  \\
HCN        & 1026 & 1.3$\times$10$^{-10}$ & 4.35  \\
C$^+$      & 181  & 1.6$\times$10$^{-10}$ & 3.60  \\
N$^+$      & 810  & 1.4$\times$10$^{-5}$  & 1.68  \\
N$_2^+$    & 1026 & 7.4$\times$10$^{-7}$  & 1.94  \\
H$^+$      & 196  & 5.5$\times$10$^{-12}$ & 5.92  \\
$e^-$      & 212  & 1.0$\times$10$^{-8}$  & 3.15  \\ \hline
      \end{tabular}
   \end{center}
   \caption[]{Mean abundances $\Bar{y_i}$ and associated uncertainty factors $F(\Bar{y_i})$ for the main atmospheric species. These values are computed at the altitude level where the uncertainty on the abundance of the considered species is maximum. $F(\Bar{y})$ gives the interval $\left [ \frac{\Bar{y}}{F(\Bar{y})}, \Bar{y}\times F(\Bar{y}) \right ]$ at 1$\sigma$.}
   \label{table_means&F_250tirages}
\end{table}

We see that very few species have a standard deviation lower than two at the level where their uncertainty is maximum, as it is only the case for N$_2$, H$_2$, CO, N$^+$ and N$_2^+$. The maximum uncertainty factor is obtained for O($^3$P) and gives a ratio between the high and the low value of the 1-$\sigma$ interval of 1.4$\times$10$^3$. We can observe that highly reactive species such as atomic nitrogen or CH$_4$ also have large uncertainty factors. 
For the majority of the studied species, high uncertainties emerge at the altitude level where their mole fractions vary strongly. This can be seen for O($^3$P) in Fig. \ref{histos_fm_250tirages}.
\newline

By plotting the histograms of the abundances of these species at the considered levels, we can highlight bimodalities in some of the distributions. This subject was studied in \citet{dobrijevic_epistemic_2008} for Titan's atmosphere. They are due to uncertainties on reaction rates and show that two distinct paths are explored by the model, which gives a bimodal distribution instead of the expected unimodal one as reaction rates follow a log-normal distribution. We call them epistemic bimodalities as they do not correspond to any real phenomenon but are artifacts arising from the large uncertainties of some reaction rates in Triton's conditions \citep{dobrijevic_epistemic_2008}.
As an example, in Fig. \ref{histos_fm_250tirages} the histogram of O($^3$P) shows a bimodality. To cancel out these effects, we have to find which reactions strongly impact the model uncertainties. 

\subsection{Identifying key uncertainty reactions}

In order to have more significant results, we need to reduce the chemical uncertainties. To do this, we must identify the key uncertainty reactions. This kind of key reaction must not be confused with the key chemical reactions: these reactions are defined as those that have the most important influence on the chemical scheme, whereas key uncertainty reactions are defined as those that have the most important contribution to the overall uncertainty on species abundances.

Thus, we have to identify these key uncertainty reactions for each species in order to see if we can reduce the uncertainty over the abundance profiles by improving our knowledge about these reactions. 
\citet{dobrijevic_comparison_2010} presented different methods to determine key uncertainty reactions. In our case, we performed global sensitivity analyzes, as presented in the following. 

\subsubsection{Global sensitivity analysis}

This type of analysis allows us to vary all the input factors at each run (here the chemical reaction rates) and study the link between these input factors and the uncertainty on the outputs, which are the abundance profiles of the studied species obtained with the Monte-Carlo procedure.
It also allows us to conserve the non-linearity and coupling of the model, resulting from the use of a high number of species and reactions.

To do this, we use Rank Correlation Coefficients (RCCs). As shown in \citet{carrasco_influence_2007}, \citet{hebrard_how_2009}, \citet{dobrijevic_key_2010} and \citet{dobrijevic_comparison_2010}, these coefficients convert a nonlinear but monotonic relationship between the input factors and the outputs into a linear relationship. To do so, it replaces the values of the sampled inputs and outputs by their respective ranks \citep{helton_survey_2006}. The outcome of this procedure is a coefficient between -1 and 1 for each input-output couple. If the coefficient is positive, it means that the two correlated parameters vary in the same way. Thus, in our case, each coefficient links a reaction to the uncertainty on the abundance profile of a particular species. Reactions with high RCCs (in absolute value) contribute strongly to the uncertainty on the abundance of this species and are therefore key uncertainty reactions. 

We analyze RCCs in two different ways: first, we perform the analysis for each of the main species at the altitude where their uncertainty is maximum (one species at a time). Second, we choose some characteristic atmospheric levels and perform an analysis over all the species at a time. Coupling the results of these two studies allows us to determine the key uncertainty reactions of our model.

\subsubsection{Results for the study for one species at a time}

For this study, we chose to focus on reactions that have a RCC higher than 0.2 in absolute value. We ran this sensitivity analysis for the 18 main species at the altitude where their uncertainty is maximum, as given in Table \ref{table_means&F_250tirages}. Reactions that were found for more than one of these species are given in Appendix \ref{Appendix_key_unc_rea}, in Table \ref{key_react_RCCsolo_250tirages}. We identified 35 reactions: 12 neutral-neutral reactions, 10 ion-neutral reactions, 4 dissociative recombinations, 3 photodissociations, 3 photoionizations and 3 reactions with magnetospheric electrons. 

We also identified which reactions gave high RCC absolute values, even if they do not necessarily appear for more than one species. These reactions are given in Table \ref{high_RCCsolo_250tirages} of Appendix \ref{Appendix_key_unc_rea}, with the species associated to a high RCC value. One additional reaction appears in this table. 

For the majority of these reactions, we always have a high value of $F_i$ or $g_i$ and sometimes both. This confirms that for this study, high RCCs are often linked to a lack of knowledge about reaction rates. 

\subsubsection{Results for the study of all species at a given altitude}

We also performed the sensitivity analysis at a given altitude level and for all the species of our model at the same time. In this case, we count the number of times that a reaction has a RCC higher than 0.2 (in absolute value) over the total number of species. For each level, we then rank the reactions that appear the most and therefore contribute the most to the overall uncertainty at this level. We chose to perform this test at seven different levels to sample diverse altitudes: 0, 86, 220, 334, 502, 758 and 1026\,km. Reactions that appear for at least one quarter of the species of our chemical scheme at these levels are given in Table \ref{RCC_All_6lvls_supQuarter_250tirages} of Appendix \ref{Appendix_key_unc_rea}. 

We notice that the reaction N($^2$D) + CO $\longrightarrow$ N($^4$S) + CO appears at each studied level. We also find that the main key uncertainty reactions are different depending on the considered altitude. In the lower atmosphere (0 and 86\,km), neutral-neutral reactions are dominant. At 86\,km, we highlight three different three-body reactions. At 220\,km, in the lower ionosphere, photolysis of CH$_4$ and CO is important. At higher levels, de-excitation of N($^2$D) through collisions with CO and C are the main key uncertainty reactions. For levels higher than 500\,km, the charge exchange between N$^+$ and C giving N$(^4$S) + C$^+$ also contributes significantly.

We also performed a complementary study focusing on the main atmospheric species to avoid biases from species with negligible abundances. This study highlights key uncertainty reactions that we already found in the previous analyzes, confirming their role in the overall uncertainty. 

Finally, as we did for the study with one species at a time, we listed the reactions with high RCC values. These reactions are given in Table \ref{RCC_main_6lvls_RCCsup0,5_250tirages} with the involved species at each of the seven levels studied here. Again, we find reactions that were previously highlighted but also some new ones, in general important for only one species at one or more levels, for example reactions between N$^+$ and CO that are important for the uncertainty of N$^+$. This also confirms that key uncertainty reactions depend on the studied altitude.
   
\subsection{Discussion - uncertainties and key uncertainty reactions}

By computing the RCCs in  various cases, we were able to identify reactions that were responsible for large uncertainties for a particular species or more globally at a given level. If we look at all the results presented in Appendix \ref{Appendix_key_unc_rea}, we see that some reactions are involved in all (or nearly all) the treated cases. These reactions are presented in Table \ref{Main_key_reactions_final_set}. 

\begin{table}[!h]     
\begin{center}
   \begin{tabular}{l l l l}     
\hline                    
Reaction & Rate coefficients & \begin{tabular}[c]{@{}l@{}}$F_i(300\text{K})$\\ or \textit{\textbf{F}}\end{tabular} & $g_i$ \\ \hline
N($^2$D) + CO $\longrightarrow$ N($^4$S) + CO & 1.9$\times$10$^{-12}$ & 1.6 & 300.0 \\
N($^2$D) + C $\longrightarrow$ N($^4$S) + C  & 4.0$\times$10$^{-12}$$\times \exp{\left ( \frac{-259}{T(z)}  \right )}$  & 3.0  & 200.0 \\
C + N$_2$ $\longrightarrow$ CNN  & $k_0$ =  3.1$\times$10$^{-33}$ $\times\left( \frac{T(z)}{300} \right)^{-1.5}$ & 1.8 & 100.0 \\
 & $k_{\infty}$ = 1.0$\times$10$^{-11}$ & 10.0 & 0.0   \\
O($^3$P) + CNN $\longrightarrow$ N$_2$ + CO & 1.0$\times$10$^{-10}$ & 3.0 & 7.0   \\
N$^+$ + C $\longrightarrow$ N($^4$S) + C$^+$ & 4.0$\times$10$^{-12}$ & 10.0 & 0.0   \\
C$^+$ + $e^-$ $\longrightarrow$ C & 4.4$\times$10$^{-12}$ $\times\left( \frac{T_e(z)}{300} \right)^{0.61}$ & 1.6 & 0.0   \\
N($^4$S) + CNN $\longrightarrow$ CN + N$_2$ & 1.0$\times$10$^{-10}$ & 3.0 & 7.0   \\
N$_2^+$ + C $\longrightarrow$ N$_2$ + C$^+$ & 1.0$\times$10$^{-10}$ & 3.0 & 0.0   \\
C + H$_2$ $\longrightarrow$ $^3$CH$_2$ & $k_0$ = 7.0$\times$10$^{-32}$ $\times\left( \frac{T(z)}{300} \right)^{-1.5}$ & 2.0  & 100.0 \\
 & $k_{\infty}$ = 2.06$\times$10$^{-11}$$\times \exp{\left ( \frac{-57}{T(z)}  \right )}$ & 3.0 & 100.0 \\
CH$_4$ + $h\nu$ $\longrightarrow$ $^1$CH$_2$ + H$_2$ & \textit{Photodissociation}  & \textbf{1.2} & -     \\ \hline
\end{tabular}
\end{center}     
\caption{Main key uncertainty reactions identified through sensitivity analyzes. These reactions are the ones that appeared the most in our sensitivity analyzes and thus contribute the most to the model uncertainties. $T(z)$ is the atmospheric neutral temperature at altitude $z$ and $T_e(z)$ the electronic temperature. $F_i(300K)$ is the uncertainty factor at 300\,K and its value is given in the corresponding column by non bold numbers, while \textit{\textbf{F}} is the global uncertainty factor that is computed with $F_i$ and $g_i$ but has a fixed value for photodissociations and reactions with ME. When \textit{\textbf{F}} is given, the value is in bold.} 
\label{Main_key_reactions_final_set}     
\end{table}

We also find that many of the reactions that were identified through the sensitivity analyzes appeared as key chemical reactions for the main species of the atmosphere.
Thus, it appears even more important to improve our knowledge of the kinetics of these key reactions as they have the biggest influence on the number density profiles of the main species. Without this supplementary knowledge at low temperature, it seems difficult to obtain a clearer image of the composition of Triton's atmosphere. 
\newline

We also note that O($^3$P) appears as a reactant in key chemical reactions for some of the main species and also in key uncertainty reactions. At the moment, we do not consider any interplanetary dust flux. But \citet{moses_dust_2017} showed that the external flux of water in Neptune's atmosphere is likely coming from interplanetary dust. If true, this flux should also impact Triton's atmospheric chemistry, as stated in \citet{poppe_interplanetary_2018}. Thus, it will be interesting to add this dust delivery and observe its impact on the number density profiles and also on chemical uncertainties. 
\newline

In Sect. \ref{results}, we observed that the electronic number density profile was higher than expected when comparing to observations presented in \citet{tyler_voyager_1989}. If we examine the electronic profiles of the 250 runs of the Monte-Carlo procedure, we find that the mean number density at the electronic peak is $\Bar{y_i}(e^{-}_{peak})$ = 1.1$\times$10$^5$ cm$^{-3}$ with an uncertainty factor of 1.38 at 1-$\sigma$.
Thus, the peak concentration of electrons found with our model is consistently higher than the measured value of (3.5$\pm$1)$\times$10$^4$ cm$^{-3}$ \citep{tyler_voyager_1989,krasnopolsky_photochemistry_1995}. 

The altitude of the electronic peak also varies at each run. For 250 runs, we find $z(e^{-}_{peak})$=(338$\pm$9)\,km, which matches the interval of (340-350)\,km given in \citet{krasnopolsky_photochemistry_1995}. 

However the electronic production, as well as the electronic density, depends mostly on N$_2$ impact ionization by magnetospheric electrons. Therefore, it depends on how the interaction between Triton's atmosphere and Neptune's magnetosphere is modeled. In our case, we used the arbitrarily modified ionization profile from \citet{strobel_magnetospheric_1990}. Thus, the results may be significantly affected when changing these arbitrary parameters. Moreover, the uncertainty factors of reactions involving magnetospheric electrons is constant and equal to 1.2. These uncertainties may be underestimated and thus the observed difference between Voyager's data and our results may not be significant. 
\newline 

In Sect. \ref{results}, we failed to find a number density of atomic nitrogen that matches the value given in \citet{krasnopolsky_photochemistry_1995} at 200\,km, that is (5$\pm$2.5)$\times$10$^8$ cm$^{-3}$, when using nominal reaction rates. With 250 runs, the number density of N is (11$\pm$4)$\times$10$^8$ cm$^{-3}$ at 196\,km and (10$\pm$4)$\times$10$^8$ cm$^{-3}$ at 204\,km. Thus, these intervals and the one of \citet{krasnopolsky_photochemistry_1995} overlap. But as for electrons, the production of atomic nitrogen depends mostly on the interaction between magnetospheric electrons and N$_2$ (directly or indirectly), reactions for which we consider a constant and probably underestimated uncertainty factor. Therefore, the small differences observed between our results and the measured data are not significant. 

\section{Discussion and Conclusion}
\label{conclu}

Our goal was to create a 1D photochemical model of Triton's atmosphere with an up-to-date chemical scheme in order to determine the atmospheric composition, the main parameters having an impact on it and also highlight which studies are necessary in order to reduce uncertainties on model results. 

To do this, we used an existing model of Titan's atmosphere from \citet{dobrijevic_1d-coupled_2016} and adapted it to Triton's conditions. We used input data from \citet{strobel_comparative_2017} for our initial conditions (temperature and number density profiles, initial abundance of CO), the eddy diffusion profile of \citet{herbert_ch4_1991}, took a solar flux corresponding to a maximum solar activity, added interplanetary radiation 
and considered the interaction between Triton and Neptune's magnetosphere through reactions between N$_2$ and magnetospheric electrons. Considering the latter processes was crucial in order to find nominal results for CH$_4$ and atomic nitrogen in agreement with the Voyager data. 

We also improved our chemical scheme by comparing the results with the Voyager data and previous articles about the photochemistry of Triton such as those by \citet{krasnopolsky_photochemistry_1995} or \citet{strobel_tritons_1995}. By doing this, we added nearly 40 atmospheric species and 500 reactions to our initial scheme, giving a total of 220 species and 1764 reactions. 
\newline

With this updated model, we studied 18 species in particular which are the main neutral species (N$_2$, N($^4$S), N($^2$D), CH$_4$, CO, H$_2$, H, C, O($^3$P)), hydrocarbons (C$_2$H$_2$, C$_2$H$_4$, C$_2$H$_6$), nitriles (HCN) and ions (C$^+$, N$^+$, N$_2^+$, H$^+$, electrons), highlighting their main production and loss processes through the identification of key chemical reactions.
Nitrogen chemistry is triggered by ionization and dissociation of N$_2$ by solar radiation and magnetospheric electrons, while methane chemistry results from its photolysis by Lyman-$\alpha$ radiation from the Sun and the interstellar medium, creating radicals and hydrogen (atomic and molecular). These radicals then react to form hydrocarbons that condense due to the very low temperature in the lower atmosphere, the most abundant C$_2$H$_x$ being C$_2$H$_4$. Molecular hydrogen reacts mostly with ions, while H reacts mainly with HCNN or radicals. Atomic nitrogen produced in the N($^2$D) state is converted to ground state N($^4$S) through collisions with CO, C and O($^3$P). N($^4$S) is then converted back to N$_2$ through the CNN cycle. 
CO is depleted in the higher atmosphere by reacting with N$^+$ ions. N$_2^+$ ions recombine rapidly with electrons producing atomic nitrogen. 

We find a dense ionosphere with a cutoff around 175\,km, where the abundance of C$^+$ increases strongly. This ion is the most abundant in the majority of the ionosphere and is produced by charge exchange reactions between N$_2^+$, CO$^+$, N$^+$ and C. The other abundant ions are N$^+$, H$^+$ and N$_2^+$. Below 175\,km, the main ions are heavier but their abundances are low. 
The electronic peak is located at 334\,km, which is consistent with Voyager data presented in \citet{tyler_voyager_1989}. Our number density profile is higher than the one given in the latter article, even considering chemical uncertainties. However, these results are very dependent on the modeling of the interaction between Triton's atmosphere and Neptune's magnetosphere. 
\newline

Using a Monte-Carlo procedure over all the chemical reaction rates, we studied uncertainties on the abundance profiles. We performed 250 runs and found that for the majority of species, the uncertainty factors on these profiles are significant. This is due to our lack of knowledge about chemical reaction rates at the very low temperatures of Triton's atmosphere. These uncertainties also lead to epistemic bimodalities in the abundance distributions of some species, at the level where the associated uncertainty is maximum. 
We determined the key uncertainty reactions responsible for the observed uncertainties and bimodalities. To do this, we performed global sensitivity analyzes using Rank Correlation Coefficients to determine correlations between the outputs (number density profiles) and the input factors (chemical reaction rates) of the model. We did this study for each of the 18 most abundant species at the level where the uncertainty on their abundance is maximum, but also for all the species of our model at different altitudes. We identified the main key uncertainty reactions for which we absolutely need to improve our knowledge if we want to improve the significance of the model results. 
Finally, we found that a great number of key uncertainty reactions also appeared as key chemical reactions for the main species, which confirms the need for new studies about these reactions. 
\newline

In the end, the use of an up-to-date chemical scheme allows a much deeper comprehension of the composition and of the mechanisms governing the chemistry of Triton's atmosphere in comparison to the work of \citet{krasnopolsky_photochemistry_1995} and \citet{strobel_tritons_1995}, while the study of the model uncertainty is the first of its kind for Triton's atmosphere.
\newline

However, the current model of Triton's atmosphere can be improved in many areas. As oxygenated species appeared in several key uncertainty reactions, we need to study how the interplanetary water flux could impact the results of our model. As the precipitation of magnetospheric electrons has a strong impact on the atmospheric chemistry, having a better model which really considers the temporal variation of the precipitation depending on the position of Triton in Neptune's complex magnetosphere should improve the significance of the model results.

The chemical scheme can also be simplified in order to decrease the computation time (cf \citealt{dobrijevic_methodology_2011}). This is not feasible before obtaining the uncertainties on model results as they give an objective criterion to verify if the results given by the reduced scheme are in agreement with the nominal model.

\section*{Acknowledgments}
B.B, M.D and T.C. acknowledge funding from CNES.
B.B, M.D, T.C and K.M.H acknowledge funding from the ‘‘Programme National Planétologie’’ (PNP) of CNRS/INSU.
K.M.H. acknowledges support from the French program ‘‘Physique et Chimie du Milieu Interstellaire’’ (PCMI) of the CNRS/INSU with the INC/INP co-funded by the CEA and CNES.

\newpage

\bibliography{biblio_final}  

\begin{thebibliography}{61}
\providecommand{\natexlab}[1]{#1}
\providecommand{\url}[1]{\texttt{#1}}
\expandafter\ifx\csname urlstyle\endcsname\relax
  \providecommand{\doi}[1]{doi: #1}\else
  \providecommand{\doi}{doi: \begingroup \urlstyle{rm}\Url}\fi

\bibitem[Agnor and Hamilton(2006)]{agnor_neptunes_2006}
C.~B. Agnor and D.~P. Hamilton.
\newblock Neptune's capture of its moon {Triton} in a binary–planet
  gravitational encounter.
\newblock \emph{Nature}, 441\penalty0 (7090):\penalty0 192--194, May 2006.
\newblock ISSN 0028-0836, 1476-4687.
\newblock \doi{10.1038/nature04792}.
\newblock URL \url{http://www.nature.com/articles/nature04792}.

\bibitem[Anicich(2003)]{anicich_index_2003}
V.~G. Anicich.
\newblock An index of the literature for bimolecular gas phase cation-molecule
  reaction kinetics.
\newblock \emph{JPL Publication-03-19, Pasadena, CA, USA}, 2003.

\bibitem[Broadfoot et~al.(1989)Broadfoot, Atreya, Bertaux, Blamont, Dessler,
  Donahue, Forrester, Hall, Herbert, Holberg, Hunter, Krasnopolsky, Linick,
  Lunine, McConnell, Moos, Sandel, Schneider, Shemansky, Smith, Strobel, and
  Yelle]{broadfoot_ultraviolet_1989}
A.~L. Broadfoot, S.~K. Atreya, J.~L. Bertaux, J.~E. Blamont, A.~J. Dessler,
  T.~M. Donahue, W.~T. Forrester, D.~T. Hall, F.~Herbert, J.~B. Holberg, D.~M.
  Hunter, V.~A. Krasnopolsky, S.~Linick, J.~I. Lunine, J.~C. McConnell, H.~W.
  Moos, B.~R. Sandel, N.~M. Schneider, D.~E. Shemansky, G.~R. Smith, D.~F.
  Strobel, and R.~V. Yelle.
\newblock Ultraviolet {Spectrometer} {Observations} of {Neptune} and {Triton}.
\newblock \emph{Science}, 246\penalty0 (4936):\penalty0 1459--1466, Dec. 1989.
\newblock ISSN 0036-8075, 1095-9203.
\newblock \doi{10.1126/science.246.4936.1459}.
\newblock URL
  \url{https://www.sciencemag.org/lookup/doi/10.1126/science.246.4936.1459}.

\bibitem[Brown et~al.(1995)Brown, Cruikshank, Veverka, Helfenstein, and
  Eluszkiewicz]{brown_surface_1995}
R.~H. Brown, D.~P. Cruikshank, J.~Veverka, P.~Helfenstein, and J.~Eluszkiewicz.
\newblock Surface composition and photometric properties of {Triton}.
\newblock Jan. 1995.
\newblock URL \url{https://ui.adsabs.harvard.edu/abs/1995netr.conf..991B}.
\newblock Conference Name: Neptune and Triton Pages: 991-1030 ADS Bibcode:
  1995netr.conf..991B.

\bibitem[Carrasco et~al.(2007)Carrasco, Hébrard, Banaszkiewicz, Dobrijevic,
  and Pernot]{carrasco_influence_2007}
N.~Carrasco, E.~Hébrard, M.~Banaszkiewicz, M.~Dobrijevic, and P.~Pernot.
\newblock Influence of neutral transport on ion chemistry uncertainties in
  {Titan} ionosphere.
\newblock \emph{Icarus}, 192\penalty0 (2):\penalty0 519--526, Dec. 2007.
\newblock ISSN 00191035.
\newblock \doi{10.1016/j.icarus.2007.08.016}.
\newblock URL
  \url{https://linkinghub.elsevier.com/retrieve/pii/S0019103507003533}.

\bibitem[{Committee on the Planetary Science and Astrobiology Decadal Survey}
  et~al.(2022){Committee on the Planetary Science and Astrobiology Decadal
  Survey}, {Space Studies Board}, {Division on Engineering and Physical
  Sciences}, and {National Academies of Sciences, Engineering, and
  Medicine}]{committee_on_the_planetary_science_and_astrobiology_decadal_survey_origins_2022}
{Committee on the Planetary Science and Astrobiology Decadal Survey}, {Space
  Studies Board}, {Division on Engineering and Physical Sciences}, and
  {National Academies of Sciences, Engineering, and Medicine}.
\newblock \emph{Origins, {Worlds}, and {Life}: {A} {Decadal} {Strategy} for
  {Planetary} {Science} and {Astrobiology} 2023-2032}.
\newblock National Academies Press, Washington, D.C., 2022.
\newblock ISBN 978-0-309-47578-5.
\newblock \doi{10.17226/26522}.
\newblock URL \url{https://www.nap.edu/catalog/26522}.
\newblock Pages: 26522.

\bibitem[Cruikshank et~al.(1995)Cruikshank, Matthews, and
  Schumann]{cruikshank_neptune_1995}
D.~P. Cruikshank, M.~S. Matthews, and A.~M. Schumann.
\newblock Neptune and {Triton}.
\newblock 1995.
\newblock URL \url{http://adsabs.harvard.edu/abs/1995netr.conf.....C}.
\newblock Conference Name: Neptune and Triton.

\bibitem[Curdt et~al.(2001)Curdt, Brekke, Feldman, Wilhelm, Dwivedi, Schühle,
  and Lemaire]{curdt_sumer_2001}
W.~Curdt, P.~Brekke, U.~Feldman, K.~Wilhelm, B.~N. Dwivedi, U.~Schühle, and
  P.~Lemaire.
\newblock The {SUMER} spectral atlas of solar-disk features.
\newblock \emph{Astronomy \& Astrophysics}, 375\penalty0 (2):\penalty0
  591--613, Aug. 2001.
\newblock ISSN 0004-6361, 1432-0746.
\newblock \doi{10.1051/0004-6361:20010364}.
\newblock URL \url{http://www.aanda.org/10.1051/0004-6361:20010364}.

\bibitem[Curdt et~al.(2004)Curdt, Landi, and Feldman]{curdt_sumer_2004}
W.~Curdt, E.~Landi, and U.~Feldman.
\newblock The {SUMER} spectral atlas of solar coronal features.
\newblock \emph{Astronomy \& Astrophysics}, 427\penalty0 (3):\penalty0
  1045--1054, Dec. 2004.
\newblock ISSN 0004-6361, 1432-0746.
\newblock \doi{10.1051/0004-6361:20041278}.
\newblock URL \url{http://www.aanda.org/10.1051/0004-6361:20041278}.

\bibitem[Dobrijevic(1996)]{dobrijevic_etude_1996}
M.~Dobrijevic.
\newblock \emph{Etude de la physico-chimie de l'atmosphère de {Neptune}}.
\newblock These de doctorat, Bordeaux 1, Jan. 1996.
\newblock URL \url{http://theses.fr/1996BOR10539}.

\bibitem[Dobrijevic and Parisot(1998)]{dobrijevic_effect_1998}
M.~Dobrijevic and J.~Parisot.
\newblock Effect of chemical kinetics uncertainties on hydrocarbon production
  in the stratosphere of neptune.
\newblock \emph{Planetary and Space Science}, 46\penalty0 (5):\penalty0
  491--505, May 1998.
\newblock ISSN 00320633.
\newblock \doi{10.1016/S0032-0633(97)00176-1}.
\newblock URL
  \url{https://linkinghub.elsevier.com/retrieve/pii/S0032063397001761}.

\bibitem[Dobrijevic et~al.(2003)Dobrijevic, Ollivier, Billebaud, Brillet, and
  Parisot]{dobrijevic_effect_2003}
M.~Dobrijevic, J.~L. Ollivier, F.~Billebaud, J.~Brillet, and J.~P. Parisot.
\newblock Effect of chemical kinetic uncertainties on photochemical modeling
  results: {Application} to {Saturn}'s atmosphere.
\newblock \emph{Astronomy \& Astrophysics}, 398\penalty0 (1):\penalty0
  335--344, Jan. 2003.
\newblock ISSN 0004-6361, 1432-0746.
\newblock \doi{10.1051/0004-6361:20021659}.
\newblock URL \url{http://www.aanda.org/10.1051/0004-6361:20021659}.

\bibitem[Dobrijevic et~al.(2008{\natexlab{a}})Dobrijevic, Carrasco, Hébrard,
  and Pernot]{dobrijevic_epistemic_2008}
M.~Dobrijevic, N.~Carrasco, E.~Hébrard, and P.~Pernot.
\newblock Epistemic bimodality and kinetic hypersensitivity in photochemical
  models of {Titan}'s atmosphere.
\newblock \emph{Planetary and Space Science}, 56\penalty0 (12):\penalty0
  1630--1643, Nov. 2008{\natexlab{a}}.
\newblock ISSN 00320633.
\newblock \doi{10.1016/j.pss.2008.05.016}.
\newblock URL
  \url{https://linkinghub.elsevier.com/retrieve/pii/S0032063308001475}.

\bibitem[Dobrijevic et~al.(2008{\natexlab{b}})Dobrijevic, Claeys-Bruno,
  Sergent, and Phan-Tan-Luu]{dobrijevic_experimental_2008}
M.~Dobrijevic, M.~Claeys-Bruno, M.~Sergent, and R.~Phan-Tan-Luu.
\newblock Experimental designs for the determination of key reactions in
  photochemical models: {Application} to the photochemistry of hydrocarbons in
  the atmosphere of {Titan}.
\newblock \emph{Planetary and Space Science}, 56\penalty0 (3-4):\penalty0
  519--529, Mar. 2008{\natexlab{b}}.
\newblock ISSN 00320633.
\newblock \doi{10.1016/j.pss.2007.10.005}.
\newblock URL
  \url{https://linkinghub.elsevier.com/retrieve/pii/S0032063307003200}.

\bibitem[Dobrijevic et~al.(2010{\natexlab{a}})Dobrijevic, Cavalié, Hébrard,
  Billebaud, Hersant, and Selsis]{dobrijevic_key_2010}
M.~Dobrijevic, T.~Cavalié, E.~Hébrard, F.~Billebaud, F.~Hersant, and
  F.~Selsis.
\newblock Key reactions in the photochemistry of hydrocarbons in {Neptune}'s
  stratosphere.
\newblock \emph{Planetary and Space Science}, 58\penalty0 (12):\penalty0
  1555--1566, Oct. 2010{\natexlab{a}}.
\newblock ISSN 00320633.
\newblock \doi{10.1016/j.pss.2010.07.024}.
\newblock URL
  \url{https://linkinghub.elsevier.com/retrieve/pii/S0032063310002266}.

\bibitem[Dobrijevic et~al.(2010{\natexlab{b}})Dobrijevic, Hébrard, Plessis,
  Carrasco, Pernot, and Bruno-Claeys]{dobrijevic_comparison_2010}
M.~Dobrijevic, E.~Hébrard, S.~Plessis, N.~Carrasco, P.~Pernot, and
  M.~Bruno-Claeys.
\newblock Comparison of methods for the determination of key reactions in
  chemical systems: {Application} to {Titan}’s atmosphere.
\newblock \emph{Advances in Space Research}, 45\penalty0 (1):\penalty0 77--91,
  Jan. 2010{\natexlab{b}}.
\newblock ISSN 02731177.
\newblock \doi{10.1016/j.asr.2009.06.005}.
\newblock URL
  \url{https://linkinghub.elsevier.com/retrieve/pii/S0273117709003834}.

\bibitem[Dobrijevic et~al.(2011)Dobrijevic, Cavalié, and
  Billebaud]{dobrijevic_methodology_2011}
M.~Dobrijevic, T.~Cavalié, and F.~Billebaud.
\newblock A methodology to construct a reduced chemical scheme for {2D}–{3D}
  photochemical models: {Application} to {Saturn}.
\newblock \emph{Icarus}, 214\penalty0 (1):\penalty0 275--285, July 2011.
\newblock ISSN 00191035.
\newblock \doi{10.1016/j.icarus.2011.04.027}.
\newblock URL
  \url{https://linkinghub.elsevier.com/retrieve/pii/S0019103511001631}.

\bibitem[Dobrijevic et~al.(2016)Dobrijevic, Loison, Hickson, and
  Gronoff]{dobrijevic_1d-coupled_2016}
M.~Dobrijevic, J.~Loison, K.~Hickson, and G.~Gronoff.
\newblock {1D}-coupled photochemical model of neutrals, cations and anions in
  the atmosphere of {Titan}.
\newblock \emph{Icarus}, 268:\penalty0 313--339, Apr. 2016.
\newblock ISSN 00191035.
\newblock \doi{10.1016/j.icarus.2015.12.045}.
\newblock URL
  \url{https://linkinghub.elsevier.com/retrieve/pii/S0019103515006119}.

\bibitem[Fletcher et~al.(2020)Fletcher, Simon, Hofstadter, Arridge, Cohen,
  Masters, Mandt, and Coustenis]{fletcher_ice_2020}
L.~N. Fletcher, A.~A. Simon, M.~D. Hofstadter, C.~S. Arridge, I.~Cohen,
  A.~Masters, K.~Mandt, and A.~Coustenis.
\newblock Ice {Giant} {System} {Exploration} in the 2020s: {An} {Introduction}.
\newblock \emph{arXiv:2008.12125 [astro-ph]}, Aug. 2020.
\newblock URL \url{http://arxiv.org/abs/2008.12125}.
\newblock arXiv: 2008.12125.

\bibitem[Fox and Victor(1988)]{fox_electron_1988}
J.~Fox and G.~Victor.
\newblock Electron energy deposition in {N2} gas.
\newblock \emph{Planetary and Space Science}, 36\penalty0 (4):\penalty0
  329--352, Apr. 1988.
\newblock ISSN 00320633.
\newblock \doi{10.1016/0032-0633(88)90123-7}.
\newblock URL
  \url{https://linkinghub.elsevier.com/retrieve/pii/0032063388901237}.

\bibitem[Fray and Schmitt(2009)]{fray_sublimation_2009}
N.~Fray and B.~Schmitt.
\newblock Sublimation of ices of astrophysical interest: {A} bibliographic
  review.
\newblock \emph{Planetary and Space Science}, 57\penalty0 (14-15):\penalty0
  2053--2080, Dec. 2009.
\newblock ISSN 00320633.
\newblock \doi{10.1016/j.pss.2009.09.011}.
\newblock URL
  \url{https://linkinghub.elsevier.com/retrieve/pii/S0032063309002736}.

\bibitem[Haynes(2012)]{haynes_crc_2012}
W.~Haynes.
\newblock \emph{{CRC} {Handbook} of {Chemistry} and {Physics}, 93rd {Edition}}.
\newblock 100 {Key} {Points}. Taylor \& Francis, 2012.
\newblock ISBN 978-1-4398-8049-4.
\newblock URL \url{https://books.google.fr/books?id=-BzP7Rkl7WkC}.

\bibitem[Helton et~al.(2006)Helton, Johnson, Sallaberry, and
  Storlie]{helton_survey_2006}
J.~Helton, J.~Johnson, C.~Sallaberry, and C.~Storlie.
\newblock Survey of sampling-based methods for uncertainty and sensitivity
  analysis.
\newblock \emph{Reliability Engineering \& System Safety}, 91\penalty0
  (10-11):\penalty0 1175--1209, Oct. 2006.
\newblock ISSN 09518320.
\newblock \doi{10.1016/j.ress.2005.11.017}.
\newblock URL
  \url{https://linkinghub.elsevier.com/retrieve/pii/S0951832005002292}.

\bibitem[Herbert and Sandel(1991)]{herbert_ch4_1991}
F.~Herbert and B.~R. Sandel.
\newblock {CH4} and haze in {Triton}'s lower atmosphere.
\newblock \emph{Journal of Geophysical Research}, 96:\penalty0 19, Oct. 1991.
\newblock \doi{10.1029/91JA01821}.
\newblock URL \url{http://adsabs.harvard.edu/abs/1991JGR....9619241H}.

\bibitem[Hickson et~al.(2020)Hickson, Bray, Loison, and
  Dobrijevic]{hickson_kinetic_2020}
K.~M. Hickson, C.~Bray, J.-C. Loison, and M.~Dobrijevic.
\newblock A kinetic study of the {N}({2D}) + {C2H4} reaction at low
  temperature.
\newblock \emph{Physical Chemistry Chemical Physics}, 22\penalty0
  (25):\penalty0 14026--14035, July 2020.
\newblock ISSN 1463-9084.
\newblock \doi{10.1039/D0CP02083D}.
\newblock URL
  \url{https://pubs.rsc.org/en/content/articlelanding/2020/cp/d0cp02083d}.
\newblock Publisher: The Royal Society of Chemistry.

\bibitem[Husain and Kirsch(1971)]{husain_kirsch_1971}
D.~Husain and L.~J. Kirsch.
\newblock Reactions of atomic carbon c(23pj) by kinetic absorption spectroscopy
  in the vacuum ultra-violet.
\newblock \emph{J. Chem. Soc. Faraday Trans.}, 67:\penalty0 2025--2035, 1971.

\bibitem[Hébrard et~al.(2006)Hébrard, Dobrijevic, Bénilan, and
  Raulin]{hebrard_photochemical_2006}
E.~Hébrard, M.~Dobrijevic, Y.~Bénilan, and F.~Raulin.
\newblock Photochemical kinetics uncertainties in modeling {Titan}’s
  atmosphere: {A} review.
\newblock \emph{Journal of Photochemistry and Photobiology C: Photochemistry
  Reviews}, 7\penalty0 (4):\penalty0 211--230, Dec. 2006.
\newblock ISSN 13895567.
\newblock \doi{10.1016/j.jphotochemrev.2006.12.004}.
\newblock URL
  \url{https://linkinghub.elsevier.com/retrieve/pii/S138955670700007X}.

\bibitem[Hébrard et~al.(2007)Hébrard, Dobrijevic, Bénilan, and
  Raulin]{hebrard_photochemical_2007}
E.~Hébrard, M.~Dobrijevic, Y.~Bénilan, and F.~Raulin.
\newblock Photochemical kinetics uncertainties in modeling {Titan}'s
  atmosphere: {First} consequences.
\newblock \emph{Planetary and Space Science}, 55\penalty0 (10):\penalty0
  1470--1489, July 2007.
\newblock ISSN 00320633.
\newblock \doi{10.1016/j.pss.2007.04.006}.
\newblock URL
  \url{https://linkinghub.elsevier.com/retrieve/pii/S0032063307001213}.

\bibitem[Hébrard et~al.(2009)Hébrard, Dobrijevic, Pernot, Carrasco, Bergeat,
  Hickson, Canosa, Le~Picard, and Sims]{hebrard_how_2009}
E.~Hébrard, M.~Dobrijevic, P.~Pernot, N.~Carrasco, A.~Bergeat, K.~M. Hickson,
  A.~Canosa, S.~D. Le~Picard, and I.~R. Sims.
\newblock How {Measurements} of {Rate} {Coefficients} at {Low} {Temperature}
  {Increase} the {Predictivity} of {Photochemical} {Models} of {Titan}’s
  {Atmosphere}.
\newblock \emph{The Journal of Physical Chemistry A}, 113\penalty0
  (42):\penalty0 11227--11237, Oct. 2009.
\newblock ISSN 1089-5639, 1520-5215.
\newblock \doi{10.1021/jp905524e}.
\newblock URL \url{https://pubs.acs.org/doi/10.1021/jp905524e}.

\bibitem[Hébrard et~al.(2012)Hébrard, Dobrijevic, Loison, Bergeat, and
  Hickson]{hebrard_neutral_2012}
E.~Hébrard, M.~Dobrijevic, J.~C. Loison, A.~Bergeat, and K.~M. Hickson.
\newblock Neutral production of hydrogen isocyanide ({HNC}) and hydrogen
  cyanide ({HCN}) in {Titan}’s upper atmosphere.
\newblock \emph{Astronomy \& Astrophysics}, 541:\penalty0 A21, May 2012.
\newblock ISSN 0004-6361, 1432-0746.
\newblock \doi{10.1051/0004-6361/201218837}.
\newblock URL \url{http://www.aanda.org/10.1051/0004-6361/201218837}.

\bibitem[Krasnopolsky and Cruikshank(1995)]{krasnopolsky_photochemistry_1995}
V.~A. Krasnopolsky and D.~P. Cruikshank.
\newblock Photochemistry of {Triton}'s atmosphere and ionosphere.
\newblock \emph{Journal of Geophysical Research}, 100\penalty0 (E10):\penalty0
  21271, 1995.
\newblock ISSN 0148-0227.
\newblock \doi{10.1029/95JE01904}.
\newblock URL \url{http://doi.wiley.com/10.1029/95JE01904}.

\bibitem[Krasnopolsky et~al.(1992)Krasnopolsky, Sandel, and
  Herbert]{krasnopolsky_properties_1992}
V.~A. Krasnopolsky, B.~R. Sandel, and F.~Herbert.
\newblock Properties of haze in the atmosphere of {Triton}.
\newblock \emph{Journal of Geophysical Research}, 97\penalty0 (E7):\penalty0
  11695, 1992.
\newblock ISSN 0148-0227.
\newblock \doi{10.1029/92JE00945}.
\newblock URL \url{http://doi.wiley.com/10.1029/92JE00945}.

\bibitem[Krasnopolsky et~al.(1993)Krasnopolsky, Sandel, Herbert, and
  Vervack]{krasnopolsky_temperature_1993}
V.~A. Krasnopolsky, B.~R. Sandel, F.~Herbert, and R.~J. Vervack.
\newblock Temperature, {N2}, and {N} density profiles of {Triton}'s atmosphere
  - {Observations} and model.
\newblock \emph{Journal of Geophysical Research}, 98:\penalty0 3065--3078, Feb.
  1993.
\newblock \doi{10.1029/92JE02680}.
\newblock URL \url{http://adsabs.harvard.edu/abs/1993JGR....98.3065K}.

\bibitem[Krimigis et~al.(1989)Krimigis, Armstrong, Axford, Bostrom, Cheng,
  Gloeckler, Hamilton, Keath, Lanzerotti, Mauk, and
  Van~Allen]{krimigis_hot_1989}
S.~M. Krimigis, T.~P. Armstrong, W.~I. Axford, C.~O. Bostrom, A.~F. Cheng,
  G.~Gloeckler, D.~C. Hamilton, E.~P. Keath, L.~J. Lanzerotti, B.~H. Mauk, and
  J.~A. Van~Allen.
\newblock Hot {Plasma} and {Energetic} {Particles} in {Neptune}'s
  {Magnetosphere}.
\newblock \emph{Science}, 246\penalty0 (4936):\penalty0 1483--1489, Dec. 1989.
\newblock ISSN 0036-8075, 1095-9203.
\newblock \doi{10.1126/science.246.4936.1483}.
\newblock URL
  \url{https://www.sciencemag.org/lookup/doi/10.1126/science.246.4936.1483}.

\bibitem[Lara et~al.(1996)Lara, Lellouch, López-Moreno, and
  Rodrigo]{lara_vertical_1996}
L.~M. Lara, E.~Lellouch, J.~J. López-Moreno, and R.~Rodrigo.
\newblock Vertical distribution of {Titan}'s atmospheric neutral constituents.
\newblock \emph{Journal of Geophysical Research: Planets}, 101\penalty0
  (E10):\penalty0 23261--23283, Oct. 1996.
\newblock ISSN 01480227.
\newblock \doi{10.1029/96JE02036}.
\newblock URL \url{http://doi.wiley.com/10.1029/96JE02036}.

\bibitem[Lellouch et~al.(2010)Lellouch, de~Bergh, Sicardy, Ferron, and
  Käufl]{lellouch_detection_2010}
E.~Lellouch, C.~de~Bergh, B.~Sicardy, S.~Ferron, and H.-U. Käufl.
\newblock Detection of {CO} in {Triton}'s atmosphere and the nature of
  surface-atmosphere interactions.
\newblock \emph{Astronomy and Astrophysics}, 512:\penalty0 L8, Mar. 2010.
\newblock ISSN 0004-6361, 1432-0746.
\newblock \doi{10.1051/0004-6361/201014339}.
\newblock URL \url{http://www.aanda.org/10.1051/0004-6361/201014339}.

\bibitem[Loison et~al.(2015)Loison, Hébrard, Dobrijevic, Hickson, Caralp, Hue,
  Gronoff, Venot, and Bénilan]{loison_neutral_2015}
J.~Loison, E.~Hébrard, M.~Dobrijevic, K.~Hickson, F.~Caralp, V.~Hue,
  G.~Gronoff, O.~Venot, and Y.~Bénilan.
\newblock The neutral photochemistry of nitriles, amines and imines in the
  atmosphere of {Titan}.
\newblock \emph{Icarus}, 247:\penalty0 218--247, Feb. 2015.
\newblock ISSN 00191035.
\newblock \doi{10.1016/j.icarus.2014.09.039}.
\newblock URL
  \url{https://linkinghub.elsevier.com/retrieve/pii/S0019103514005144}.

\bibitem[Loison et~al.(2019)Loison, Dobrijevic, and
  Hickson]{loison_photochemical_2019}
J.~Loison, M.~Dobrijevic, and K.~Hickson.
\newblock The photochemical production of aromatics in the atmosphere of
  {Titan}.
\newblock \emph{Icarus}, 329:\penalty0 55--71, Sept. 2019.
\newblock ISSN 00191035.
\newblock \doi{10.1016/j.icarus.2019.03.024}.
\newblock URL
  \url{https://linkinghub.elsevier.com/retrieve/pii/S0019103518305190}.

\bibitem[{Lyons} et~al.(1992){Lyons}, Yung, and Allen]{lyons_solar_1992}
{Lyons}, Y.~Yung, and M.~Allen.
\newblock Solar control of the upper atmosphere of {Triton}.
\newblock \emph{Science}, 256\penalty0 (5054):\penalty0 204--206, Apr. 1992.
\newblock ISSN 0036-8075, 1095-9203.
\newblock \doi{10.1126/science.11540928}.
\newblock URL
  \url{https://www.sciencemag.org/lookup/doi/10.1126/science.11540928}.

\bibitem[Majeed et~al.(1990)Majeed, McConnell, Strobel, and
  Summers]{majeed_ionosphere_1990}
T.~Majeed, J.~C. McConnell, D.~P. Strobel, and M.~E. Summers.
\newblock The ionosphere of {Triton}.
\newblock \emph{Geophysical Research Letters}, 17\penalty0 (10):\penalty0
  1721--1724, 1990.
\newblock ISSN 1944-8007.
\newblock \doi{10.1029/GL017i010p01721}.
\newblock URL
  \url{https://agupubs.onlinelibrary.wiley.com/doi/abs/10.1029/GL017i010p01721}.
\newblock \_eprint:
  https://agupubs.onlinelibrary.wiley.com/doi/pdf/10.1029/GL017i010p01721.

\bibitem[McKinnon et~al.(1995)McKinnon, Lunine, and
  Banfield]{mckinnon_origin_1995}
W.~B. McKinnon, J.~I. Lunine, and D.~Banfield.
\newblock Origin and evolution of {Triton}.
\newblock pages 807--877, 1995.
\newblock URL \url{http://adsabs.harvard.edu/abs/1995netr.conf..807M}.
\newblock Conference Name: Neptune and Triton.

\bibitem[Moses and Poppe(2017)]{moses_dust_2017}
J.~I. Moses and A.~R. Poppe.
\newblock Dust ablation on the giant planets: {Consequences} for stratospheric
  photochemistry.
\newblock \emph{Icarus}, 297:\penalty0 33--58, Nov. 2017.
\newblock ISSN 00191035.
\newblock \doi{10.1016/j.icarus.2017.06.002}.
\newblock URL
  \url{https://linkinghub.elsevier.com/retrieve/pii/S001910351730180X}.

\bibitem[Nuñez-Reyes et~al.(2019{\natexlab{a}})Nuñez-Reyes, Loison, Hickson,
  and Dobrijevic]{nunez-reyes_low_2019}
D.~Nuñez-Reyes, J.-C. Loison, K.~M. Hickson, and M.~Dobrijevic.
\newblock A low temperature investigation of the {N}( $^{\textrm{2}}$ {D}) +
  {CH} $_{\textrm{4}}$ , {C} $_{\textrm{2}}$ {H} $_{\textrm{6}}$ and {C}
  $_{\textrm{3}}$ {H} $_{\textrm{8}}$ reactions.
\newblock \emph{Physical Chemistry Chemical Physics}, 21\penalty0
  (12):\penalty0 6574--6581, 2019{\natexlab{a}}.
\newblock ISSN 1463-9076, 1463-9084.
\newblock \doi{10.1039/C9CP00798A}.
\newblock URL \url{http://xlink.rsc.org/?DOI=C9CP00798A}.

\bibitem[Nuñez-Reyes et~al.(2019{\natexlab{b}})Nuñez-Reyes, Loison, Hickson,
  and Dobrijevic]{nunez-reyes_rate_2019}
D.~Nuñez-Reyes, J.-C. Loison, K.~M. Hickson, and M.~Dobrijevic.
\newblock Rate constants for the {N}({2D}) + {C2H2} reaction over the 50–296
  {K} temperature range.
\newblock \emph{Physical Chemistry Chemical Physics}, 21\penalty0
  (40):\penalty0 22230--22237, Oct. 2019{\natexlab{b}}.
\newblock ISSN 1463-9084.
\newblock \doi{10.1039/C9CP04170B}.
\newblock URL
  \url{https://pubs.rsc.org/en/content/articlelanding/2019/cp/c9cp04170b}.
\newblock Publisher: The Royal Society of Chemistry.

\bibitem[Oliveira et~al.(2022)Oliveira, Sicardy, Gomes-Júnior, Ortiz, Strobel,
  Bertrand, Forget, Lellouch, Desmars, Bérard, Doressoundiram, Lecacheux,
  Leiva, Meza, Roques, Souami, Widemann, Santos-Sanz, Morales, Duffard,
  Fernández-Valenzuela, Castro-Tirado, Braga-Ribas, Morgado, Assafin, Camargo,
  Vieira-Martins, Benedetti-Rossi, Santos-Filho, Banda-Huarca, Quispe-Huaynasi,
  Pereira, Rommel, Margoti, Dias-Oliveira, Colas, Berthier, Renner, Hueso,
  Pérez-Hoyos, Sánchez-Lavega, Rojas, Beisker, Kretlow, Herald, Gault, Bath,
  Bode, Bredner, Guhl, Haymes, Hummel, Kattentidt, Klös, Pratt, Thome,
  Avdellidou, Gazeas, Karampotsiou, Tzouganatos, Kardasis, Christou, Xilouris,
  Alikakos, Gourzelas, Liakos, Charmandaris, Jelínek, Štrobl, Eberle, Rapp,
  Gährken, Klemt, Kowollik, Bitzer, Miller, Herzogenrath, Frangenberg,
  Brandis, Pütz, Perdelwitz, Piehler, Riepe, von Poschinger, Baruffetti,
  Cenadelli, Christille, Ciabattari, Di~Luca, Alboresi, Leto, Sanchez, Bruno,
  Occhipinti, Morrone, Cupolino, Noschese, Vecchione, Scalia, Savio, Giardina,
  Kamoun, Barbosa, Behrend, Spano, Bouchet, Cottier, Falco, Gallego,
  Tortorelli, Sposetti, Sussenbach, Abbeel, André, Llibre, Pailler, Ardissone,
  Boutet, Sanchez, Bretton, Cailleau, Pic, Granier, Chauvet, Conjat, Dauvergne,
  Dechambre, Delay, Delcroix, Rousselot, Ferreira, Machado, Tanga, Rivet,
  Frappa, Irzyk, Jabet, Kaschinski, Klotz, Rieugnie, Klotz, Labrevoir,
  Lavandier, Walliang, Leroy, Bouley, Lisciandra, Coliac, Metz, Erpelding,
  Nougayrède, Midavaine, Miniou, Moindrot, Morel, Reginato, Reginato, Rudelle,
  Tregon, Tanguy, David, Thuillot, Hestroffer, Vaudescal, Aissa, Grigahcene,
  Briggs, Broadbent, Denyer, Haigh, Quinn, Thurston, Fossey, Arena, Jennings,
  Talbot, Alonso, Reche, Casanova, Briggs, Iglesias-Marzoa, Ibáñez, Martín,
  González, García, Marchant, Ordonez-Etxeberria, Martorell, Salamero,
  Organero, Ana, Fonseca, Peris, Brevia, Selva, Perello, Cabedo, Gonçalves,
  Ferreira, Dias, Daassou, Barkaoui, Benkhaldoun, Guennoun, Chouqar, Jehin,
  Rinner, Lloyd, Moutamid, Lamarche, Pollock, Caton, Kouprianov, Timerson,
  Blanchard, Payet, Peyrot, Teng-Chuen-Yu, Françoise, Mondon, Payet, Boissel,
  Castets, Hubbard, Hill, Reitsema, Mousis, Ball, Neilsen, Hutcheon, Lay,
  Anderson, Moy, Jonsen, Pink, Walters, and Downs]{oliveira_constraints_2022}
J.~M. Oliveira, B.~Sicardy, A.~R. Gomes-Júnior, J.~L. Ortiz, D.~F. Strobel,
  T.~Bertrand, F.~Forget, E.~Lellouch, J.~Desmars, D.~Bérard,
  A.~Doressoundiram, J.~Lecacheux, R.~Leiva, E.~Meza, F.~Roques, D.~Souami,
  T.~Widemann, P.~Santos-Sanz, N.~Morales, R.~Duffard,
  E.~Fernández-Valenzuela, A.~J. Castro-Tirado, F.~Braga-Ribas, B.~E. Morgado,
  M.~Assafin, J.~I.~B. Camargo, R.~Vieira-Martins, G.~Benedetti-Rossi,
  S.~Santos-Filho, M.~V. Banda-Huarca, F.~Quispe-Huaynasi, C.~L. Pereira, F.~L.
  Rommel, G.~Margoti, A.~Dias-Oliveira, F.~Colas, J.~Berthier, S.~Renner,
  R.~Hueso, S.~Pérez-Hoyos, A.~Sánchez-Lavega, J.~F. Rojas, W.~Beisker,
  M.~Kretlow, D.~Herald, D.~Gault, K.-L. Bath, H.-J. Bode, E.~Bredner, K.~Guhl,
  T.~V. Haymes, E.~Hummel, B.~Kattentidt, O.~Klös, A.~Pratt, B.~Thome,
  C.~Avdellidou, K.~Gazeas, E.~Karampotsiou, L.~Tzouganatos, E.~Kardasis, A.~A.
  Christou, E.~M. Xilouris, I.~Alikakos, A.~Gourzelas, A.~Liakos,
  V.~Charmandaris, M.~Jelínek, J.~Štrobl, A.~Eberle, K.~Rapp, B.~Gährken,
  B.~Klemt, S.~Kowollik, R.~Bitzer, M.~Miller, G.~Herzogenrath, D.~Frangenberg,
  L.~Brandis, I.~Pütz, V.~Perdelwitz, G.~M. Piehler, P.~Riepe, K.~von
  Poschinger, P.~Baruffetti, D.~Cenadelli, J.-M. Christille, F.~Ciabattari,
  R.~Di~Luca, D.~Alboresi, G.~Leto, R.~Z. Sanchez, P.~Bruno, G.~Occhipinti,
  L.~Morrone, L.~Cupolino, A.~Noschese, A.~Vecchione, C.~Scalia, R.~L. Savio,
  G.~Giardina, S.~Kamoun, R.~Barbosa, R.~Behrend, M.~Spano, E.~Bouchet,
  M.~Cottier, L.~Falco, S.~Gallego, L.~Tortorelli, S.~Sposetti, J.~Sussenbach,
  F.~V.~D. Abbeel, P.~André, M.~Llibre, F.~Pailler, J.~Ardissone, M.~Boutet,
  J.~Sanchez, M.~Bretton, A.~Cailleau, V.~Pic, L.~Granier, R.~Chauvet,
  M.~Conjat, J.~L. Dauvergne, O.~Dechambre, P.~Delay, M.~Delcroix,
  L.~Rousselot, J.~Ferreira, P.~Machado, P.~Tanga, J.-P. Rivet, E.~Frappa,
  M.~Irzyk, F.~Jabet, M.~Kaschinski, A.~Klotz, Y.~Rieugnie, A.~N. Klotz,
  O.~Labrevoir, D.~Lavandier, D.~Walliang, A.~Leroy, S.~Bouley, S.~Lisciandra,
  J.-F. Coliac, F.~Metz, D.~Erpelding, P.~Nougayrède, T.~Midavaine, M.~Miniou,
  S.~Moindrot, P.~Morel, B.~Reginato, E.~Reginato, J.~Rudelle, B.~Tregon,
  R.~Tanguy, J.~David, W.~Thuillot, D.~Hestroffer, G.~Vaudescal, D.~B. Aissa,
  Z.~Grigahcene, D.~Briggs, S.~Broadbent, P.~Denyer, N.~J. Haigh, N.~Quinn,
  G.~Thurston, S.~J. Fossey, C.~Arena, M.~Jennings, J.~Talbot, S.~Alonso, A.~R.
  Reche, V.~Casanova, E.~Briggs, R.~Iglesias-Marzoa, J.~A. Ibáñez, M.~C.~D.
  Martín, H.~González, J.~L.~M. García, J.~Marchant, I.~Ordonez-Etxeberria,
  P.~Martorell, J.~Salamero, F.~Organero, L.~Ana, F.~Fonseca, V.~Peris,
  O.~Brevia, A.~Selva, C.~Perello, V.~Cabedo, R.~Gonçalves, M.~Ferreira, F.~M.
  Dias, A.~Daassou, K.~Barkaoui, Z.~Benkhaldoun, M.~Guennoun, J.~Chouqar,
  E.~Jehin, C.~Rinner, J.~Lloyd, M.~E. Moutamid, C.~Lamarche, J.~T. Pollock,
  D.~B. Caton, V.~Kouprianov, B.~W. Timerson, G.~Blanchard, B.~Payet,
  A.~Peyrot, J.-P. Teng-Chuen-Yu, J.~Françoise, B.~Mondon, T.~Payet,
  C.~Boissel, M.~Castets, W.~B. Hubbard, R.~Hill, H.~J. Reitsema, O.~Mousis,
  L.~Ball, G.~Neilsen, S.~Hutcheon, K.~Lay, P.~Anderson, M.~Moy, M.~Jonsen,
  I.~Pink, R.~Walters, and B.~Downs.
\newblock Constraints on the structure and seasonal variations of {Triton}'s
  atmosphere from the 5 {October} 2017 stellar occultation and previous
  observations.
\newblock \emph{arXiv:2201.10450 [astro-ph]}, Jan. 2022.
\newblock URL \url{http://arxiv.org/abs/2201.10450}.
\newblock arXiv: 2201.10450.

\bibitem[Poling et~al.(2001)Poling, Prausnitz, and O'Connell]{Poling}
B.~E. Poling, J.~M. Prausnitz, and J.~P. O'Connell.
\newblock \emph{The Properties of Gases and Liquids}.
\newblock 2001.

\bibitem[Poppe and Horányi(2018)]{poppe_interplanetary_2018}
A.~R. Poppe and M.~Horányi.
\newblock Interplanetary dust delivery of water to the atmospheres of {Pluto}
  and {Triton}.
\newblock \emph{Astronomy \& Astrophysics}, 617:\penalty0 L5, Sept. 2018.
\newblock ISSN 0004-6361, 1432-0746.
\newblock \doi{10.1051/0004-6361/201833980}.
\newblock URL \url{https://www.aanda.org/10.1051/0004-6361/201833980}.

\bibitem[Rymer et~al.(2021)Rymer, Runyon, Clyde, Núñez, Nikoukar, Soderlund,
  Sayanagi, Hofstadter, Quick, Stern, Becker, Hedman, Cohen, Crary, Fortney,
  Vertesi, Hansen, de~Pater, Paty, Spilker, Stallard, Hospodarsky, Smith,
  Wakeford, Moran, Annex, Schenk, Ozimek, Arrieta, McNutt, Masters, Simon,
  Ensor, Apland, Bruzzi, Patthoff, Scott, Campo, Krupiarz, Cochrane, Gantz,
  Rodriguez, Gallagher, Hurley, Crowley, Abel, Provornikova, Turtle, Clark,
  Wilkes, Hunt, Roberts, Rehm, Murray, Wolfarth, Fletcher, Spilker, Martin,
  Parisi, Norkus, Izenberg, Stough, Vervack, Mandt, Stevenson, Kijewski, Cheng,
  Feldman, Allen, Prabhu, Dutta, Young, and Williams]{rymer_neptune_2021}
A.~M. Rymer, K.~D. Runyon, B.~Clyde, J.~I. Núñez, R.~Nikoukar, K.~M.
  Soderlund, K.~Sayanagi, M.~Hofstadter, L.~C. Quick, S.~A. Stern, T.~Becker,
  M.~Hedman, I.~Cohen, F.~Crary, J.~J. Fortney, J.~Vertesi, C.~Hansen,
  I.~de~Pater, C.~Paty, T.~Spilker, T.~Stallard, G.~B. Hospodarsky, H.~T.
  Smith, H.~Wakeford, S.~E. Moran, A.~Annex, P.~Schenk, M.~Ozimek, J.~Arrieta,
  R.~L. McNutt, A.~Masters, A.~A. Simon, S.~Ensor, C.~T. Apland, J.~Bruzzi,
  D.~A. Patthoff, C.~Scott, C.~Campo, C.~Krupiarz, C.~J. Cochrane, C.~Gantz,
  D.~Rodriguez, D.~Gallagher, D.~Hurley, D.~Crowley, E.~Abel, E.~Provornikova,
  E.~P. Turtle, G.~Clark, J.~Wilkes, J.~Hunt, J.~H. Roberts, J.~Rehm,
  K.~Murray, L.~Wolfarth, L.~N. Fletcher, L.~Spilker, E.~S. Martin, M.~Parisi,
  M.~Norkus, N.~Izenberg, R.~Stough, R.~J. Vervack, K.~Mandt, K.~B. Stevenson,
  S.~Kijewski, W.~Cheng, J.~D. Feldman, G.~Allen, D.~Prabhu, S.~Dutta,
  C.~Young, and J.~Williams.
\newblock Neptune {Odyssey}: {A} {Flagship} {Concept} for the {Exploration} of
  the {Neptune}–{Triton} {System}.
\newblock \emph{The Planetary Science Journal}, 2\penalty0 (5):\penalty0 184,
  Oct. 2021.
\newblock ISSN 2632-3338.
\newblock \doi{10.3847/PSJ/abf654}.
\newblock URL \url{https://iopscience.iop.org/article/10.3847/PSJ/abf654}.

\bibitem[Sander et~al.(2006)Sander, Rvishankara, Golden, Kolb, Kurylo, Molina,
  Moortgat, Finlayson-Pitts, H., and Huie]{sander_notitle_2006}
S.~Sander, A.~Rvishankara, D.~Golden, C.~Kolb, M.~Kurylo, M.~Molina,
  G.~Moortgat, B.~Finlayson-Pitts, W.~H., and R.~Huie.
\newblock 06-2:\penalty0 1--522, 2006.

\bibitem[Stevens et~al.(1992)Stevens, Strobel, Summers, and
  Yelle]{stevens_thermal_1992}
M.~H. Stevens, D.~F. Strobel, M.~E. Summers, and R.~V. Yelle.
\newblock On the thermal structure of {Triton}'s thermosphere.
\newblock \emph{Geophysical Research Letters}, 19\penalty0 (7):\penalty0
  669--672, Apr. 1992.
\newblock ISSN 00948276.
\newblock \doi{10.1029/92GL00651}.
\newblock URL \url{http://doi.wiley.com/10.1029/92GL00651}.

\bibitem[Strobel and Summers(1995)]{strobel_tritons_1995}
D.~F. Strobel and M.~E. Summers.
\newblock Triton's upper atmosphere and ionosphere.
\newblock pages 1107--1148, 1995.
\newblock URL \url{http://adsabs.harvard.edu/abs/1995netr.conf.1107S}.
\newblock Conference Name: Neptune and Triton.

\bibitem[Strobel and Zhu(2017)]{strobel_comparative_2017}
D.~F. Strobel and X.~Zhu.
\newblock Comparative planetary nitrogen atmospheres: {Density} and thermal
  structures of {Pluto} and {Triton}.
\newblock \emph{Icarus}, 291:\penalty0 55--64, July 2017.
\newblock ISSN 00191035.
\newblock \doi{10.1016/j.icarus.2017.03.013}.
\newblock URL
  \url{https://linkinghub.elsevier.com/retrieve/pii/S0019103516306583}.

\bibitem[Strobel et~al.(1990{\natexlab{a}})Strobel, Cheng, Summers, and
  Strickland]{strobel_magnetospheric_1990}
D.~F. Strobel, A.~F. Cheng, M.~E. Summers, and D.~J. Strickland.
\newblock Magnetospheric interaction with {Triton}'s ionosphere.
\newblock \emph{Geophysical Research Letters}, 17\penalty0 (10):\penalty0
  1661--1664, 1990{\natexlab{a}}.
\newblock ISSN 1944-8007.
\newblock \doi{10.1029/GL017i010p01661}.
\newblock URL
  \url{https://agupubs.onlinelibrary.wiley.com/doi/abs/10.1029/GL017i010p01661}.
\newblock \_eprint:
  https://agupubs.onlinelibrary.wiley.com/doi/pdf/10.1029/GL017i010p01661.

\bibitem[Strobel et~al.(1990{\natexlab{b}})Strobel, Simmers, Herbert, and
  Sandel]{strobel_photochemistry_1990}
D.~F. Strobel, M.~E. Simmers, F.~Herbert, and B.~R. Sandel.
\newblock The photochemistry of methane in the atmosphere of {Triton}.
\newblock \emph{Geophysical Research Letters}, 17\penalty0 (10):\penalty0
  1729--1732, Sept. 1990{\natexlab{b}}.
\newblock ISSN 00948276.
\newblock \doi{10.1029/GL017i010p01729}.
\newblock URL \url{http://doi.wiley.com/10.1029/GL017i010p01729}.

\bibitem[Summers and Strobel(1991)]{summers_tritons_1991}
M.~E. Summers and D.~F. Strobel.
\newblock Triton's atmosphere: {A} source of {N} and {H} for {Neptune}'s
  magnetosphere.
\newblock page~4, 1991.

\bibitem[Thuillier et~al.(2004)Thuillier, Floyd, Woods, Cebula, Hilsenrath,
  Hersé, and Labs]{thuillier_solar_2004}
G.~Thuillier, L.~Floyd, T.~Woods, R.~Cebula, E.~Hilsenrath, M.~Hersé, and
  D.~Labs.
\newblock Solar irradiance reference spectra for two solar active levels.
\newblock \emph{Advances in Space Research}, 34\penalty0 (2):\penalty0
  256--261, Jan. 2004.
\newblock ISSN 02731177.
\newblock \doi{10.1016/j.asr.2002.12.004}.
\newblock URL
  \url{https://linkinghub.elsevier.com/retrieve/pii/S0273117704002388}.

\bibitem[Tyler et~al.(1989)Tyler, Sweetnam, Anderson, Borutzki, Campbell,
  Eshleman, Gresh, Gurrola, Hinson, Kawashima, Kursinski, Levy, Lindal, Lyons,
  Marouf, Rosen, Simpson, and Wood]{tyler_voyager_1989}
G.~L. Tyler, D.~N. Sweetnam, J.~D. Anderson, S.~E. Borutzki, J.~K. Campbell,
  V.~R. Eshleman, D.~L. Gresh, E.~M. Gurrola, D.~P. Hinson, N.~Kawashima, E.~R.
  Kursinski, G.~S. Levy, G.~F. Lindal, J.~R. Lyons, E.~A. Marouf, P.~A. Rosen,
  R.~A. Simpson, and G.~E. Wood.
\newblock Voyager {Radio} {Science} {Observations} of {Neptune} and {Triton}.
\newblock \emph{Science}, 246\penalty0 (4936):\penalty0 1466--1473, Dec. 1989.
\newblock ISSN 0036-8075, 1095-9203.
\newblock \doi{10.1126/science.246.4936.1466}.
\newblock URL
  \url{https://www.sciencemag.org/lookup/doi/10.1126/science.246.4936.1466}.

\bibitem[Vuitton et~al.(2019)Vuitton, Yelle, Klippenstein, Hörst, and
  Lavvas]{vuitton_simulating_2019}
V.~Vuitton, R.~Yelle, S.~Klippenstein, S.~Hörst, and P.~Lavvas.
\newblock Simulating the density of organic species in the atmosphere of
  {Titan} with a coupled ion-neutral photochemical model.
\newblock \emph{Icarus}, 324:\penalty0 120--197, May 2019.
\newblock ISSN 00191035.
\newblock \doi{10.1016/j.icarus.2018.06.013}.
\newblock URL
  \url{https://linkinghub.elsevier.com/retrieve/pii/S0019103517307522}.

\bibitem[Yelle et~al.(1991)Yelle, Lunine, and Hunten]{yelle_energy_1991}
R.~V. Yelle, J.~I. Lunine, and D.~M. Hunten.
\newblock Energy balance and plume dynamics in {Triton}'s lower atmosphere.
\newblock \emph{Icarus}, 89\penalty0 (2):\penalty0 347--358, Feb. 1991.
\newblock ISSN 00191035.
\newblock \doi{10.1016/0019-1035(91)90182-S}.
\newblock URL
  \url{https://linkinghub.elsevier.com/retrieve/pii/001910359190182S}.

\bibitem[Yelle et~al.(1995)Yelle, Lunine, Pollack, and Brown]{yelle_lower_1995}
R.~V. Yelle, J.~I. Lunine, J.~B. Pollack, and R.~H. Brown.
\newblock Lower atmospheric structure and surface-atmosphere interactions on
  {Triton}.
\newblock pages 1031--1105, 1995.
\newblock URL \url{http://adsabs.harvard.edu/abs/1995netr.conf.1031Y}.
\newblock Conference Name: Neptune and Triton.

\bibitem[Zhu et~al.(2014)Zhu, Strobel, and Erwin]{zhu_density_2014}
X.~Zhu, D.~F. Strobel, and J.~T. Erwin.
\newblock The density and thermal structure of {Pluto}’s atmosphere and
  associated escape processes and rates.
\newblock \emph{Icarus}, 228:\penalty0 301--314, Jan. 2014.
\newblock ISSN 00191035.
\newblock \doi{10.1016/j.icarus.2013.10.011}.
\newblock URL
  \url{https://linkinghub.elsevier.com/retrieve/pii/S0019103513004302}.

\end{thebibliography}

\newpage

\begin{appendix} 

   \section{Identification of key uncertainty reactions}
   \label{Appendix_key_unc_rea}

   \begin{table}[!h]     
   \begin{center}
      \small
      \begin{tabular}{l c l l c}     
   \hline                    
   \multicolumn{1}{c}{Reaction}                                 & \begin{tabular}[c]{@{}l@{}}Number of\\ occurrences\end{tabular} & \begin{tabular}[c]{@{}l@{}}Maximum\\ RCC\end{tabular} & $F_i$        & $g_i$ \\ \hline
   O($^3$P) + CNN $\longrightarrow$ CO + N$_2$                  & 6                                                               & 0.43                                                  & 3.0          & 7.0   \\
   C + N$_2$ $\longrightarrow$ CNN                              & 6                                                               & 0.51                                                  & 1.8          & 10.0  \\
                                                                &                                                                 &                                                       & 10.0         & 0.0   \\
                                                                &                                                                 &                                                       & 30.0         & 0.0   \\
   N($^4$S) + N($^4$S) $\longrightarrow$ N$_2$                  & 5                                                               & 0.42                                                  & 2.5          & 100.0 \\
                                                                &                                                                 &                                                       & 2.0          & 100.0 \\
                                                                &                                                                 &                                                       & 1.0          & 0.0   \\
   CH$_4$ + $h\nu$ $\longrightarrow$ CH$_3$ + H                 & 4                                                               & 0.33                                                  & \textbf{1.2} & -     \\
   CH + CH$_4$ $\longrightarrow$ C$_2$H$_4$ + H                 & 4                                                               & 0.44                                                  & 1.3          & 4.45  \\
   N($^2$D) + CO $\longrightarrow$ N($^4$S) + CO                & 4                                                               & 1.0                                                   & 1.6          & 300.0 \\
   CO + $h\nu$ $\longrightarrow$ CO$^+$ + $e^-$                     & 4                                                               & 0.34                                                  & \textbf{1.2} & -     \\
   N($^4$S) + H $\longrightarrow$ NH                            & 3                                                               & 0.55                                                  & 3.16         & 100.0 \\
                                                                &                                                                 &                                                       & 2.0          & 100.0 \\
                                                                &                                                                 &                                                       & 30.0         & 0.0   \\
   H + iC$_4$H$_7$ $\longrightarrow$ CH$_3$ + CH$_2$CHCH$_2$    & 3                                                               & 0.22                                                  & 0.0          & 0.0   \\
                                                                &                                                                 &                                                       & 0.0          & 0.0   \\
                                                                &                                                                 &                                                       & 1.0          & 0.0   \\
   C$_{10}$H$_8$ + $h\nu$ $\longrightarrow$ C$_{10}$H$_8^+$ + $e^-$ & 3                                                               & 0.29                                                  & \textbf{1.2} & -     \\
   CH$_3^+$ + CH$_4$ $\longrightarrow$ C$_2$H$_5^+$ + H$_2$     & 3                                                               & 0.46                                                  & 1.4          & 0.0   \\
   CO$^+$ + H$_2$ $\longrightarrow$ HOC$^+$ + H                 & 3                                                               & 0.23                                                  & 1.15         & 0.0   \\
   CO$^+$ + C $\longrightarrow$ C$^+$ + CO                      & 3                                                               & 0.28                                                  & 1.4          & 0.0   \\
   N$^+$ + C $\longrightarrow$ N($^4$S) + C$^+$                 & 3                                                               & 0.80                                                  & 10.0         & 0.0   \\
   N$_2$ + ME $\longrightarrow$ N$_2^+$ + $e^-$                     & 3                                                               & 0.61                                                  & \textbf{1.2} & -     \\
   CH$_4$ + $h\nu$ $\longrightarrow$ $^1$CH$_2$ + H$_2$         & 2                                                               & 0.28                                                  & \textbf{1.2} & -     \\
   CH$_3$CH$_2$CCH + $h\nu$ $\longrightarrow$ C$_4$H$_5$ + H    & 2                                                               & 0.23                                                  & \textbf{1.2} & -     \\
   N($^2$D) + C$_6$H$_5$C$_6$H$_5$ $\longrightarrow$ AROM       & 2                                                               & 0.21                                                  & 3.0          & 0.0   \\
   N($^2$D) + C$_2$H$_3$ $\longrightarrow$ NH + C$_2$H$_2$      & 2                                                               & 0.2                                                   & 2.0          & 7.0   \\
   N($^4$S) + CH $\longrightarrow$ CN + H                       & 2                                                               & 0.26                                                  & 1.6          & 7.0   \\
   O($^3$P) + CN $\longrightarrow$ CO + N($^4$S)                & 2                                                               & 0.27                                                  & 2.0          & 0.0   \\
   C + H$_2$ $\longrightarrow$ $^3$CH$_2$                       & 2                                                               & 0.30                                                  & 2.0          & 100.0 \\
                                                                &                                                                 &                                                       & 3.0          & 100.0 \\
                                                                &                                                                 &                                                       & 1.0          & 0.0   \\
                                                                N$_2$ + $h\nu$ $\longrightarrow$ N$_2^+$ + $e^-$                                       & 2                                                               & 0.39                                                  & \textbf{1.2} & -     \\
   H$_3^+$ + C$_6$H$_5$C$_2$H$_5$ $\longrightarrow$ C$_6$H$_5^+$ + C$_2$H$_6$ + H$_2$ & 2                                                               & 0.24                                                  & 1.4          & 0.0   \\
   CH$_3$CNH$^+$ + CH$_3$NH$_2$ $\longrightarrow$ CH$_3$NH$_3^+$ + CH$_3$CN           & 2                                                               & 0.22                                                  & 3.0          & 0.0   \\
   C$_2$H$_5^+$ + C$_4$H$_2$ $\longrightarrow$ C$_4$H$_3^+$ + C$_2$H$_4$              & 2                                                               & 0.21                                                  & 2.0          & 0.0   \\
   N$_2^+$ + H$_2$ $\longrightarrow$ N$_2$H$^+$ + H                                   & 2                                                               & 0.22                                                  & 1.25         & 0.0   \\
   N$_2^+$ + N($^4$S) $\longrightarrow$ N$_2$ + N$^+$                                 & 2                                                               & 0.29                                                  & 3.0          & 0.0   \\
   C$^+$ + H$_2$ $\longrightarrow$ CH$_2^+$                                           & 2                                                               & 0.31                                                  & 2.0          & 100.0 \\
                                                                                      &                                                                 &                                                       & 3.0          & 100.0 \\
                                                                                      &                                                                 &                                                       & 1.6          & 0.0   \\
   NH$_2^+$ + $e^-$ $\longrightarrow$ NH + H                                              & 2                                                               & 0.22                                                  & 1.6          & 0.0   \\
   NH$_2^+$ + $e^-$ $\longrightarrow$ N($^4$S) + H + H                                    & 2                                                               & 0.21                                                  & 1.6          & 0.0   \\
   N$_2^+$ + $e^-$ $\longrightarrow$ N($^4$S) + N($^2$D)                                  & 2                                                               & 0.71                                                  & 2.0          & 0.0   \\
   N$_2^+$ + $e^-$ $\longrightarrow$ N($^2$D) + N($^2$D)                                  & 2                                                               & 0.72                                                  & 2.0          & 0.0   \\
   N$_2$ + ME $\longrightarrow$ N$^+$ + N($^2$D) + $e^-$                                  & 2                                                               & 0.36                                                  & \textbf{1.2} & -     \\
   N$_2$ + ME $\longrightarrow$ N($^4$S) + N($^2$D)                                   & 2                                                               & 0.46                                                  & \textbf{1.2} & -     \\ \hline                  
   \end{tabular}
   \end{center}          
   \caption{\small Reactions with $\lvert$ RCC $\rvert$ > 0.2 for several of the main atmospheric species. These reactions are identified through the sensitivity analyzes done for each of the main atmospheric species at the altitude where the uncertainty on this species abundance is maximum, as given in Table \ref{table_means&F_250tirages}.
   These analyzes are performed for 250 iterations of the Monte Carlo procedure. The number of occurrences is the number of main species for which the considered reaction has a RCC higher than 0.2 in absolute value. $F_i$ is the uncertainty factor at 300\,K (=$F_i(300\text{K})$) for all reactions except for photodissociations, photoionizations and reactions with ME (Magnetospheric Electrons) where it is the global uncertainty factor (in this case the value is given in bold). $g_i$ is the coefficient allowing to compute $F_i$ at different temperatures following Eq. \eqref{calc_Fi(T)}. For three-body reactions, we give the uncertainty factors corresponding respectively to $k_0$, $k_{\infty}$ and $k_r$ (in this order, cf Eq. \eqref{three_body_ki}). } 
   \label{key_react_RCCsolo_250tirages} 
   \end{table}

\begin{table}[!h]
   \begin{center}
      \begin{tabular}{l l c}
        \hline
\multicolumn{1}{c}{Reaction}                      & Species    & \begin{tabular}[c]{@{}c@{}}Maximum\\ RCC\end{tabular} \\ \hline
N($^2$D) + CO $\longrightarrow$ N($^4$S) + CO     & N($^2$D)   & -1.0                                                  \\
                                                  & H$^+$      & -0.67                                                 \\
                                                  & C$_2$H$_6$ & 0.52                                                  \\
N$^+$ + C $\longrightarrow$ N($^4$S) + C$^+$      & N$^+$      & -0.80                                                 \\
                                                  & N$_2^+$    & -0.55                                                 \\
N$_2$ + ME $\longrightarrow$ N$_2^+$ + $e^-$          & N$_2$      & -0.61                                                 \\
                                                  & CO         & -0.5                                                  \\
H + N($^4$S) $\longrightarrow$ NH                 & N($^4$S)   & -0.55                                                 \\
C + N$_2$ $\longrightarrow$ CNN                   & C$_2$H$_2$ & -0.51                                                 \\
CH + N$_2$ $\longrightarrow$ HCNN                 & H          & -0.52                                                 \\
N$_2^+$ + $e^-$ $\longrightarrow$ N($^2$D) + N($^2$D) & N$_2^+$    & -0.72                                                 \\
N$_2^+$ + $e^-$ $\longrightarrow$ N($^4$S) + N($^2$D) & N$_2^+$    & -0.71                                                 \\ \hline
\end{tabular}
\end{center}
\caption[]{Reactions with $\lvert$ RCC $\rvert$ > 0.5 for the main atmospheric species. These reactions are identified through the sensitivity analyzes done for each of the main atmospheric species at the altitude where the uncertainty on this species abundance is maximum, as given in Table \ref{table_means&F_250tirages}. These analyzes are done for 250 iterations of the Monte Carlo procedure. The value of the RCC is given along with the species for which it is reached.}
\label{high_RCCsolo_250tirages}
\end{table}

\begin{table}[!h]
   \begin{center}
      \begin{tabular}{c c c}
     \hline
\multicolumn{1}{c}{\begin{tabular}[c]{@{}c@{}}Altitude\\ level\\ {[}km{]}\end{tabular}} & \multicolumn{1}{c}{Reaction}                             & \begin{tabular}[c]{@{}c@{}}Number\\ of\\ occurrences\end{tabular} \\ \hline
0                                                                                       & N($^2$D) + CO $\longrightarrow$ N($^4$S) + CO            & 69                                                                \\ \hline
86                                                                     & O($^3$P) + CNN $\longrightarrow$ CO + N$_2$              & 82                                                                \\
                                                                                     & C + N$_2$ $\longrightarrow$ CNN                          & 76                                                                \\
                                                                                     & N($^2$D) + CO $\longrightarrow$ N($^4$S) + CO            & 66                                                                \\
                                                                                     & C + H$_2$ $\longrightarrow$ $^3$CH$_2$                   & 66                                                                \\
                                                                                     & HCO$^+$ + C $\longrightarrow$ CH$^+$ + CO                & 66                                                                \\
                                                                                     & N($^4$S) + N($^4$S) $\longrightarrow$ N$_2$              & 57                                                                \\ \hline
220                                                                    & N($^2$D) + CO $\longrightarrow$ N($^4$S) + CO            & 99                                                                \\
                                                                                     & CH$_3^+$ + CH$_4$ $\longrightarrow$ C$_2$H$_5^+$ + H$_2$ & 83                                                                \\
                                                                                     & CH$_4$ + $h\nu$ $\longrightarrow$ CH$_3$ + H             & 76                                                                \\
                                                                                     & CO + $h\nu$ $\longrightarrow$ CO$^+$ + $e^-$                 & 62                                                                \\ \hline
334                                                                    & N($^2$D) + CO $\longrightarrow$ N($^4$S) + CO            & 121                                                               \\
                                                                                     & N($^2$D) + C $\longrightarrow$ N($^4$S) + C              & 61                                                                \\ \hline
502                                                                    & N($^2$D) + CO $\longrightarrow$ N($^4$S) + CO            & 127                                                               \\
                                                                                     & N($^2$D) + C $\longrightarrow$ N($^4$S) + C              & 68                                                                \\
                                                                                     & N$^+$ + C $\longrightarrow$ N($^4$S) + C$^+$             & 58                                                                \\ \hline
758                                                                    & N($^2$D) + CO $\longrightarrow$ N($^4$S) + CO            & 128                                                               \\
                                                                                     & N$^+$ + C $\longrightarrow$ N($^4$S) + C$^+$             & 78                                                                \\
                                                                                     & N($^2$D) + C $\longrightarrow$ N($^4$S) + C              & 65                                                                \\ \hline
1026                                                                   & N($^2$D) + CO $\longrightarrow$ N($^4$S) + CO            & 121                                                               \\
                                                                                     & N$^+$ + C $\longrightarrow$ N($^4$S) + C$^+$             & 74                                                                \\
                                                                                     & N($^2$D) + C $\longrightarrow$ N($^4$S) + C              & 64                                                                \\ \hline
\end{tabular}
   \end{center}
   \caption[]{Reactions with $\lvert$ RCC $\rvert$ > 0.2 for at least one quarter of the species considered in our chemical scheme (=55). Each of these sensitivity analyzes is performed for all the atmospheric species at once. We perform an analysis at seven different altitude levels in order to sample different zones of the atmosphere. 250 iterations of the Monte-Carlo procedure are used here.}
   \label{RCC_All_6lvls_supQuarter_250tirages}
  
\end{table}

\begin{table}[!h]     
   \begin{center}
      \begin{tabular}{l l l l l l l l}     
         \hline                    
         Reaction                                             & 0 km                                                                                          & 86 km                                                                                  & 220 km                                                                & 334 km                                                                                  & 502 km                                                                                           & 758 km                                                                                           & 1026km                                                                                           \\ \hline
         N($^2$D) + CO $\longrightarrow$ N($^4$S) + CO        & N($^2$D)                                                                                      & N($^2$D)                                                                               & \begin{tabular}[c]{@{}l@{}}N($^2$D)\\ C$_2$H$_6$\\ H$^+$\end{tabular} & \begin{tabular}[c]{@{}l@{}}N($^2$D)\\ C$_2$H$_2$\\ C$_2$H$_4$\\ C$_2$H$_6$\end{tabular} & \begin{tabular}[c]{@{}l@{}}N($^2$D)\\ CH$_4$\\ C$_2$H$_2$\\ C$_2$H$_4$\\ C$_2$H$_6$\end{tabular} & \begin{tabular}[c]{@{}l@{}}N($^2$D)\\ CH$_4$\\ C$_2$H$_2$\\ C$_2$H$_4$\\ C$_2$H$_6$\end{tabular} & \begin{tabular}[c]{@{}l@{}}N($^2$D)\\ CH$_4$\\ C$_2$H$_2$\\ C$_2$H$_4$\\ C$_2$H$_6$\end{tabular} \\ \hline
         O($^3$P) + CNN $\longrightarrow$ CO + N$_2$          &                                                                                               & \begin{tabular}[c]{@{}l@{}}N($^4$S)\\ N$_2$\end{tabular}                               & N($^4$S)                                                              & N($^4$S)                                                                                & N($^4$S)                                                                                         & N($^4$S)                                                                                         & N($^4$S)                                                                                         \\ \hline
         CH$_4$ + $h\nu$ $\longrightarrow$ $^1$CH$_2$ + H$_2$ & \begin{tabular}[c]{@{}l@{}}H$_2$\\ CH$_4$\\ C$_2$H$_2$\\ C$_2$H$_4$\\ C$_2$H$_6$\end{tabular} & \begin{tabular}[c]{@{}l@{}}H$_2$\\ CH$_4$\end{tabular}                                 & H$_2$                                                                 & H$_2$                                                                                   &                                                                                                  &                                                                                                  &                                                                                                  \\ \hline
         N$_2^+$ + $e^-$ $\longrightarrow$ N($^4$S) + N($^2$D)    &                                                                                               &                                                                                        &                                                                       & N$_2^+$                                                                                 & \begin{tabular}[c]{@{}l@{}}H\\ N$_2^+$\end{tabular}                                              & \begin{tabular}[c]{@{}l@{}}H\\ N$_2^+$\end{tabular}                                              & \begin{tabular}[c]{@{}l@{}}H\\ N$_2^+$\end{tabular}                                              \\ \hline
         N$_2^+$ + $e^-$ $\longrightarrow$ N($^2$D) + N($^2$D)    &                                                                                               &                                                                                        &                                                                       & N$_2^+$                                                                                 & \begin{tabular}[c]{@{}l@{}}H\\ N$_2^+$\end{tabular}                                              & \begin{tabular}[c]{@{}l@{}}H\\ N$_2^+$\end{tabular}                                              & \begin{tabular}[c]{@{}l@{}}H\\ N$_2^+$\end{tabular}                                              \\ \hline
         N$_2$ + ME $\longrightarrow$ N$_2^+$ + $e^-$             &                                                                                               &                                                                                        &                                                                       & N$_2$                                                                                   & N$_2$                                                                                            & N$_2$                                                                                            & \begin{tabular}[c]{@{}l@{}}N$_2$\\ CO\end{tabular}                                               \\ \hline
         N$^+$ + CO $\longrightarrow$ C + NO$^+$              &                                                                                               &                                                                                        & N$^+$                                                                 & N$^+$                                                                                   & N$^+$                                                                                            &                                                                                                  &                                                                                                  \\ \hline
         N$^+$ + CO $\longrightarrow$ N($^4$S) + CO$^+$       &                                                                                               &                                                                                        & N$^+$                                                                 & N$^+$                                                                                   & N$^+$                                                                                            &                                                                                                  &                                                                                                  \\ \hline
         N($^2$D) + C $\longrightarrow$ N($^4$S) + C          &                                                                                               &                                                                                        &                                                                       &                                                                                         & N($^2$D)                                                                                         & N($^2$D)                                                                                         & N($^2$D)                                                                                         \\ \hline
         N$^+$ + C $\longrightarrow$ N($^4$S) + C             &                                                                                               &                                                                                        &                                                                       &                                                                                         &                                                                                                  & \begin{tabular}[c]{@{}l@{}}H$^+$\\ C$^+$\\ N$^+$\\ N$_2^+$\\ E\end{tabular}                      & \begin{tabular}[c]{@{}l@{}}C$^+$\\ N$^+$\\ N$_2^+$\\ E\end{tabular}                              \\ \hline
         C + N$_2$ $\longrightarrow$ CNN                      &                                                                                               & \begin{tabular}[c]{@{}l@{}}C\\ O($^3$P)\\ C$_2$H$_2$\\ C$_2$H$_4$\\ C$^+$\end{tabular} & \begin{tabular}[c]{@{}l@{}}C\\ O($^3$P)\end{tabular}                  &                                                                                         &                                                                                                  &                                                                                                  &                                                                                                  \\ \hline
         N$_2^+$ + C $\longrightarrow$ N$_2$ + C$^+$          &                                                                                               &                                                                                        &                                                                       & \begin{tabular}[c]{@{}l@{}}C$^+$\\ N$_2^+$\\ E\end{tabular}                             & \begin{tabular}[c]{@{}l@{}}H\\ H$^+$\\ C$^+$\\ E\end{tabular}                                    &                                                                                                  &                                                                                                  \\ \hline
         C + H$_2$ $\longrightarrow$ $^3$CH$_2$               & CO                                                                                            & \begin{tabular}[c]{@{}l@{}}CO\\ O($^3$P)\end{tabular}                                  &                                                                       &                                                                                         &                                                                                                  &                                                                                                  &                                                                                                  \\ \hline
         H$^+$ + $e^-$ $\longrightarrow$ H                        &                                                                                               &                                                                                        &                                                                       &                                                                                         &                                                                                                  & H$^+$                                                                                            & H$^+$                                                                                            \\ \hline
         N$_2^+$ + N($^4$S) $\longrightarrow$ N$_2$ + N$^+$   &                                                                                               &                                                                                        &                                                                       & \begin{tabular}[c]{@{}l@{}}CO\\ N$^+$\end{tabular}                                      &                                                                                                  &                                                                                                  &                                                                                                  \\ \hline
         H + N($^4$S) $\longrightarrow$ NH                    & N$_2$                                                                                         &                                                                                        &                                                                       &                                                                                         &                                                                                                  &                                                                                                  &                                                                                                  \\ \hline
         CH + N$_2$ $\longrightarrow$ HCNN                    & H                                                                                             &                                                                                        &                                                                       &                                                                                         &                                                                                                  &                                                                                                  &                                                                                                  \\ \hline
         CO + $h\nu$ $\longrightarrow$ CO$^+$ + $e^-$             &                                                                                               & E                                                                                      &                                                                       &                                                                                         &                                                                                                  &                                                                                                  &                                                                                                  \\ \hline
         CH + N$_2$ $\longrightarrow$ HCNN                    &                                                                                               & E                                                                                      &                                                                       &                                                                                         &                                                                                                  &                                                                                                  &                                                                                                  \\ \hline
         CH$_4$ + $h\nu$ $\longrightarrow$ CH$_3$ + H         &                                                                                               & CH$_4$                                                                                 &                                                                       &                                                                                         &                                                                                                  &                                                                                                  &                                                                                                  \\ \hline
         N$_2$+ $h\nu$ $\longrightarrow$ N$_2^+$ + $e^-$          &                                                                                               &                                                                                        & N$_2^+$                                                               &                                                                                         &                                                                                                  &                                                                                                  &                                                                                                  \\
         \hline                  
         \end{tabular}
   \end{center}     
   \caption{Reactions with $\lvert$ RCC $\rvert$ > 0.5 for at least one of the main atmospheric species at one of the seven considered altitude levels. Data from the sensitivity analyzes performed for all the atmospheric species at once are used. These analyzes are done for 250 iterations of the Monte-Carlo procedure. The main species for which the absolute value of the RCC of the reaction is higher than 0.5 at a given level are displayed. ME stands for "Magnetospheric Electrons".
   } 
   \label{RCC_main_6lvls_RCCsup0,5_250tirages}      
   \end{table}

\end{appendix}

\end{document}